\documentclass[12pt]{JHEP3}

\usepackage{amsmath}
\usepackage{amssymb} 

\usepackage{xspace}
\usepackage{epsfig}
\usepackage{graphics}
\usepackage{subfigure}

\usepackage{amsthm}
\theoremstyle{definition}

\newcommand{\color}[2][x]{}

\newcommand{\ie}{\emph{i.e.}\ }
\newcommand{\eg}{\emph{e.g.}\ }
\newcommand{\cnf}{\emph{cf.}\ }

\newcommand{\ptilde}{{\tilde p}}

\def\order#1{{\cal{O}}\left(#1\right)}

\newcommand{\Tr}{\mathop{\mathrm{Tr}}}

\newcommand{\cH}{{\cal H}}

\newcommand{\cN}{{\cal N}}
\newcommand{\hc}{\cal{H_C}}
\newcommand{\hr}{\cal{H_R}}

\newcommand{\jets}{\;\mathrm{jets}}

\newcommand{\vProb}{f}
\newcommand{\vhProb}{{\hat f}}
\newcommand{\momConf}{{\cal B}}

\newcommand{\subProc}{\delta}

\newcommand{\muf}{\mu_\textsc{f}}

\newcommand{\caesar}{\textsc{caesar}\xspace}

\def\qbar{{\bar q}}

\newcommand{\sbar}{{\bar s}}

\def\be{\beta}

\def\lam{\lambda}

\def\cF{{\cal{F}}}    
\def\cP{{\cal{P}}}
\def\cR{{\cal{R}}}

\def\cR{{\cal{R}}}               

\def\MSbar{\overline{\mbox{\scriptsize MS}}}

\def\CF{C_F}
\def\TR{T_R}
\def\CA{C_A}
\def\NC{N_c}
\def\nf{n_{\!f}}

\def\as{\alpha_{{\textsc{s}}}}
\def\gae{{\gamma_{\textsc{e}}}}
\def\asb{{\bar \alpha}_{{\textsc{s}}}}

\def\cO#1{{\cal{O}}\left(#1\right)}

\def\ee{e^+e^-}

\def\shat{\hat s}
\def\that{\hat t}

\title{Principles of general final-state resummation and automated
  implementation}

\author{Andrea Banfi\\
  NIKHEF Theory Group, P.O. Box 41882, 1009 DB Amsterdam, The
  Netherlands.\\
  Cavendish Laboratory,
  University of Cambridge, Madingley Road, Cambridge, CB3 0HE, UK.
}

\author{Gavin P. Salam\\
  LPTHE, Universities of Paris VI and VII and CNRS UMR 7589, Paris,
  France.}

\author{Giulia Zanderighi\\
  Fermilab, P.O. Box 500, Batavia, IL, US.}

\abstract{
  Next-to-leading logarithmic final-state resummed predictions have
  traditionally been calculated, manually, separately for each
  observable. In this article we derive NLL resummed results for
  generic observables. We highlight and discuss the conditions that
  the observable should satisfy for the approach to be valid, in
  particular continuous globalness and \emph{recursive} infrared and
  collinear safety. The resulting resummation formula is expressed in
  terms of certain well-defined characteristics of the observable.  We
  have written a computer program, \caesar, which, given a subroutine for an
  arbitrary observable, determines those characteristics, enabling
  full automation of a large class of final-state resummations, in a
  range of processes.
}

\keywords{QCD, NLO Computations, Jets, Hadronic Colliders}

\preprint
{
  Cavendish-HEP-05/08\\
  FERMILAB-PUB-04--116-T\\
  LPTHE--04--16\\
  NIKHEF/2004--005\\
  hep-ph/0407286 \\
  Revised version of May 2005
}

\begin{document}

\section{Introduction}
\label{sec:intro}
It is a well known feature of QCD, and gauge theories in general, that
final-state properties of the bulk of events in high-energy collisions
cannot be predicted by standard fixed-order perturbative calculations.
The very concept of `bulk', or `typical' events implies that in the
expression for their probability, each power of the formally small
coupling, $\as$, is compensated by a coefficient of order $1/\as$.
These large coefficients are generally associated with logarithms
($L$) of widely disparate scales in the problem, and fixed-order
truncations of the perturbative series often give unreliable answers.

So it is necessary to reorganise the perturbative series in terms of
sets of dominant logarithmically enhanced classes of terms, \ie a
class of leading logarithmic (LL) terms (which might for example go as
$\as^n L^{2n}$), next-to-leading logarithmic (NLL) terms (\eg $\as^n
L^{2n-1}$) and so on. For an appropriate range of (large) values of
the logarithm $L$, it can be shown that this \emph{resummed} hierarchy
is convergent,\footnote{Strictly it will be an asymptotic series whose
  first few orders converge.} \ie that NLL terms are truly smaller
than LL terms, and that next-to-next-to-leading logarithms (NNLL) are
smaller than NLL terms, etc.

Despite the considerable practical importance of resummed results, the
methods for making resummed final-state predictions suffer from
significant limitations. On one hand there exist purely analytical
approaches, such as \cite{CSS,CTTW,Bonciani:2003nt}, that give
state-of-the-art
accuracy, but which must be repeated manually for each new observable,
often requiring considerable understanding of the underlying physics,
as well as mathematical ingenuity. On the other hand, there are Monte
Carlo event generators, such as Herwig \cite{Herwig} or Pythia
\cite{Pythia}, whose predictions can be applied to any observable, but
without any formal guarantees as to the accuracy of the prediction,
other than leading double logarithms. Often, the accuracy will
actually be higher, but this can only be established given a detailed
understanding of the observable. Additionally, event generator
predictions are difficult to match with fixed-order results (though
progress is being made~\cite{FrixWeb}), and they are always
`contaminated' by non-perturbative corrections, even at parton level.

This situation is quite unsatisfactory, especially compared to that
for fixed-order predictions. There, one has access to a range of
programs (fixed-order Monte Carlos ---
FOMCs, \eg~\cite{CataniSeymour,Disaster,NLOJET,MCFM}) which, given a
subroutine that 
calculates an observable for arbitrary final-state configurations,
return the coefficients of the first few (currently two for most
processes) orders of the perturbative prediction for the observable. A
user wanting a prediction for some new observable can in this way
easily obtain it, without having to understand any of the subtleties
of higher-order calculations or real-virtual cancellations, all hidden
inside the FOMC.

The purpose of the current paper is to show how one can automate
resummed calculations of final-states, while maintaining the `quality'
associated with analytical resummations: guaranteed\footnote{Except in
certain pathological contrived cases, as discussed later.} state-of-the-art
accuracy (NLL, as discussed below), a purely
perturbative answer, clean separation of LL, NLL contributions without
spurious contamination from uncontrolled higher-orders, and the
ability to obtain the order-by-order expansion for comparison and
matching with fixed-order predictions.

These requirements imply a quite different approach compared to FOMCs
or event generators, in that the result will not simply be a weighted
average over return values from the computer routine for the
observable: to obtain `analytic' quality in the result, one needs to
know something about the analytical properties of the observable.  It
is up to the automated resummation program to establish those
properties, by probing the observable-subroutine with suitable
configurations, generally involving very soft and collinear emissions
--- high-precision computer arithmetic making it possible to take
nearly asymptotic limits.  Having established certain analytical
properties of the observable the program can then use Monte Carlo
methods over \emph{specifically chosen} sets of final states to
cleanly determine the remaining information needed for the
resummation.

One of the characteristics of such a program is that it may reach the
conclusion that the observable under consideration is outside the
class of supported observables.  While seemingly a limitation --- it
implies that the program cannot resum all observables --- it is
actually an essential feature, since it is only for certain classes of
observable that we have a good understanding of the approximations
that are legitimate when seeking a given accuracy.\footnote{One could
  also envisage using such an approach to establish the accuracy that
  will be achieved for a given observable when using normal event
  generators such as Herwig \cite{Herwig} or Pythia \cite{Pythia}. For
  example, specifically for two-jet events, our understanding is that
  Herwig, which uses a two-loop, CMW scheme \cite{CMW} running
  coupling, and exact angular ordering, should implicitly contain\ the
  full NLL resummed result for all global, exponentiating observables,
  though it is also accompanied by unavoidable (potentially spurious)
  subleading and non-perturbative contributions.}

Let us now examine in more detail the problem that we treat.

\subsection{Problem specification}
\label{sec:problem}

We consider an observable $V(q_1,q_2,\ldots)$, some non-negative
function of the momenta $q_1,q_2,\ldots$ in the final state. We assume
that it is infrared and collinear safe, and, furthermore, that there
is some number $n$ (we will explicitly discuss $2\le n \le 4$) such
that the observable goes smoothly to zero for momentum configurations
that approach the limit of $n$ narrow jets. We call this an
($n+1$)-jet observable.  Any incoming beam jets ($n_i$ of them), as
well as the outgoing jets, are included in this counting.

We start by introducing a procedure that selects events with $n$ or
more hard jets. This could, for instance, be through a jet algorithm
that counts the number of well separated hard jets in the event,
or through a cut on some secondary, $n$-jet, observable.\footnote{For
  example, if one wishes to resum the thrust minor, a $4$-jet
  observable in $\ee$, possible ways of selecting $3$-jet events would
  be to use a jet algorithm, or to place a cut on
  some $3$-jet observable such as the thrust.} %
This selection procedure is expressed mathematically in terms of a
function $\cH(q_1,q_2,\ldots)$ that is $1$ for events that pass the
selection cuts, and zero otherwise. This allows us to define a hard
$n$-jet cross section,
\begin{equation}
  \label{eq:sigmacut}
  \sigma_\cH = \sum_{N=n-n_i}^\infty \int d\Phi_N\,
  \frac{d\sigma_N}{d\Phi_N}\,
  \cH(q_1,\ldots,q_N)\,,
\end{equation}
where $d\sigma_N/ d\Phi_N$ is the differential cross section for
producing $N$ final-state particles.

We consider the integrated cross section, $\Sigma_\cH(v)$, for events
satisfying the hard $n$-jet cut, $\cH$, and for which, additionally,
the observable is smaller than some value $v$,
\begin{equation}
  \label{eq:SigmaIntcut}
  \Sigma_\cH(v) = \sum_{N} \int d\Phi_N
  \frac{d\sigma_N}{d\Phi_N} \,\Theta(v - V(q_1,\ldots,q_N))\,
  \cH(q_1,\ldots,q_N)\,,
\end{equation}
from which one can obtain $(1/\sigma_{\cH})d \Sigma_\cH(v)/dv$, the
differential distribution for the observable.

It is convenient to rewrite eq.~(\ref{eq:SigmaIntcut}) in a factorised
form
\begin{equation}
  \label{eq:Sigmacut_resummed}
  \Sigma_\cH(v) = \sum_{\subProc} \int d\momConf\,
  \frac{d\sigma_\subProc}{d\momConf} \,\vProb_{\momConf,\subProc}(v)\,
  \cH(p_{n_i+1},\ldots,p_{n})\,,
\end{equation}
involving, on one hand, the leading order differential cross section,
$d\sigma_\subProc/d\momConf$, for producing a `Born' event,
$\momConf$, that consists of $n-n_i$ outgoing hard momenta
$p_{n_i+1},\ldots p_{n}$ in a given scattering channel $\subProc$ (for
example $qq \to qq$ or $qg \to qg$); and on the other hand an
`observable-dependent' function $f_{\momConf,\subProc}(v)$, which can
roughly be understood as representing the fraction of events, for the
given subprocess and Born configuration, for which the observable is
smaller than $v$. Note, however, that this fraction is normalised to
the leading-order Born differential cross section.

One can write $\Sigma_\cH(v)$ in the form
eq.~(\ref{eq:Sigmacut_resummed}) for any value of $v$. However the
factorisation property of eq.~(\ref{eq:Sigmacut_resummed}), namely
that $\vProb_{\momConf,\subProc}(v)$ is independent of the procedure
used to select $n$-jet events, holds only in the limit of small $v$
and for global observables (those affected by radiation in any
direction \cite{NG1}). %
It is a consequence of the factorisation properties of soft and
collinear radiation, and of our choice to normalise
$\vProb_{\momConf,\subProc}(v)$ to the leading order Born cross
section.
In contrast, for $v\sim1$ the factorisation, understood in this
manner, is in general not
possible: $\vProb_{\momConf,\subProc}(v)$ depends implicitly also
on the behaviour of $\cH$ for final states with arbitrarily large
numbers of partons. This is related to the fact that there is no unique
prescription for mapping an arbitrary number of hard momenta onto
a $n-n_i$ parton structure.

\subsection{Structure of result and nature of approach}

For all ($n+1$)-jet global observables that have so far been resummed
in the $n$-jet limit \cite{CTTW,thr_res,mh_ee, cpar_res, CTWbroad,
  DLMSBroadPT, y3-kt_ee, CatDokWeb, JetsDisSchmell, BSZ, BKS03,
  ADS,DSBroad,eeKout,KoutZ0,KoutDIS,
  AzimDIS,GRthrust,GRmass,GardMan,BergerMagnea}, %
$f_{\momConf,\subProc}(v)$ has been found to have the property that,
for small $v$, it can be written (dropping the $\momConf$ and
$\subProc$ indexes, for compactness) \cite{CTTW},
\begin{equation}
  \label{eq:vProb-general}
  \vProb(v) \simeq (1 + C(\as))
  \exp\left[ L g_1(\as L) + g_2(\as L) 
  + \as g_3(\as L) + \cdots\right]\,, \quad L = \ln \frac1v\,,
\end{equation}
to within corrections usually\footnote{See footnote~\ref{foot:extra},
  p.~\pageref{foot:extra}.}
suppressed by powers of $v$. The function 
$Lg_1(\as L)$ resums Sudakov leading (or `double') logarithms in the
exponent, $\as^n L^{n+1}$; $g_2(\as L)$ resums next-to-leading (or
`single') logarithms in the exponent, $\as^n L^{n}$; and so forth. The
term $C(\as)$ has the expansion $C_1 \as/2\pi + C_2
(\as/2\pi)^2 + \ldots$, where the $C_n$ are constants.

It is non-trivial that $\vProb(v)$ should have an `exponentiated' form
such as eq.~(\ref{eq:vProb-general}), since its expansion contains
terms with much stronger logarithmic dependence $\as^n L^{2n}$, $\as^n
L^{2n-1}$, etc., than is present in the exponent. All of these
strongly logarithmically enhanced terms should be consistent with the
exponential form.\footnote{Sometimes confusion arises as to whether
  one defines the logarithmic accuracy for the expansion or the
  exponent. Here we shall always refer to the accuracy in the
  exponent.} %
Certain observables, notably JADE jet-resolution thresholds
\cite{JADE}, for which the first logarithmically enhanced terms
have been calculated \cite{JadeDL,CataniEtAlJets,Leder}, have been
explicitly found to be inconsistent with exponentiation. So far no
observable of this kind has been resummed, even at LL accuracy.

Here, rather than attempting to resum some given specific
observable, we will consider (in section~\ref{sec:masterderiv}) the
derivation of the final-state resummation for a \emph{generic}
continuously-global \cite{NG1,DiscontGlobal} observable. We find it
helpful to enter into somewhat more detail than is usually provided
for observable-specific resummations (nowadays quite standard),
because it allows us to isolate the characteristics of the
observable that are necessary so as to arrive at the form
eq.~(\ref{eq:vProb-general}).

The main new condition that emerges from this derivation is one that
we call \emph{recursive} infrared and collinear (rIRC) safety,
eqs.~(\ref{eq:rIRClimit1},\ref{eq:rIRClimit2}), because it involves
two nested, ordered, infrared and collinear limits. It essentially
states that when there are emissions on multiple widely separated
scales, it should always be possible to remove all but the hardest
emissions without affecting the value of the observable.\footnote{If
  this sounds suspiciously like normal infrared
  collinear safety, then (a) think hard and (b) read on!} %
It is sufficient in order to guarantee, up to NLL accuracy (and
beyond, we believe), that the resummed result will be of the form
eq.~(\ref{eq:vProb-general}).

Given rIRC safety, the resummed result is given by a master formula,
eq.~(\ref{eq:Master}), where the LL and NLL terms, $g_1(\as L)$ and
$g_2(\as L)$, are expressed in terms of a variety of well-identifiable
characteristics of the observable.  For example the LL contribution,
as well as part of the NLL contribution, are just related to the
manner in which the observable scales as one takes a single emission
and makes it soft and/or collinear, eq.~(\ref{eq:simple-second-time}).
The remaining part of the NLL contribution depends instead on the
value of the observable when multiple emissions are simultaneously
made soft and collinear. It is obtained by integrating over a
suitable subset of such configurations,
eq.~(\ref{eq:cF-with-vto0-limit-anyn}).

The strength of this approach is that the relevant characteristics of
the observable are sufficiently well-defined that they can be
determined numerically given just a subroutine for the observable.
Some general features of the computer program that we have written to
carry out the procedure, the `Computer Automated Expert
Semi-Analytical Resummer' (\caesar) are described in
section~\ref{sec:caesar}. It will be made publicly available in the
coming future. It makes use of high-precision arithmetic \cite{MP} to
reliably take infrared and collinear limits, and behaves in a manner
somewhat reminiscent of an expert system, insofar as it poses (and
answers) a set of questions about the observable, so as to
establish the suitability of the observable for resummation, and
determine the best strategies for the numerical integrations that are
to be carried out. Thus new observables can be resummed without a user
having any resummation expertise.

One should be aware that not all observables are suited to this
approach. For example, recursively IRC unsafe observables cannot be
dealt with, and 
often lead to a result for $g_2(\as L)$ that is divergent
logarithmically in an infrared regulator, much as occurs for NLO
coefficients with (plain) IRC unsafe observables. One of the main
characteristics of \caesar is that it establishes whether an
observable is within its scope.

There also exist observables that are rIRC safe, but for which
$g_2(\as L)$ diverges above some fixed value of $\as L$. This is akin
to divergences of fixed-order coefficients that can occur close to
specific kinematic boundaries, and is a sign of a need for further
resummation. In our case the problem arises for observables whose
value can be small due to cancellations between contributions from
different emissions, and it can in some situations be resolved with a 
transform-based general approach such as \cite{Bonciani:2003nt}. It
often occurs \cite{DSBroad} that such divergences are in a
sufficiently suppressed region that they can in practice be ignored.

Despite the existence of these partial limitations, the method is
suitable for a wide variety of observables, reproducing existing
results and having already produced a number of new predictions. In the
form discussed here, it is suitable for $\ee \to 2\jets$, $\ee \to
3\jets$, DIS $1+1 \jets $ and $2+1 \jets$, hadron-hadron $1+2\jets$
with an additional hard boson ($\gamma, W^\pm, Z^{0}, H$, not all
implemented numerically yet) and hadron-hadron $2+2\jets$, the latter
involving also the 
colour-evolution soft anomalous dimension matrices of
\cite{BottsSterman,KS,KOS,Oderda,KidonakisOwens}.

A companion paper \cite{BSZhh}, which discusses a range of possible
continuously global event shapes for hadron-hadron dijet events,
provides an illustration of the power of the method, insofar as all
resummed results presented there have been obtained with \caesar. Some
results for continuous classes of $\ee$ observables, such as those of
\cite{BKS03,BS03} are also discussed here, in
appendix~\ref{sec:ee-observables}.

\subsection{Guide to reading the article}
\label{sec:guide}

The table of contents provides an overview of the different sections
in this paper. In view of the length of the paper however we provide
here also some guidance for readers wishing to concentrate on certain
specific issues.

For a reader not too familiar with resummations and interested in
understanding the physical principles behind the approach, or one who
wishes to study in detail the assumptions that are being made in the
derivation of the master resummation formula,
section~\ref{sec:masterderiv} should be read first. Accompanying
material is given in appendices~\ref{sec:highorder-multemsn} and
\ref{sec:hardcollinear}. 

In any case we recommend that at some stage the reader take a look at
section~\ref{sec:summary-master}, which contains the main analytical
results and applicability conditions for a general resummation.  In
the event that this appears too abstract, section~\ref{sec:thrust}
provides a detailed worked example, within our approach, of the
canonical event shape resummation, that for the $\ee$ thrust. 

The question of how to translate the analytical results into a
computer automated approach is the subject of
section~\ref{sec:caesar}. An overview of the implementation is given
as a flowchart, figure~\ref{fig:flowchart}, while the text discusses a
combination of general and more technical issues that arise in
practice.  For readers interested in the details, or in implementing
the approach themselves, explicit formulae are given in
appendices~\ref{sec:analytics} and \ref{sec:softlargeangle},
including, for completeness, a number of expressions that exist
already in the literature. Important subtleties that arise for the
consistent insertion of multiple emissions are discussed
appendix~\ref{sec:recoil}.

For readers interested especially in certain specific physics issues,
we recommend a more transversal reading. This is especially the case
for recursive IRC safety, whose origins are to be found in
section~\ref{sec:allorders}. Section~\ref{sec:rIRC-deriv} is designed
to bridge between the conditions of recursive IRC safety as they
naturally arise in the derivation of the master formula and its
central definition presented in section~\ref{sec:summary-master}.  An
intuitive understanding of the rIRC conditions may be helped by a number
of examples, in section~\ref{sec:rIRC} and appendix
\ref{sec:further-rIRC-unsafe}, of rIRC safe observables that are rIRC
unsafe, while numerical tests of rIRC safety are discussed in
section~\ref{sec:rIRCtests}.
Of related interest is appendix~\ref{sec:IRC}, which discusses the
difficulties in finding a mathematically rigorous definition of normal
IRC safety.

The NLL term in the resummation, $\cF$, that accounts for the
observable's sensitivity to multiple emissions is also discussed at
various points in the paper. The initial derivation is in
section~\ref{sec:mult-indep-emsn}, while two final forms for it are
given in the master-formula, section \ref{sec:summary-master}. A
number of issues arise in its general practical determination, as
presented in section~\ref{sec:efficiency_for_cF}.

A number of more specialised issues arise for observables whose $\cF$
diverges at finite values of $\as L$.  The origin of the problem is
reviewed in sections~\ref{sec:div} and \ref{sec:gen-div-cF}, together
with a discussion of the
location of potential divergences.  The question of divergences is of
interest also from the point of view of the practical implementation in
\caesar, because of numerical convergence issues that arise when a
divergence is present. This has led to our developing various
techniques to probe the cancellations that lead to the divergences in
the first place and semi-analytical integration methods to improve the
Monte Carlo convergence. These issues are discussed in
appendices~\ref{sec:analysis-divergent-cF} and \ref{app:MCDetails}.

As we have already mentioned, readers interested in applications of
the method should consult the companion paper \cite{BSZhh} for
examples in hadronic dijet events, as well as
appendix~\ref{sec:ee-observables} for a discussion of two continuous
classes (one proposed in \cite{BKS03}, the other new) of $\ee$
observables.

Finally, we invite the reader to consult the web site
\cite{qcd-caesar.org}, which contains a range of extra
resources, including results from automated analyses of a large number
of observables in a range of processes, far more than could reasonably
be discussed here and in \cite{BSZhh}.

\section{Derivation of master resummation formula}
\label{sec:masterderiv}

The master formula that we shall here derive was originally presented
in \cite{BSZ03}. Numerous considerations enter into its derivation.
First we will examine a little more closely the general problem that
we wish to solve; we 
will then show how to obtain the solution in a simple case,
progressively introducing the elements needed to obtain the final
general result. 

We consider a
hard event consisting of $n$ hard partons, all massless, having
four-momenta $p_1$, \ldots, $p_n$. 
We shall call this our `Born' event and each of the hard Born partons
will be referred to as `legs'. For brevity we will use $\{ p\}$ to
denote the set of all the Born momenta.  An index $\ell$ will be used
when we refer to a particular leg.

Given such a system, we shall consider an observable (or variable)
$V$, which is a function of the momenta in the event.  The observable
should be positive definite and vanish for the Born event, $V(\{ p\}) =
0$.  Furthermore it should give a continuous measure of the extent to
which the energy-momentum flow in the event differs from that of the
Born event, or equivalently a measure of the departure from the
$n$-jet limit.

Observables of this kind, such as event-shapes and jet-resolution
parameters, usually have the property that in the presence of a single
emission $k$ that is soft and collinear to a leg $\ell$, the value of
the observable can be parametrised as
\begin{equation}
  \label{eq:parametricform}
  V(\{{\tilde p}\}, k)=
  d_{\ell}\left(\frac{k_t^{(\ell)}}{Q}\right)^{a_\ell}
  e^{-b_\ell\eta^{(\ell)}}\, 
  g_\ell(\phi^{(\ell)})\>.
\end{equation}
The $\{{\tilde p}\}$ denote the Born momenta after recoil from the
emission $k$; $Q$ is what we shall call the hard scale of the problem,
though in practice there may not be a unique way of defining it.
The observable's dependence on the momentum $k$ is expressed in terms
of $k_t^{(\ell)}$, $\eta^{(\ell)}$ and $\phi^{(\ell)}$, respectively
the transverse momentum, rapidity and azimuthal angle of the emission,
as measured with respect to the hard leg $\ell$. To fully specify the
azimuthal angle (where relevant) one needs additionally to define a
suitable reference plane, for example that containing $p_\ell$ and
some second (non-parallel) leg.

The precise parametric dependence of the observable on the momentum
$k$ is specified through the values of the coefficients $a_\ell$,
$b_\ell$ and the combination $d_\ell g_\ell(\phi^{(\ell)})$.  For
example for the thrust $T$ in $\ee \to 2$~jets~\cite{ThrustDef}, one
has \cite{CTTW}
\begin{equation}\label{eq:thrustexample}
\tau = 1-T\,,\qquad\quad 
  \tau(\{{\tilde p}\}, k) = \frac{k_t^{(\ell)}}{Q}
  e^{-\eta^{(\ell)}},
  \end{equation}
giving $a_\ell = b_\ell = d_\ell = g_\ell(\phi) = 1$, for $\ell =
1,2$.  Though the dependence on $d_\ell$ and $g_\ell(\phi)$ arises
only through the product $d_\ell g_\ell(\phi)$, we will find it
convenient to give a standard normalisation to the $g_\ell(\phi)$,
such as $g_\ell(\pi/2) = 1$, leaving the observable-dependent
normalisation in $d_\ell$.

The form (\ref{eq:parametricform}) is sufficiently
common~\cite{thr_res, mh_ee,cpar_res,CTWbroad,DLMSBroadPT,
  y3-kt_ee,CatDokWeb,BSZ,BKS03,ADS, DSBroad,eeKout,KoutZ0,
  KoutDIS,AzimDIS,DiscontGlobal,ml_ee,FoxWolfram,JetratesHH}
that we can safely make it a prerequisite of our approach without
unduly losing in generality.

Note that the coefficients $a_\ell$, $b_\ell$, $d_\ell$ and the
function $g_\ell$ can depend on the Born configuration under
consideration, \ie they may be a function of the $\{p\}$. Here we
shall carry out our analysis for a specific Born configuration, and
leave to section~\ref{sec:Born-integrate} the discussion of how to
integrate over the Born configurations.

Knowledge of the above coefficients for each leg is of course not
sufficient to fully specify the observable's dependence on a single
emission, since eq.~(\ref{eq:parametricform}) is relevant only to the
limit of a soft \emph{and} collinear emission (a LL, or double
logarithmic region). One may legitimately worry that for a NLL (single
logarithmic) resummation one might also need some information on the
large-angle soft limit or on the hard collinear limit. We shall return
to this issue in a while.

\subsection[Single-emission results (${q\bar q}$ case)]{Single-emission results ($\boldsymbol{q\bar q}$ case)}
\label{sec:single-emission}

Having parametrised the observable's dependence on a single emission,
let us now examine how that information can be used to determine the
logarithmic structure of a first order calculation --- this is a
convenient first step on the way to a full resummation. We will
initially consider the simple case of a colour-singlet quark-antiquark
system, but 
with the feature that the quark ($p_1$) and anti-quark ($p_2$), both
outgoing, are not 
necessarily back-to-back, nor of the same energy. This will make it
easier to generalise the answer subsequently.

\subsubsection{Single-gluon emission pattern}
\label{sec:singlegluonemission}

Let us decompose the momentum of the emitted gluon $k$ into its
Sudakov components:
\begin{equation}
  \label{eq:SudakovForK}
  k = z^{(1)} p_1 + z^{(2)} p_2 + k_t \cos \phi\, n_\mathrm{in} + k_t \sin
  \phi\, n_\mathrm{out}\,,
\end{equation}
where $n_\mathrm{in}$ and $n_\mathrm{out}$ are space-like unit
vectors, orthogonal to $p_1$ and $p_2$ and whose vector components are
respectively in and perpendicular to the $\vec p_1$-$\vec p_2$ plane,
\begin{equation}
  n_\mathrm{in} = \left(\cot \frac{\theta_{12}}{2}\,;\, \frac{1}{\sin
      \theta_{12}}\left(\frac{\vec p_1}{E_1} + \frac{\vec
        p_2}{E_2}\right)
  \right)\,,\qquad\quad
  n_\mathrm{out} = \left(0\,;\, \frac{\vec p_1 \times \vec p_2}{E_1
      E_2 \sin \theta_{12}}\right)\,, 
\end{equation}
where $p_{\ell}= (E_{\ell}; \vec p_{\ell})$,
and $\theta_{ab}$ is the angle between momenta $a$ and $b$.
The condition that the emission be massless implies
$k_t^2 = z^{(1)} z^{(2)} Q_{12}^2$, where $Q_{12}^2$ is the invariant squared
mass of the $q\bar q$ dipole, $Q_{12}^2 = 2p_1.p_2$; $k_t$ is the
relativistically invariant transverse momentum of the emission with
respect to the dipole,
\begin{equation}
  \label{eq:ktinv}
  k_t^2 = \frac{(2k.p_1)(2k.p_2)}{(2p_1.p_2)}\,.
\end{equation}
Note that for an emission sufficiently collinear to leg $1$, the
invariant transverse momentum $k_t$, and azimuthal angle $\phi$,
coincide with those defined relative to leg $1$, $k_t^{(1)}$ and
$\phi^{(1)}$, that appear in eq.~(\ref{eq:parametricform}). This holds
as long as $\tan \frac{\theta_{1k}}2 \ll \tan \frac{\theta_{12}}{2}$.
Furthermore, in this region the emission's rapidity with respect to
leg $1$ is,
\begin{equation}
  \label{eq:etacoll}
  \eta^{(1)} = \ln \frac{2z^{(1)} E_1}{k_t} 
  = \eta + \ln \frac{2E_1}{Q_{12}}\,,\qquad\quad \eta = \frac{1}{2}\ln
  \frac{z^{(1)}}{z^{(2)}}\,,
\end{equation}
where $\eta$ is the rapidity of the emission in the dipole
centre-of-mass system.  Analogous statements hold for emissions
collinear to leg $2$.

To calculate the distribution for the observable in the one-gluon
approximation, one also needs the matrix element for the emission of a
single gluon that is soft or collinear to either of the hard legs. Let
us first recall its form (see \eg \cite{ESWbook}) for collinear gluon
emission with respect to leg $\ell$,\footnote{Subtleties arise in
  specifying the matrix element and phase space, insofar as our
  definition of the gluon momentum, eq.~(\ref{eq:SudakovForK}), does
  not uniquely specify the final state, notably in the hard collinear
  limit --- to do so requires additionally that one give a
  prescription for the relation between the Born momenta before
  ($\{p\}$) and after emission ($\{\tilde p\}$). As discussed in
  appendix~\ref{sec:recoil}, for a single emission, the details of the
  prescription are however irrelevant at our accuracy.}
\begin{equation}
  \label{eq:matrixelement_ell}
  |M^2_\ell(k)| = \frac{\as C_F}{2\pi}\, \frac{ z^{(\ell)}
    p_{gq}(z^{(\ell)}) }{k_t^2}\,, 
\end{equation}
where $p_{gq}$ is the quark to gluon splitting
function (with colour
factors removed), $z p_{gq}(z) = 1 + (1-z)^2$. A factor of $16\pi^2$
has been extracted from the matrix element and is included instead in
the phase space for integration $[dk]$, which can be written as
\begin{equation}
  \label{eq:phasespace}
  [dk] = \frac{dz^{(1)}}{z^{(1)}} \frac{d\phi}{2\pi} dk_t^2\,.
\end{equation}
The generalisation of eq.~(\ref{eq:matrixelement_ell}), so that it is
valid for any soft and/or collinear emission from the $12$ dipole, is
\begin{equation}
  \label{eq:matrixelement}
  |M^2(k)| = \frac{\as C_F}{4\pi}\, \frac{z^{(1)}
    p_{gq}(z^{(1)})\,\cdot\, z^{(2)} 
    p_{gq}(z^{(2)}) }{k_t^2}\,.
\end{equation}

With this notation, the first-order expression for the fraction of events,
$\vProb(v)$, for which the final-state observable $V$ is smaller than a
given value $v$ is:
\begin{subequations}
\begin{align}
  \label{eq:vProbOrderAlphaFirstGo1}
  \vProb(v) &= 1 + \int [dk] \,|M^2(k)|\, \big(\Theta(v - V(\{\tilde p\},
  k)) - 1\big)\,,\\
  \label{eq:vProbOrderAlphaFirstGo2}
  & = 1 - \int [dk] \,|M^2(k)|\, \Theta(V(\{\tilde p\}, k) - v)\,.
\end{align}
\end{subequations}
In the upper line, the first term in the bracket corresponds to
the real emission of a gluon, which contributes to $\vProb(v)$ only if
$V(\{\tilde p\}, k)$ is smaller than $v$. The second term represents the
order $\as$ virtual contribution, whose matrix-element is identical
(modulo the sign) to that for the real emission, because of unitarity.
Since virtual contributions do not affect the value of the observable,
this term contributes over the whole integration region.

\subsubsection{Further requirements on the observable}
\label{sec:further-requirements}

\FIGURE{
  \includegraphics[width=0.95\textwidth]{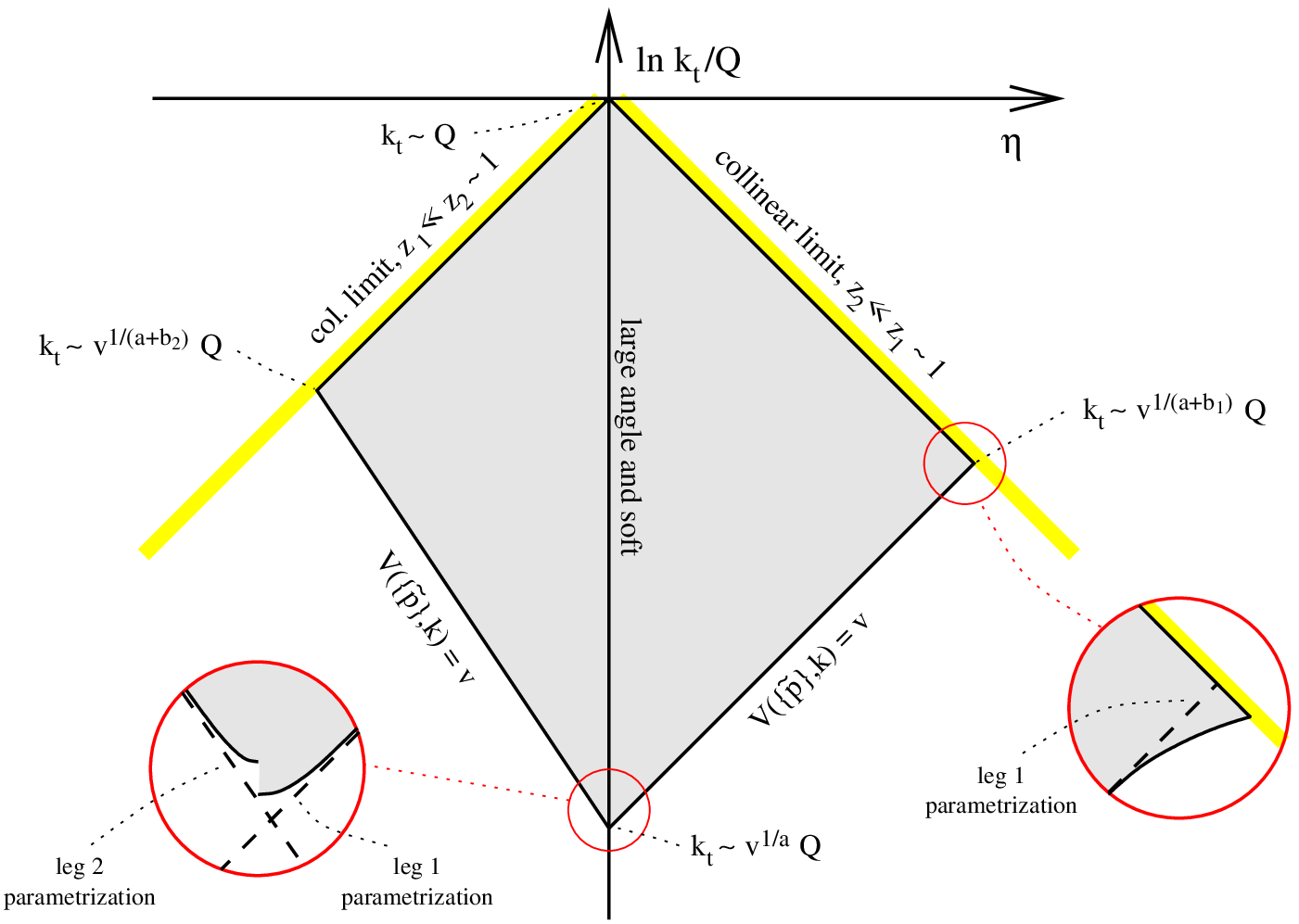}
  \caption{The $\eta$--$\ln (k_t/Q)$ plane for a single emission,
    together with a representation (shaded area) of the region in
    $k_t$ and $\eta$ over which the integrand of
    eq.~(\ref{eq:vProbOrderAlphaFirstGo2}) is non zero. The specific
    positions of the lines correspond to the case of an observable
    with $a_1 = a_2 \equiv a = 1$ and $b_1 = 1$, $b_2 = 3/2$. For
    simplicity, the $\phi$-dependence of the problem has been
    neglected.  The insets correspond to a magnification by a factor
    of order $\ln 1/v$. Further details are given in the text. }
  \label{fig:lozenge}
}

To help us consider the issues that arise in the evaluation of
eq.~(\ref{eq:vProbOrderAlphaFirstGo2}), figure~\ref{fig:lozenge} shows
in the $\eta$--$\ln (k_t/Q)$ plane, the region (shaded area) in which the
integrand of eq.~(\ref{eq:vProbOrderAlphaFirstGo2}) is non-zero, for
some value of $v$. This region is delimited by two kinds of
boundaries.  Firstly, there are kinematic boundaries associated with
the requirements $z^{(1)} < 1$ and $z^{(2)} < 1$. These give the upper edges
of the shaded region.  Secondly there are boundaries at $V(\{\tilde
p\}, k) = v$ associated with the $\Theta$-function in
eq.~(\ref{eq:vProbOrderAlphaFirstGo2}).

The intersections of the various boundaries set the characteristic
scales (transverse momenta) of the problem. Firstly, the scale at the
point where the two hard boundaries meet is of the order of the hard
scale, $Q$, of the problem. In this corner, $z^{(1)}\sim z^{(2)} \sim 1$,
eq.~(\ref{eq:matrixelement}) is a poor approximation to the true real
and virtual matrix elements. But the region $z^{(1)}\sim z^{(2)} \sim1$
contributes at most at $\cO{\as(Q)}$ (without logarithmic enhancements) to the
integral, so the `error' is NNLL and can accordingly be neglected.

Another scale arises, for each leg $\ell$, from the intersection
between the kinematic boundary and the $\Theta$-function boundary,
\ie 
the left and right-hand corners of the shaded region. If one makes the
assumption that one can extend the soft and collinear
parametrisation~(\ref{eq:parametricform}) into the hard collinear
region, then one finds, using eq.~(\ref{eq:etacoll}), that for a given
fixed $z^{(\ell)}$, the observable scales as $k_t^{a_\ell + b_\ell}$.  The
scales associated with the lateral corners of the shaded region are
then
\begin{equation}
  \label{eq:Kin+Thet}
  k_t \sim v^{1/(a_\ell + b_\ell)} Q\,.
\end{equation}
In practice, in the hard collinear region, the observable $V(\{\tilde
p\}, k)$ may depart from its soft and collinear parametrisation
(\ref{eq:parametricform}). Such a situation is illustrated in the
right-hand inset of fig.~\ref{fig:lozenge}, which represents the true
boundary of the shaded region (solid line), $V(\{\tilde p\}, k) = v$,
and the boundary that would be obtained based on the soft-collinear
parametrised form for $V$ (dashed line). As long as the difference
between the true form of the observable and the parametrisation is
just a non-zero $z^{(\ell)}$-dependent factor of order $1$, then
eq.~(\ref{eq:Kin+Thet}) remains valid.  Furthermore, when evaluating
eq.~(\ref{eq:vProbOrderAlphaFirstGo2}), replacing the true observable
$V(\{\tilde p\}, k)$ with its parametrised form leads to a difference
of order $\as$, which is a NNLL correction.\footnote{Strictly
  speaking, for this to be true, one needs also to ensure that the
  difference compared to the parametrisation is truly limited to the
  collinear region. Defining $\xi(z)$ as ratio of the true value of
  the observable to its parametrisation, this requirement can be
  expressed by saying that $\int_0^1 \frac{dz}{z} \ln \xi(z)$ should
  be finite. \label{foot:extra} In addition, if
  eq.~(\ref{eq:vProb-general}) is to hold to within corrections
  suppressed by powers of $v$, $\ln \xi(z)$ should vanish, for small $z$,
  at least as fast as a positive power of $z$.  }

From a practical (numerical) point of view, it is rather difficult to
establish whether a departure from the parametrised form is of order
$1$. However the condition can be formulated equivalently by requiring
that for collinear emissions, almost everywhere, $V$ be non-zero and
that
\begin{equation}
  \label{eq:HardCollRequirement}
  \left. \frac {\partial \ln V(\{\tilde p\}, k)}{\partial \ln k_t^{(\ell)}} 
  \right|_{\mathrm{fixed}\;z^{(\ell)},\,\phi^{(\ell)}} \!\!\!\!=
  \;\;a_\ell + b_\ell\,. 
\end{equation}
Here the expression `almost everywhere' should be taken in its usual
mathematical sense of everywhere except possibly a region of zero
measure. An important point about eq.~(\ref{eq:HardCollRequirement})
concerns collinear safety: the observable must vanish as $k_t$ is
taken to zero. Accordingly we have the condition $a_\ell + b_\ell >
0$.  A similar condition has been noted also in \cite{BKS03}.

As a final source of characteristic scales of the problem, we have the
intersection between the $\Theta$-function in
eq.~(\ref{eq:vProbOrderAlphaFirstGo2}) and the large-angle boundary
between the hard legs. Let us temporarily assume that we can extend
the soft and collinear parametrisation to the soft large-angle region.
Then for leg $\ell$ the characteristic scale that emerges is
\begin{equation}
  \label{eq:largeanglescale}
  k_t \sim v^{1/a_\ell} Q\,.
\end{equation}
We immediately see that a problem will arise if $a_1 \neq a_2$: the
knowledge that we have so far gathered about the observable does not
tell us where, in $\eta$, the transition occurs between the parametrised
forms for the different legs. This ambiguity corresponds to a single
logarithmic integration from $k_t \sim v^{1/a_1} Q$ to $k_t \sim
v^{1/a_2} Q$ over an unknown region of angle. Since the boundary
between the legs may be determined by some potentially quite complex
procedure, such as a jet algorithm, in a first instance it is
preferable not to require any understanding of it.

One partial solution to this problem is to consider only observables
for which $a_1 = a_2$. This ensures that the ambiguity in the boundary
between the two jets leads at most to an uncertainty in
eq.~(\ref{eq:vProbOrderAlphaFirstGo2}) of order $\alpha_s$ (NNLL).
Fig.~\ref{fig:lozenge} illustrates this in the left-hand inset, in a
case where additionally the true behaviour of the observable (solid lines)
does not exactly follow the parameterisations (dashed lines). As long
as this deviation from the parametrisation is by a factor of order
$1$, in a limited region in angle, then it too will only affect
eq.~(\ref{eq:vProbOrderAlphaFirstGo2}) by a NNLL
correction.\footnote{The precise requirements are analogous to those
  for deviations in the hard collinear region,
  footnote~\ref{foot:extra}, with $z$ replaced by $e^{-\eta}$.} %
Technically, it is most convenient to formulate the requirement as
being that, for soft emissions, almost everywhere, $V$ should be
non-zero and that
\begin{equation}
  \label{eq:SoftLARequirement}
  \left. \frac {\partial \ln V(\{\tilde p\}, k)}{\partial \ln k_t^{(\ell)}} 
  \right|_{\mathrm{fixed}\;\eta^{(\ell)},\,\phi^{(\ell)}} \!\!\!\!\equiv
  \;a \; = \; a_1\; =\; a_2\,. 
\end{equation}
This coincides with the condition for continuous globalness
\cite{NG1,DiscontGlobal},
and ensures, at higher orders, the absence also of so-called
non-global logarithms. Finally, we note that infrared safety implies
$a > 0$.

\subsubsection{Evaluation of single-emission integrals}
\label{sec:single-gluon-eval}

Given the extra requirements on the observable,
eqs.~(\ref{eq:HardCollRequirement}) and (\ref{eq:SoftLARequirement}),
we are now in a position to carry out the integrations of
eq.~(\ref{eq:vProbOrderAlphaFirstGo2}), replacing $V(\{\tilde p\}, k)$
with its parametrised form, eq.~(\ref{eq:parametricform}). As a
shorthand, we introduce $R(v)$, (minus) the single-gluon contribution to
$\vProb$,
\begin{equation}
  \label{eq:sigma1_def}
  R(v) = \int [dk] |M^2(k)| \Theta(V(\{\tilde p\}, k) - v)\,,
\end{equation}
which can be written as
\begin{multline}
  \label{eq:sigma1_explicit}
  R(v) = \sum_{\ell=1}^2 C_F  \int^{Q^2} \frac{dk_t^2}{k_t^2} 
    \int d\eta \, \frac{d\phi}{2\pi}\; \frac{\as(k_t^2)}{2\pi} \;
     z^{(\ell)} p_{gq}(z^{(\ell)}) \; \times \\ \;\times
    \Theta(\eta)\,
    \Theta(1-z^{(\ell)}) \,\Theta\left(v - d_\ell \left(\frac{k_t}{Q}\right)^a
        e^{-b_{\ell} \eta^{(\ell)}} g_\ell(\phi)\right)\,.
\end{multline}
We recall that the relations between $z^{(\ell)}$, $\eta$ and
$\eta^{(\ell)}$ were given in section~\ref{sec:singlegluonemission}.
Only one splitting function, $p_{gq}(z^{(\ell)})$, appears because the
splitting function from the other leg has a very small argument and
one can replace $zp_{gq}(z)=2$. The separation between the two legs
has been arbitrarily placed at $\eta=0$.

We note the introduction of the scale $k_t^2$ for the coupling: though
the scale of the coupling has no relevance at first order, it is
useful to keep track of it in anticipation of what follows later.

For observables with $b_\ell\ne0$, the $k_t$ integration in
eq.~(\ref{eq:sigma1_explicit}) can be separated into two parts,
according to whether the upper limit on $\eta$ stems from the
$\Theta$-function of $1-z^{(\ell)}$, or from that associated with the
observable. The boundary between the two regions occurs for $k_t \sim
Q v^{\frac{1}{a + b_\ell}}$. We perform the $\eta$ integration
separately in each of the two regions and write
\begin{multline}
  \label{eq:sigma1_etadone}
  R(v) = \sum_{\ell=1}^2 C_F \left[\int^{Q^2}_{Q^2
    v^{\frac{2}{a+b_\ell}}} \frac{dk_t^2}{k_t^2}
    \; \frac{\as(k_t^2)}{\pi}
    \left(\ln \frac{Q_{12}}{k_t} + B_\ell\right) + 
    \right.\\\left. +
      \int^{Q^2v^{\frac{2}{a+b_\ell}}}_{Q^2
        v^{\frac{2}{a}}} \frac{dk_t^2}{k_t^2}\, \frac{d\phi}{2\pi}
      \; \frac{\as(k_t^2)}{\pi} \left(\ln \frac{Q_{12}}{2E_\ell}
        + \frac{1}{b_\ell} \ln \left[\left(\frac{k_t}{Q}\right)^a\frac{d_\ell
          g_\ell(\phi)}{v} 
      \right]\right)
  \right]\,,
\end{multline}
where we have neglected NNLL contributions associated with the exact
position of the boundary between the two regions. In the upper $k_t$
region, the constant $B_\ell$ is associated with the large-$\eta$ part
of the integration over the $p_{qg}$ splitting function,
\begin{equation}
  \label{eq:Bell}
  B_\ell = \int_0^1 \frac{dz}{z} \left(\frac{z p_{gq}(z)}{2} -
    1\right) = -\frac34\,.
\end{equation}
In the lower $k_t$ region, the upper limit on $\eta$ comes from the
condition on the observable, and it is implicitly assumed that the
observable (specifically, $d_\ell g_\ell(\phi)$) is positive definite.

It is convenient to express eq.~(\ref{eq:sigma1_etadone}) in terms
of certain `standard building-blocks',
\begin{multline}
  \label{eq:sigma1_standard}
  R(v) = \sum_{\ell=1}^2 C_F \left[r_\ell(L) + r_\ell'(L)
    \left(\ln {\bar d}_\ell - b_\ell \ln\frac{2E_\ell}{Q}\right) +
                                \right.\\\left. +
    B_\ell\, T\!\left(\frac{L}{a+b_\ell} \right) \right] 
  + 2C_F \, T\!\left(\frac{L}{a}\right) \ln \frac{Q_{12}}{Q}\,,
  \qquad\quad L \equiv \ln \frac1v\,,
\end{multline}
where 
\begin{equation}
  \label{eq:dbarell}
  \ln {\bar d}_\ell = \ln d_\ell + \int_0^{2\pi} \frac{d\phi}{2\pi}
  \ln g_\ell(\phi)\,.
\end{equation}
The `standard building blocks' are the double logarithmic piece
$r_\ell$ (containing all the LL and some NLL contributions), 
\begin{equation}
  \label{eq:rell}
  r_\ell(L) =
  \int_{Q^2
    e^{-\frac{2L}{a+b_\ell}}
  }^{Q^2} \frac{dk_t^2}{k_t^2}\frac{\as(k_t^2)}{\pi}\ln\frac{Q}{k_t}
  +
  \int_{Q^2 e^{-\frac{2L}a}}^{Q^2
    e^{-\frac{2L}{a+b_\ell}}
  }\frac{dk_t^2}{k_t^2}\frac{\as(k_t^2)}{\pi}
  \left(\frac{L}{b_\ell} + \ln\left(\frac{k_t}{Q}\right)^{a/b_\ell}\right),
\end{equation}
as well as various purely single logarithmic (NLL) pieces,
\begin{equation}
  \label{eq:T}
  T(L) = \int_{Q^2 e^{-2L}
  }^{Q^2}\frac{dk_t^2}{k_t^2}\frac{\as(k_t^2)}{\pi}\>,
\end{equation}
and $r_\ell' = \partial_L r_\ell$, which can be expressed in terms of
the $T(L)$ as
\begin{equation}
  \label{eq:rpell}
  r_\ell'(L) = \frac{1}{b_\ell}\left[T\left(\frac{L}{a}\right)
    - T\left(\frac{L}{a+b_\ell}\right)
  \right]\,.
\end{equation}
Though the results here have been derived for $b_\ell\ne0$, their
$b_\ell\to0$ limit is finite and well-defined, as can straightforwardly be
verified.

Several remarks are in order concerning
eq.~(\ref{eq:sigma1_standard}). Firstly, in the sum over legs, the
contributions all depend just on $Q$ and the properties of the given
leg --- dependence on the invariant mass of the two legs, $Q_{12}$,
has been placed outside the sum, and is independent of the $b_\ell$.
Such a structure will be useful when extending the result to
configurations with several hard legs.

Another point concerns frame dependence and $Q$ dependence of
eq.~(\ref{eq:sigma1_standard}). The derivation has been carried out in
a specific Lorentz frame and with some arbitrary value for $Q$.
The result should not however depend on the choice of frame or of $Q$.
To see that it truly does not, we observe that a change of frame
corresponds simply to a change in the values of the leg energies,
$E_\ell \to E_\ell'$. For a given emission this corresponds to change
in rapidity with respect to the leg $\eta^{(\ell)} \to
{\eta^{(\ell)}}'$ and an associated change in the coefficients $d_\ell
\to d_\ell'$ (such that the observable remains frame-independent):
\begin{subequations}
\begin{align}
  {\eta^{(\ell)}}' &= \eta^{(\ell)} + \ln \frac{E_\ell'}{E_{\ell}}\,,
  \\
  d_\ell' &= d_\ell +  b_\ell\ln \frac{E_\ell'}{E_{\ell}}\,.
\end{align}
\end{subequations}
Inserting the change in $d_\ell$ into eq.~(\ref{eq:sigma1_standard}),
leads to the result that $R(v)$ is frame-independent.

The demonstration that eq.~(\ref{eq:sigma1_standard}) is independent
of the choice of $Q$ is only slightly more involved: at NLL accuracy,
$r_\ell(v)$ depends on $Q$ as follows,
\begin{equation}
  \label{eq:rellQdep}
  \frac{\partial r_\ell(L)}{\partial \ln Q} =
  T\left(\frac{L}{a}\right) - (a+b_\ell)r_\ell' 
  + \cO{\mathrm{NNLL}}\,,
\end{equation}
while $T(L)$ and $r_\ell'$ have $Q$ dependence only at NNLL accuracy.
The $Q$-independence of the observable implies $\partial_{\ln Q}
\ln d_\ell = a$. Inserting this into eq.~(\ref{eq:sigma1_standard}), one
finds that $R(v)$ is $Q$-dependent only at NNLL accuracy
(strictly speaking, the NNLL terms arise only in the running-coupling
case, so for the first-order, fixed-coupling result there is no
$Q$-dependence at all).

\subsection[All-order treatment (${q\bar q}$ case)]{All-order
  treatment ($\boldsymbol{q\bar q}$ case)}
\label{sec:allorders}

For the continuously global observables that we discuss in this
article, the extension of the previous section's treatment to all
(NLL) orders involves two main ingredients: the running of the
coupling, with its associated scheme dependence; and the treatment of
multiple `independent' emissions that are widely separated in rapidity.

This separation can be explained at second order for example by noting
that in the soft and collinear region one can write the squared matrix
element for two-gluon production as
\begin{equation}
  \label{eq:twogluonME}
  |M^2(k_1,k_2)| =   \left(|{M}^2(k_1)||{M}^2(k_2)| +  |{\widetilde
    M}^2(k_1,k_2)| \right)\,,
\end{equation}
where we have the product of two independent emissions, $|M^2(k)|$
being the squared matrix element for single gluon emission, as given
in eq.~(\ref{eq:matrixelement}), plus a correlated, `non-abelian' part
$|{\widetilde M}^2(k_1,k_2)|$ which contributes only when the two
gluons are close in rapidity or both at large $z$ (there is also a
corresponding part with a $q\bar q$ pair). 
This structure is a consequence of QCD coherence \cite{Coherence}:
when two gluons are emitted on very different angular scales, the one
at larger angle is emitted coherently from the combination of the
quark and the gluon at smaller angle. Since the coherent combination
of a quark and gluon has the same colour charge as a quark, the
emission of two gluons on widely different angular scales simply
behaves as independent emission. More generally, the emission of any
number of gluons, all at very different angles, behaves as independent
emission --- this will be important for the all-order extension of
eq.~(\ref{eq:twogluonME}).

However, before dealing with the all-order generalisation of the
independent emission part of eq.~(\ref{eq:twogluonME}), we need first
to consider the correlated part of the two-gluon emission, which is
inextricably linked to the running of the coupling.

\subsubsection{Correlated two-gluon emission}
\label{sec:runn-coupl-effects}

The treatment of the running coupling in resummations has been
extensively discussed in the literature
\cite{CSS,KodairaTrentadue,CataniTrentadue} and can be 
summarised essentially as follows. Firstly one considers the
non-abelian (N.A.) correlated double emission term together with the
non-abelian part of the virtual (1-loop) correction to single gluon
emission, and notes that (including the $q \bar q$ contributions, without
the $1/2!$)
\begin{multline}
  \label{eq:RealVirtCombined}
  [dk] \,|M^2_{\mathrm{1-loop,N.A.}}(k,\mu)| + \frac1{2!}
      \int [dk_1] [dk_2] |{\widetilde M}^2(k_1,k_2)| \delta^3(k - k_1 - k_2)
   \\ = [dk] \, |M^2(k)| \left(\beta_0 \ln \frac{k_t^2}{\mu^2} +
     \frac{K}{2\pi}\right) \as\,,
\end{multline}
where the $\delta^3$-function is (in analogy to \cite{Milan1}) over
the two components of the transverse momentum and the rapidity, and
$\mu$ is the renormalisation scale; $\beta_0 = (11\CA - 2\nf)/(12\pi)$
and, in the $\MSbar$ renormalisation scheme, $K =
(\frac{67}{18}-\frac{\pi^2}{6})\CA - \frac{5}{9}\nf$.
Thus one can add the $\as^2$ non-abelian terms to
eq.~(\ref{eq:vProbOrderAlphaFirstGo2}),
\begin{multline}
  \label{eq:sigma_1emsn_2loops_a}
  \vProb(v) = 1 - \int [dk] \,\left(\,|M^2(k,\as\!=\! \as(\mu^2))|\, +
    |M^2_{\mathrm{1-loop,N.A.}}(k,\mu)|\right) \Theta(V(\{\tilde p\}, k)
  - v) \\ 
  - \frac{1}{2!}\int [dk_1][dk_2] |{\widetilde M}^2(k_1,k_2)| 
  \Theta(V(\{\tilde p\}, k_1,k_2) - v)\,,
\end{multline}
and rewrite the result, using eq.~(\ref{eq:RealVirtCombined}), as
\begin{multline}
  \label{eq:sigma_1emsn_2loops_b}
  \vProb(v) = 1 - \int [dk] \,|M^2(k,\as\!=\! \as(\mu^2))|\,
  \left(1 + \left(\beta_0 \ln \frac{k_t^2}{\mu^2} +
     \frac{K}{2\pi}\right) \as \right)
  \Theta(V(\{\tilde p\}, k)
  - v) \\ 
  - \frac{1}{2!}\int [dk_1][dk_2] |{\widetilde M}^2(k_1,k_2)| 
  \left( \Theta(V(\{\tilde p\}, k_1,k_2) - v) 
    - \Theta(V(\{\tilde p\}, k) - v)\right)\,,
\end{multline}
where, in the second line, $k$ is a massless four-vector with the same
transverse momentum and rapidity as $k_1 + k_2$. 

\paragraph{Reproducing the running coupling.}
Let us initially just consider the first line of
eq.~(\ref{eq:sigma_1emsn_2loops_b}). If one takes $\mu \sim Q$, then
since $\ln k_t^2/Q^2$ is of the same order of magnitude as $\ln 1/v$,
one sees that the $\beta_0$ term will correct the leading $\as(\mu^2)
\ln^2 1/v$ contribution by an amount $\as^2 \ln^3 1/v$, also a LL
contribution. The term
involving $K$ leads to a correction of order $\as^2 \ln^2 1/v$, i.e.\ 
a NLL term. One can also choose to reabsorb these contributions into
the leading term: taking $\mu = k_t$, the $\beta_0$ term disappears;
furthermore defining $\as$ to be in the Bremsstrahlung (CMW) scheme
\cite{CMW,KodairaTrentadue}, $\alpha_{s,\mathrm{CMW}} =
\alpha_{s,\MSbar} + K \as^2/2\pi$, one can reabsorb the term proportional
to $K$.

It turns out that using $\as(k_t^2)$ (as was anticipated in
eq.~(\ref{eq:sigma1_explicit})), in the CMW scheme, is sufficient to
account for the running coupling contributions at \emph{all orders}
\cite{CTTW,CataniTrentadue}, giving an implicit resummation of terms
of the form $\beta_0^{n-1}\as^n \ln^{n+1}1/v$ and $K\beta_0^{n-2}\as^n
\ln^{n}1/v$. The only proviso is that the running of $\as(k_t)$ has to
be carried out at two-loop level, in order to properly account also
for NLL terms $\beta_1 \beta_0^{n-3} \as^n \ln^n 1/v$ ($n \ge 3$).

Strictly speaking this discussion applies to the region of soft
\emph{and} collinear gluon emission. Subtleties arise both in the
hard-collinear and large-angle soft regions. In the former, the
relation eq.~(\ref{eq:RealVirtCombined}) holds only at the accuracy of
the $\beta_0$ term, but not of $K$. However since the hard-collinear
region is single-logarithmic, the correction $K$ is associated with
terms $\as^{n} \ln^{n-1} 1/v$ and so is NNLL. For soft large-angle
emissions, the problem is instead that there may be difficulties in
identifying $k_t$: for the problems with two hard legs that we have
discussed so far, one can show that it is the invariant transverse
momentum with respect to the dipole that is the appropriate scale.
However in ensembles with several hard legs (four or more), there is,
to our knowledge,
no procedure for unambiguously associating the emission with a
particular dipole, and the appropriate definition of $k_t$ is
ambiguous to within a factor of order $1$. Again, however, this
ambiguity arises in a single-logarithmic region of integration over
transverse momenta (at large angles) --- writing such an integral as
$\int^Q_{A v^{1/a}Q} \frac{dk_t}{k_t} \as(Bk_t)$, where $A$ and $B$ are
factors of order $1$ that parametrise our ignorance, and recalling
that $\as(B  k_t) \simeq \as(k_t) + \beta_0 \ln B \as^2(k_t)/\pi$, it
should be clear that the ambiguities translate to NNLL uncertainties
proportional to $\ln A \as^n \ln^{n-1}1/v$ and $\ln B \as^n \ln^{n-1}1/v$.
\footnote{We note that NNLL corrections come also
  from the full treatment of the emission (or, more precisely, the
  corresponding virtual term) of three correlated partons,
  all soft and collinear to a hard leg. Such contributions are related
  to the $A_3$ term calculated in~\cite{Moch:2002sn,Berger:2002sv}.}

\paragraph{Observable's dependence on correlated gluon emission.}
So far we have concentrated only on the first line of
eq.~(\ref{eq:sigma_1emsn_2loops_b}), whose properties have been widely
discussed in the literature. The second line, in contrast, has
received less scrutiny, but nevertheless needs to be examined in some
detail.  Let us first consider the region where the relative
transverse momentum of $k_1$ and $k_2$ (we label this $k_{t,12}$) is
of the same order of magnitude as their transverse momenta with
respect to the hard leg, $k_{t,12} \sim k_t$.  This region of
integration is suppressed by a power of $\as$ relative to the
single-gluon emission. The question of how much it contributes to
$\vProb(v)$ depends on the observable: if $V(\{\tilde p\}, k_1,k_2)$
differs from $V(\{\tilde p\}, k)$ by no more than a factor of order
$1$ then the difference of $\Theta$-functions in the second line of
eq.~(\ref{eq:sigma_1emsn_2loops_b}) is non-zero only in a narrow band
of $k$, where $V(\{\tilde p\}, k)$ is of order $v$.  Expressing this
with reference to figure~\ref{fig:lozenge}, one has a contribution of
relative order $\as$ in a band of width $\sim 1$ (in $\ln k_t/Q$) along
the lower edges of the shaded region. This corresponds to a NNLL term,
$\as^2 \ln 1/v$, which can be neglected.  Such a contribution has been
commented before in \cite{DLMSBroadPT}.

Suppose, instead, that the observable is such that $V(\{\tilde p\},
k_1,k_2)$ differs substantially from $V(\{\tilde p\}, k)$, say by a
factor that grows as a power of $V(\{\tilde p\}, k)$ --- in this case the
band in which the difference of $\Theta$-functions is non-zero will
have a width of order $\ln 1/v$ and the second line of
eq.~(\ref{eq:sigma_1emsn_2loops_b}) will contribute an amount $\as^2
\ln^2 1/v$, \ie a NLL term. This would mean that the `correlated' part
of two-gluon emission could not simply be absorbed into the running of
the coupling, necessitating a more sophisticated resummation
treatment.

We also need to examine what happens where $k_1$ and $k_2$ are
collinear and/or one of them is soft, $k_{t,12} \ll k_t$. At first
sight it seems natural to argue that since we have an infrared and
collinear (IRC) safe observable, $V(\{\tilde p\},k_1,k_2) \simeq
V(\{\tilde p\},k)$ and so the difference of $\Theta$-functions is
zero. This is certainly true in the limit $k_{t,12}/k_t \to 0$, but
there is a question of how small the ratio $k_{t,12}/k_t$ has to
be in order for the difference $|V(\{\tilde p\},k_1,k_2) - V(\{\tilde
p\},k)|$ to be negligible (say less than $\varepsilon$). If the condition
is for example $k_{t,12}/k_t \lesssim \varepsilon^p$ where $p$ is some
arbitrary positive power, then one can show that the second line of
eq.~(\ref{eq:sigma_1emsn_2loops_b}) will contribute at most an NNLL
piece. 

But if the condition instead involves $k_t/Q$, or $e^{-\eta}$ in the
right-hand side, for example $k_{t,12}/k_t \lesssim \varepsilon^p
(k_t/Q)^{p'}$, then the difference of $\Theta$-functions will be
non-zero over a large, logarithmic integration region in $k_{t,12}$
and the second line of eq.~(\ref{eq:sigma_1emsn_2loops_b}) could lead
to contributions $\as^2 \ln^3 1/v$ or $\as^2 \ln^2 1/v$. In such a
case, we would again be in a situation where the correlated two-gluon
emission effects could not simply be absorbed into a pure
running-coupling term.

\paragraph{Remarks.} It is quite often taken for granted that the
effects of `correlated' gluon emission can be absorbed into the
running coupling in an appropriate scheme. The general analysis of
this section reveals that this is true as long as the observable meets
certain conditions --- essentially that the scaling properties of the
observable be the same whether there be one or two (or more)
emissions;\footnote{In this article we consider only global
  observables. For non-global observables the situation is more
  complex, in that there can legitimately be boundaries in angle that
  delimit regions of different scaling. One then has the condition
  that the scaling of the observable as one simultaneously varies the
  momenta of two (or more) emissions, should correspond to the weakest
  of the scalings when varying the momentum of each emission
  individually.}
 and that the IRC safety of the observable for secondary splitting of
a primary emission should manifest itself for secondary splittings of
the same order of magnitude of hardness as the primary emission.

The second of these conditions especially may seem quite
non-intuitive. One is generally used to thinking of IRC safety in
contexts where all the emissions (except the one being made collinear
or soft) are of similar hardnesses, \ie there is a single hard scale
with respect to which one defines the degree of softness or
collinearity. But when dealing with final-state resummations, one
introduces a second scale in the problem, related to the (small)
value of the observable. IRC safety merely states that the observable
should be insensitive to an extra arbitrarily infrared or collinear emission
--- it does not specify at what scale that insensitivity should set
in. It is natural to assume that it is simply the smaller of the two
scales in the problem. If that is the case then the observable is
resummable with `usual' techniques. However there are observables for
which the relevant `insensitivity scale' involves some more
complicated combination of the two scales in the problem, \eg
$k_t^2/Q$. A concrete example, which will be discussed in
appendix~\ref{sec:Geneva},
is the Geneva $y_{23}$ jet-resolution parameter. Such observables
require a more sophisticated resummation treatment, which is beyond
the scope of this paper.

While the above discussion has been framed in terms of configurations
with two correlated emissions, one should be aware that for the
all-order reconstruction of the running coupling, the observable
should have similar properties also with multiple correlated emissions
and/or secondary collinear branchings.
As we shall see, such a condition will actually turn out to form
part of a more general class of requirements, which will be called
\emph{recursive} infrared and collinear (rIRC) safety.

\subsubsection{Towards all orders}
\label{sec:towards-all-orders}

The picture that emerges from the above section is that, for suitable
observables, two-gluon emission can be separated into two parts with
distinct physical roles: there is a correlated part which provides
scheme and running-coupling corrections to single-gluon emission; and
there is an uncorrelated `independent-emission' part, which we have
yet to treat explicitly.  There were two prerequisites for such a
separation: firstly, we made use of a property of QCD matrix elements,
QCD coherence \cite{Coherence}, which guarantees that emissions widely
separated in rapidity are effectively independent; secondly we
specified a property of the observable, rIRC safety (to be further
elaborated below), which guarantees, among other things, that the
observable is sufficiently inclusive that the details of the
correlations of emissions close in rapidity affect the predictions
only at NNLL accuracy and beyond (except through the running of the
coupling).

To generalise this to any number of emissions we make use of coherence
at all orders: emissions at some given angular scale are emitted
coherently from the ensemble of emissions at much smaller angles,
meaning that emissions on disparate angular scales are effectively
emitted independently. A priori however, there is no reason to believe
that emissions are widely separated in rapidity --- on the contrary it
is straightforward to see\footnote{Neglecting non-global logarithms,
  the naive `primary-emission' probability of there being no emissions
  in a region of size $\Delta \eta$ centred on rapidity $\eta$ is
  roughly $P \sim \exp\left(-\Delta \eta \int^{Q e^{-\eta}}_{Q_0}
    \frac{dk_t}{k_t} \frac{2\CF \as(k_t^2)}{\pi} \right)$, where $Q_0$
  is a non-perturbative cutoff. This differs substantially from $1$,
  and since the neglected non-global effects generally lead to extra
  suppression \cite{NG1}, it is an upper bound on the full
  single-logarithmic expression for the probability.} %
that, for a typical event, any given region of rapidity of size
$\Delta \eta \sim 1$ is likely to contain emissions.

\FIGURE{
  \includegraphics[width=0.95\textwidth]{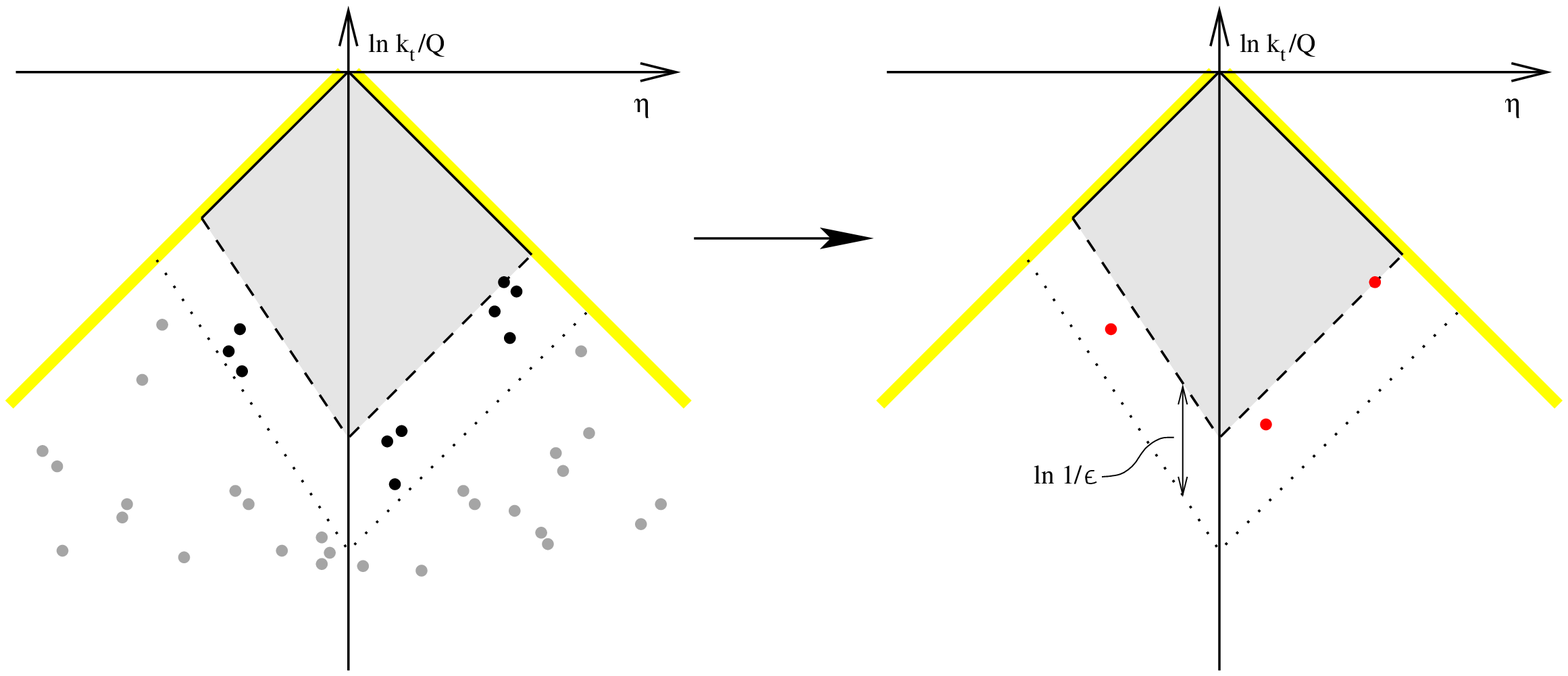}
  \caption{Illustrative pattern of emissions (dots) in the $\ln k_t
    $--$\eta$ plane (as in figure~\ref{fig:lozenge}) given that the
    event shape has a value of order $v$ (corresponding to the dashed
    boundary). Left: emissions (grey) below the dotted line would each
    individually give a value of the observable $< \epsilon v$ and can
    be ignored (see text).  Right: the remaining emissions (black) can
    be grouped into clusters local in rapidity, each cluster then
    being replaced by a single emission (red).}
  \label{fig:lozengecluster}
}

This is illustrated in the left-hand diagram of
fig.~\ref{fig:lozengecluster}, which shows a possible configuration of
emissions (dots) that gives a value of order $v$ for the observable. To work
around the problem of a `dense' distribution of emissions in rapidity,
we separate the emissions into two groups by introducing a small
parameter $\epsilon$: emissions that individually would lead to a
value of the observable smaller than $\epsilon v$ are coloured grey,
and the remaining ones are coloured black. Most observables have the
property (to be formalised below as part of the rIRC conditions) that
the `grey' emissions (dense in rapidity) can be removed from the
ensemble without significantly altering the value of the observable.
Therefore we are free to study configurations without these grey
emissions.

As a next step we group the remaining `black' emissions into clusters
that are local in rapidity.\footnote{We refer to clusters rather than
  to individual emissions in order to ensure the infrared and
  collinear safety of our discussion. The concept is made more precise
  below.} %
The number of clusters per square of unit rapidity and $\ln 1/k_t$ is
of order $\as$; given that the available rapidity is of order $\ln
1/v$ and that the range of $\ln 1/k_t$ over which the clusters are
distributed (the height of the band between the dashed and dotted
lines in figure 2) is $\ln 1/\epsilon$, the total number of clusters
will be of order $\as \ln v^{-1} \ln \epsilon^{-1}$. By imposing $ \as
\ln 1/\epsilon \ll 1$, we can ensure that the total number of clusters is
much smaller than the total available rapidity $\sim \ln
v^{-1}$. Therefore when integrating over rapidities of emissions, the
dominant contribution will come from configurations in which all
clusters are widely separated in rapidity. Accordingly we can treat
each cluster independently of the others.

We are then free, for each cluster, to proceed as we did with the
correlated ${\widetilde M}^2(k_1,k_2)$ part of the two-gluon matrix
element in section~\ref{sec:runn-coupl-effects}: for a cluster
consisting of $m$ partons, $k_{i+1},\ldots,k_{i+m}$, we can replace
$V(\{\tilde p\}, \ldots,k_{i+1},\ldots,k_{i+m},\ldots)$ with
$V(\{\tilde p\}, \ldots,k_{i+1}+\ldots+k_{i+m},\ldots)$ and integrate
\emph{inclusively} over the momenta of the cluster partons, while
keeping the total cluster momentum fixed. The fact that one can
integrate inclusively over the parton momenta is guaranteed by
continuous globalness together with the
parts of the rIRC condition outlined in
sec.~\ref{sec:runn-coupl-effects} (and to which we return in detail in
sections~\ref{sec:rIRC-deriv} and \ref{sec:master}), and is critical in
ensuring that 
one can make use of the standard result from resummation studies
\cite{CSS,CTTW,KodairaTrentadue,CataniTrentadue,CollinsSoper} that
after additionally summing over $m$ one reproduces the all-order
running coupling. 

In the graphical representation of fig.~\ref{fig:lozengecluster}, this
last step corresponds to the replacement of the clusters of black
emissions (left-hand diagram) with individual red emissions
(right-hand diagram) with the same total momentum, as if each one were
emitted independently with a coupling that runs as $\as(k_t^2)$. It is
this approximation of `independent emissions' with running coupling
(as well as a consistent treatment of virtual corrections, one that
guarantees unitarity) that we will use in the next subsection, in
order to calculate the all-order resummed distribution for our
general observable.

First though, since we aim to guarantee NLL accuracy, it is important,
in the above
arguments, to review the accuracy of the approximations made in each
step. The elimination of the `grey' emissions (those individually
giving a contribution $< \epsilon v$) leads to a correction
suppressed by a positive power of $\epsilon$. In the limit of small
$\as$, it is possible to make $\epsilon$ small,
while maintaining $\as \ln \epsilon^{-1} \ll 1$, 
therefore this part of the procedure does not produce corrections to
the logarithmic structure of the distribution.

A second potential source of inaccuracy might appear to come from the
approximation that two clusters are independent even when they are
close in rapidity.  This problem can be avoided however, by making use
of the freedom that we have in defining what is meant by a cluster:
we recall that in section~\ref{sec:runn-coupl-effects}, we decomposed
the two-gluon matrix element into two pieces. The two-correlated gluon
piece, ${\widetilde M}^2(k_1,k_2)$ can be seen as a single-cluster
term, while the independent emission piece, $M^2(k_1) M^2(k_2)$, acts
as a two-cluster term. The combination of these two contributions
reconstructs
the full matrix element, the error made in treating the two gluons as
independent (two-cluster piece) being compensated by the one-cluster
piece.
Similarly, in an $m$-gluon matrix element, as a consequence of
coherence, it is possible to identify distinct contributions which
behave as having one cluster (\ie a contribution that is suppressed
unless all partons are close in rapidity), two clusters (each local in
rapidity, and independent of the other), and so on up to $m$
single-gluon `clusters' (all independent).  By basing our
classification into clusters on such a decomposition of the matrix
element (this is outlined in more detail in
appendix~\ref{sec:mpartonDecomposition}), and summing over all
possible decompositions, one reproduces the full soft-collinear
$m$-parton matrix element, without any approximations.

Therefore, the only source of logarithmic inaccuracy, in our
approximation of independent emissions with running coupling, will
come from our
replacement (in the observable) of the individual momenta of the
cluster partons by the total cluster momentum. We showed, in
section~\ref{sec:runn-coupl-effects}, in the two gluon case, that for
an rIRC safe observable this contributed at most $\as^2 L$, \ie a NNLL
correction. 
Since, because of coherence, the details of gluon correlations inside
a cluster are independent of the properties of all other clusters of
emissions of the event, such a term will be independent of the
contributions from other emissions (each one of which, with virtual
corrections, may provide up to $\as L^2$), and 
one should immediately be able to see that in general the
correction goes as $\as^n L^{2n-3}$.

In order to additionally meet the claims stated in
section~\ref{sec:intro}, \ie to obtain a final form
eq.~(\ref{eq:vProb-general}), one should be able to show that, at all
orders, corrections are at most of the form $\as h(\as L) e^{L g_1(\as L)}$,
where $h$ is some arbitrary function. To see why this is the case,
one should first understand that the form
eq.~(\ref{eq:vProb-general}) has its origins in the exponentiated
double-logarithmic Sudakov suppression associated with the resummation
of the virtual corrections in the shaded region of
fig.~\ref{fig:lozengecluster}; factorised single-logarithmic
corrections arise because of $\order{1}$ relative modifications to the
observable's value when there are multiple emissions in a narrow
(single-logarithmic) strip along the boundary of the shaded region. A
similar mechanism is relevant for the effects of non-inclusiveness of
the observable with respect to the momenta of a cluster, except that
by requiring two emissions to have the same rapidity (\ie a cluster)
one loses a logarithm, and a single-logarithmic factorised correction
$\as^n L^n$ becomes a NNLL, $\as^n L^{n-1}$, factorised correction. A
more mathematical treatment can only be given once we have seen in detail
the origin, at single logarithmic level, of the structure of
eq.~(\ref{eq:vProb-general}).  Accordingly it is deferred to
appendix~\ref{sec:CorrelEmsn} and for now we proceed with the
independent-emission approximation, including running-coupling
corrections.

\subsubsection{Multiple independent emission}
\label{sec:mult-indep-emsn}

The result that we shall obtain here was first found in \cite{BSZ},
however the derivation given here is intended to be slightly more
direct and to highlight more fully the requirements that must be
satisfied by the observable in order for the result to be valid.

For observables for whose resummation the picture of multiple
independent emission is a good approximation (as discussed above), one
can write at NLL accuracy
\begin{multline}
  \label{eq:MultipleIndepEmsn}
  \vProb(v) = \exp\left ( - \int [dk] \,|M^2_{rc}(k)|\, \right)
  \sum_{n=0}^{\infty} \frac{1}{n!} \left( \prod_{i=1}^n
    \int [dk_i] \, |M^2_{rc}(k_i)| 
  \right)\times\\ \times
  \Theta(v - V(\{\tilde p\},k_1, \ldots, k_n))\,,
\end{multline}
where the first factor resums the virtual corrections, while the rest
of the expression accounts for real emissions.  Both real and virtual
integrations should be understood as regularised. The coupling is always
to be evaluated at scale $k_t$ and in the CMW scheme and the matrix
element has been written with a subscript `$rc$' as a reminder that
\emph{r}unning-\emph{c}oupling effects have already been resummed.

Note that in section~\ref{sec:towards-all-orders}, in order to
guarantee the independent emission approximation, we removed all
emissions below a softness and collinearity threshold, $\epsilon v$,
those shown in grey in figure~\ref{fig:lozengecluster}.  Yet here, in
our independent-emission formula, eq.~(\ref{eq:MultipleIndepEmsn}), we
have not imposed any such cut.  This is legitimate because the
condition that allowed us to remove these emissions (\ie the fact they
do not affect the value of the observable, a consequence of rIRC
safety) also allows us to put them back in with some arbitrary
distribution, in particular with an independent emission distribution.

An important point regarding eq.~(\ref{eq:MultipleIndepEmsn})
concerns the manner in which one specifies the momenta $k_i$. In the
case of a single emission we used the definition
eq.~(\ref{eq:SudakovForK}), which has the property that the $k_t$
entering the definition coincides closely with the actual transverse
momentum relative to the final Born partons (after recoil). This is
important because the $dk_t^2/k_t^2 dz/z$ divergence of the matrix
element holds for a transverse momentum $k_t$ relative to the final
Born momenta. When there are multiple emissions, the situation is more
complicated: transverse momenta defined relative to fixed axes, as in
eq.~(\ref{eq:SudakovForK}), do not necessarily coincide with the
transverse momenta relative to the final Born partons. Since it is the
latter that are of interest to us, when we refer to a given momentum
$k_i$, it should be understood as being defined through its transverse
momentum and rapidity (or energy fraction) relative to the final Born
partons. In particular the actual 4-momentum components may well
differ depending on what other emissions are present in the event.
This point, and related issues, are discussed in more detail in
appendix~\ref{sec:recoil}.

To evaluate eq.~(\ref{eq:MultipleIndepEmsn}), it will be convenient to
identify the $k_i$ with the largest value of $V(\{\tilde p\},k_i)$,
and relabel it as $k_1$. We therefore rewrite the sum in
eq.~(\ref{eq:MultipleIndepEmsn})
\begin{multline}
  \label{eq:playingwithvProbsum}
  \sum_{n=0}^{\infty} \frac{1}{n!}   \left(  \prod_{i=1}^n
    \int [dk_i] \, |M^2_{rc}(k_i)| 
  \right) \\
  = 1 + 
  \int [dk_1] \, |M^2_{rc}(k_1)| \, \sum_{m=0}^{\infty} \frac{1}{m!}
  \left( \prod_{i=2}^{m+1} 
    \int [dk_i] \, |M^2_{rc}(k_i)| \Theta(v_1 - v_i)
  \right)\,,
\end{multline}
where we have introduced the notation $ v_i \equiv V(\{\tilde
p\},k_i)$. The constant term, $1$, accounts for the case in which
there are no emissions --- because of the formally infinite
suppression associated with the virtual corrections, it can 
from now on be neglected.

A technically useful step, next, as anticipated already in
section~\ref{sec:towards-all-orders}, is to 
split the sum in eq.~(\ref{eq:playingwithvProbsum}) into two
parts, with emissions satisfying $v_i > \epsilon v_1$ and $v_i < \epsilon
v_1$ respectively; $\epsilon$ is an arbitrary small parameter,
which for suitable observables can be chosen such that $\epsilon \ll
1$, while $\ln 1/\epsilon \ll \ln 1/v$ (in the limit $v\to0$ we assume
that it is possible to choose $\epsilon$ independently of $v$). 
We thus write
\begin{multline}
  \label{eq:splittingvProbsum}
  \sum_{n=0}^{\infty} \frac{1}{n!}   \left(  \prod_{i=1}^n
    \int [dk_i] \, |M^2_{rc}(k_i)| 
  \right) 
  =\\ 
  \int [dk_1] \, |M^2_{rc}(k_1)| 
  \sum_{m=0}^{\infty} \frac{1}{m!}
  \left( \prod_{i=2}^{m+1} 
    \int_{\epsilon v_1}^{v_1} [dk_i] \, |M^2_{rc}(k_i)|
  \right) \times \\\times 
  \sum_{k=0}^{\infty} \frac{1}{k!}
  \left( \prod_{i=m+2}^{k+m+1} 
    \int^{\epsilon v_1} [dk_i] \, |M^2_{rc}(k_i)| 
  \right),
\end{multline}
where we have introduced the shorthand of integration limits that
apply not directly to the $k_i$, but to the $v_i = V(\{\tilde
p\},k_i)$.

The above separation is of interest, because we require (as part of
the rIRC safety conditions) that the emissions with $v_i < \epsilon v_1$
not contribute significantly to the final value of the observable, \ie 
\begin{equation}
  \label{eq:ignorelowvemissions}
  V(\{\tilde p\},k_1, \ldots, k_{m+1}, k_{m+2},\ldots,k_{k+m+1} )  
  =
 V(\{\tilde p\},k_1, \ldots, k_{m+1}) + \cO{\epsilon^p v_1}\,,
\end{equation}
where $p$ is some positive power. So we can sum over these emissions
without affecting the $\Theta$-function on the observable in
eq.~(\ref{eq:MultipleIndepEmsn}). This sum cancels the part of the
virtual corrections associated with values of $k$ such that
$V(\{\tilde p\},k) < \epsilon v_1$, allowing us to write
\begin{multline}
  \label{eq:MultipleIndepReg}
  \vProb(v) = 
  \int [dk_1] \, |M^2_{rc}(k_1)| 
  \exp\left ( - \int_{\epsilon v_1} [dk] \,|M^2_{rc}(k)|\, 
  \right) \times \\ \times
  \sum_{m=0}^{\infty} \frac{1}{m!}
  \left( \prod_{i=2}^{m+1} 
    \int_{\epsilon v_1}^{v_1} [dk_i] \, |M^2_{rc}(k_i)| 
  \right)
  \Theta(v - V(\{\tilde p\},k_1, \ldots, k_{m+1}))\,.
\end{multline}
We next take the virtual corrections and split them as follows
\begin{equation}
  \label{eq:splitVirtual}
  \exp\left ( - \int_{\epsilon v_1} [dk] \,|M^2_{rc}(k)|\, 
  \right) = e^{-R(\epsilon v_1)} = e^{-R(v) -  R'  \ln
    \frac{v}{\epsilon v_1}
   + \order{R''} }
  \,,\qquad
  \quad R' \equiv \frac{dR}{d\ln 1/v}\,,
\end{equation}
where $R(v)$ is the single-gluon contribution to $\vProb(v)$,
discussed in section~\ref{sec:single-gluon-eval}, and we have expanded
$R(\epsilon v_1)$ around $v$, neglecting the second order ($R''=
\partial^2_{\ln 1/v} R = \order{\as^{n+1}L^n}$) term in the expansion,
since it is NNLL (as long as eq.~(\ref{eq:MultipleIndepReg}) is
dominated by momenta $k_1$ such that $v_1\sim v$ --- this is
formalised in section~\ref{sec:div}). The resummed
distribution can therefore be written
\begin{equation}
  \label{eq:Result:expRcF}
  \vProb(v) = e^{-R(v)} \cF\,,
\end{equation}
\ie the exponential of the single gluon result, multiplied by a
correction factor $\cF$ which accounts for the details of the
observable's dependence on multiple emissions,
\begin{multline}
  \label{eq:FfirstDef}
  \cF = \int [dk_1] \, |M^2_{rc}(k_1)| e^{-  R' \ln \frac{v}{\epsilon v_1}}
  \sum_{m=0}^{\infty} \frac{1}{m!}
  \left( \prod_{i=2}^{m+1} 
    \int_{\epsilon v_1}^{v_1} [dk_i] \, |M^2_{rc}(k_i)| 
  \right)\times \\ \times
  \Theta(v - V(\{\tilde p\},k_1, \ldots, k_{m+1}))\,.
\end{multline}
The function $\cF$ can be evaluated directly in this form, by Monte
Carlo methods. However this tends not to be very efficient and it is
worthwhile manipulating the expression a little further. This will be
useful also to help us highlight the single-logarithmic nature of
$\cF$ and to eliminate subleading logarithmic contributions.

We introduce the notation $k^{(\rho)}$ for a  momentum $k$ that has
been subjected to a `generalised rescaling' defined as follows:
$V(\{\tilde p\},k^{(\rho)}) = \rho \, V (\{\tilde p\},k)$;
$\phi$ should not depend on the scaling; and the rapidity should scale
as $\ln V(\{\tilde p\},k^{(\rho)})$, so that the whole of the
phase-space remains covered after large
rescalings. Figure~\ref{fig:lozengescale} illustrates the effect of
such a rescaling on three momenta represented in the $\eta$--$\ln
(k_t/Q)$ plane (as introduced in fig.~\ref{fig:lozenge}). 
An explicit form for the scaling can be obtained by introducing the
maximum rapidity that an emission can have on leg $\ell$, for a given
value of $v$,
\begin{equation}
\eta_{{\max},\ell}(v) = \frac{\ln 1/v}{a+b_\ell} + \order{1}\,,
\label{eq:etamaxFirstTime}
\end{equation}
(the piece of $\order{1}$ depends on the leg energy, and on $d_\ell
g_\ell(\phi)$) and then parameterising an emission $k_i$'s rapidity as
a fraction $\xi_i$ of this maximum rapidity, \ie $\xi_i =
\eta_i/\eta_{{\max},\ell}(V(\{\tilde p\},k_i))$. The rescaling is then
fully specified by the requirement on the value of $V(\{\tilde
p\},k_i^{(\rho)})$, the azimuthal angle and by the condition that
$\xi_i$ be conserved,
\begin{equation}
  \label{eq:kti_rho}
  k_{ti}^{(\rho)} = k_{ti}\, \rho^{(1-\xi_i)/a + \xi_i/(a+b_\ell)}\,,\qquad
  \eta_i^{(\rho)} = \eta_i - \frac{\xi_i \ln \rho}{a+b_\ell}\,,\qquad
  \phi_i^{(\rho)} = \phi_i\,.
\end{equation}

\FIGURE{
  \includegraphics[width=0.95\textwidth]{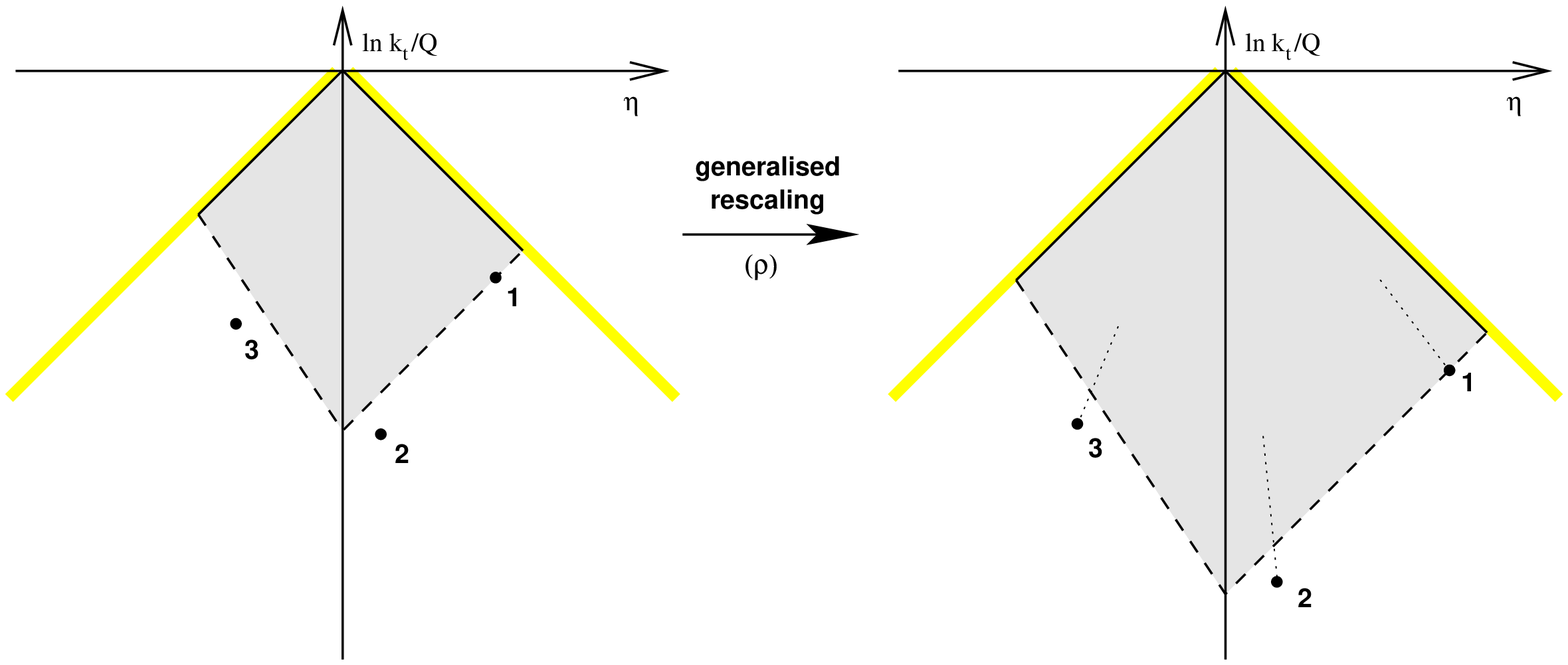}
  \caption{Left: three momenta in the $\eta$--$\ln (k_t/Q)$ plane.
    Right: those same three momenta after a common generalised
    rescaling $\rho$ has been applied to them; the dotted lines
    indicate the paths taken in the $\eta$--$\ln (k_t/Q)$ plane due to
    the rescaling. For each emission, the vertical distance to the
    dashed boundary is identical in the left and right-hand diagrams,
    consistent with a common scaling $\rho$ having been applied to all
    emissions. }
  \label{fig:lozengescale}
}

Given this rescaling, we now need to introduce a new requirement on
the observable, namely 
that when all momenta are scaled in the same fashion, the effect on
the observable should be that same scaling:
\begin{equation}
  \label{eq:scalingprop}
  V(\{\tilde p\},k_1^{(\rho)}, \ldots, k_{m+1}^{(\rho)})
   = \rho \,V(\{\tilde p\},k_1, \ldots, k_{m+1})\,.
\end{equation}
This forms yet another part of the rIRC safety
conditions.\footnote{\label{ftnt:rIRC-semiexception}Strictly 
speaking, certain exceptions are allowed to the condition as
formulated here. In particular for configurations in which two
emissions are close in rapidity (a rare occurrence) the condition, as
formulated, is not necessary because the associated correction is a
NNLL effect, of the kind already discussed in
section~\ref{sec:runn-coupl-effects}. A more general 
formulation of the condition is given below.} %
It may not be obvious why it should ever hold, nevertheless, it is
satisfied for all commonly-studied event shapes. In the case of
observables whose definitions involve just linear functions of the
momenta, it can be understood as a direct consequence of this
linearity.

The importance of eq.~(\ref{eq:scalingprop}) is, in part, that it
allows us to divide the integral over $k_1$ into an integral over the
value of $v_1$ (or rather, over $\rho = v_1/v$) and an integral over
the remaining degrees of freedom of $k_1$,
\begin{multline}
  \label{eq:FIntegralOverRho}
  \cF = \int \frac{d\rho}{\rho} \int [dk_1] \, |M^2_{rc}(k_1)|
  \,\delta\!\left(\ln \frac{v_1}{v}\right) 
  e^{ R' \ln \rho \epsilon}
  \sum_{m=0}^{\infty} \frac{1}{m!}
  \left( \prod_{i=2}^{m+1} 
    \int_{\epsilon v}^{v} [dk_i] \, |M^2_{rc}(k_i)| 
  \right)
  \times \\ \times
  \Theta(v - \rho V(\{\tilde p\},k_1, \ldots, k_{m+1}))\,,
\end{multline}
where a change of variables has been carried out, $k_i \to
k_i^{(1/\rho)}$, giving $V(\{\tilde p\},k_i) \to \rho V(\{\tilde
p\},k_i^{(1/\rho)})$, and then the $k_i^{(1/\rho)}$ have been renamed
$k_i$ so as to simplify the notation.
We assume that the integral will be dominated by values of $\rho\sim1$
(expressing the earlier assumption that $v_1\sim v$), which ensures
that the neglected corrections 
to the $[dk_i] \, |M^2_{rc}(k_i)| $ from the rescaling have at most a NNLL
effect.

Thus one can integrate analytically over $\rho$ to obtain
\begin{multline}
  \label{eq:FSemiFinalNoLimit}
  \cF = \frac{e^{R' \ln \epsilon}}{R'}
  \int [dk_1] \, |M^2_{rc}(k_1)|
  \,\delta\!\left(\ln \frac{v_1}{v}\right) 
  \sum_{m=0}^{\infty} \frac{1}{m!}
  \left( \prod_{i=2}^{m+1} 
    \int_{\epsilon v}^{v} [dk_i] \, |M^2_{rc}(k_i)| 
  \right)
  \times \\ \times
  \exp\left(-R'\ln\frac{V(\{\tilde p\},k_1, \ldots, k_{m+1})}{v}
  \right)\,.
\end{multline}
This manipulation is of course valid only if the observable is
positive definite.

That the resummation result can be expressed in terms of the product,
eq.~(\ref{eq:Result:expRcF}), of the exponential of the single-gluon
result and the above function $\cF$ was one of the main results of
\cite{BSZ}. This separation is critical for our approach: all the
double logarithmic terms are collected in the exponentiated
single-gluon result and can be treated analytically, as was done in
section~\ref{sec:single-emission}. In contrast the function $\cF$,
which will usually have to be evaluated by Monte Carlo methods, is
single-logarithmic.
To see this let us rewrite $[dk_i] \, |M^2_{rc}(k_i)|$ as follows:
\begin{multline}
  \label{eq:rewrite-dk-m2}
  \int [dk_i] \, |M^2_{rc}(k_i)| = \sum_{\ell_i=1}^2 
   \int \frac{dv_i}{v_i} \frac{C_F
     r'_{\ell_i}}{\cN_{\ell_i}(\as(Q) \ln v_i)}
   \times \\ \times
  \int_0^1 \frac{d\xi_i}{1 + \frac{a + (1-\xi_i)b_{\ell_i}}{a(a+b_{\ell_i})}
    2\beta_0 \,\as(Q) \ln v_i}
  \int_0^{2\pi} \frac{d\phi_i}{2\pi} \,,
\end{multline}
where we have taken into account the NLL correction due to the running
of the coupling, $r'_\ell$ is as defined in eq.~(\ref{eq:rpell}),
$\xi_i$ is the emission's rapidity divided by the maximum possible
rapidity for the given value of $v$, and $\cN$ normalises the integral
over $\xi_i$:
\begin{equation}
  \label{eq:xi-and-norm-defs}
  \xi_i = \frac{\eta_i}{\frac{1}{a+b_\ell}\ln\frac{1}{v_i}}\,,\quad\qquad
  \cN_{\ell}(\as \ln v) = 
  \int_0^1 \frac{d\xi}{1 + \frac{a + (1-\xi)b_{\ell}}{a(a+b_{\ell})}
    2\beta_0 \,\as \ln v}\,.
\end{equation}
In changing to an integral over the rapidity fraction $\xi_i$, and
defining $\xi_i$ as in eq.~(\ref{eq:xi-and-norm-defs}) (where we have
omitted contributions to the maximum rapidity of $\cO{1}$), we have
neglected various NNLL contributions associated with the exact upper
and lower limits of the integrals. As a result, for a given value of
$v_i$ the remaining part of the phase-space integrations and matrix
element has the property that it depends only on the
single-logarithmic quantity $\as(Q) \ln v_i$ (we recall that this is a
property also of $r_{\ell_i}'$).

Let us now take the $v\to0$ limit of eq.~(\ref{eq:FSemiFinalNoLimit})
in such a way that $ \beta_0 \as \ln v \equiv \lambda$ is kept
constant (and so also $r'_\ell$). One obtains
\begin{multline}
  \label{eq:cF-with-vto0-limit}
  \cF = 
  \frac{e^{R' \ln \epsilon}}{R'}
    \sum_{m=0}^{\infty} \frac{1}{m!}
  \left( \prod_{i=1}^{m+1} \sum_{\ell_i=1}^2 
    \int_{\epsilon}^{1} \frac{d\zeta_i}{\zeta_i} 
    \frac{C_F r_{\ell_i}'}{\cN_{\ell_i}(\lambda/\beta_0)}
    \int_0^1 \frac{d\xi_i}{1 + \frac{a + (1-\xi_i)b_{\ell_i}}{a(a+b_{\ell})}
    2\lambda} 
  \int_0^{2\pi} \frac{d\phi_i}{2\pi}
  \right)  \times \\ \times
 \delta\!\left(\ln \zeta_1\right)
  \exp\left(-R'\ln \lim_{v\to0} \frac{V(\{\tilde p\},k_1, \ldots,
      k_{m+1})}{v}\right),
  \qquad v_i \equiv \zeta_i v\,,
\end{multline}
where we have neglected the difference between $\as \ln v_i $ and $\as
\ln v$, and we emphasise that here the $k_i$ are functions of the
$\xi_i$, $\phi_i$ and $v_i$, the limit $v\to 0$ being taken with
$\xi_i$, $\phi_i$, and $\zeta_i = v_i/v$ constant.
Since the generalised scaling that we discussed above, $k_i\to
k^{(\rho)}_i$, is nothing but a scaling $v_i \to \rho v_i$ with
$\xi_i$ and $\phi_i$ kept constant, eq.~(\ref{eq:scalingprop}) ensures
that the $v\to0$ limit in eq.~(\ref{eq:cF-with-vto0-limit}) is well
defined. Strictly, eq.~(\ref{eq:scalingprop}) would suggest that no
$v\to 0 $ limit is necessary. However there is a small fraction ($\sim
1 / \ln \frac1v$) of configurations, with emissions close in rapidity
(or at the extremities of the allowed rapidity region) that are
allowed to violate eq.~(\ref{eq:scalingprop}) and which contribute a
NNLL correction to $\cF$. Taking the $v\to0$ limit ensures that they
disappear, so that $\cF$ is a purely single logarithmic function, free
of any NNLL contamination.

There exist observables (typically those referred to as event shapes),
for which a further simplification of
eq.~(\ref{eq:cF-with-vto0-limit}) is possible. They have the property
that for small $v_i$ the observable is independent of the $\xi_i$
values (except potentially, non-asymptotically, for $\xi_i$ close to
$0$ or $1$). Accordingly one can perform the $\xi_i$ integrations
analytically and write
\begin{multline}
  \label{eq:cF-evshps}
  \cF = 
  \frac{e^{R' \ln \epsilon}}{R'}
    \sum_{m=0}^{\infty} \frac{1}{m!}
  \left( \prod_{i=1}^{m+1} \sum_{\ell_i=1}^2  C_F r_{\ell_i}'
    \int_{\epsilon}^{1} \frac{d\zeta_i}{\zeta_i} 
    \int_0^{2\pi} \frac{d\phi_i}{2\pi}
  \right)   \delta\!\left(\ln \zeta_1\right)
  \times \\ \times
  \exp\left(-R'\ln \lim_{v\to0} \frac{V(\{\tilde p\},k_1, \ldots,
      k_{m+1})}{v}\right),
  \qquad v_i \equiv \zeta_i v\,,\qquad \xi_i = \mathrm{any}\,,
\end{multline}
where one is free in one's choice of the $\xi_i$ values to be used for
fixing the $k_i$. Typically one takes $\xi_i$ far from the edges
of rapidity, which nearly always ensures that any finite-$v$
corrections disappear rapidly, \eg as a power of $v$, rather
than as a power of $1/\ln \frac1v$ as is the case for
eq.~(\ref{eq:cF-with-vto0-limit}).

\subsubsection{Recursive IRC safety}
\label{sec:rIRC-deriv}

In the above derivation, we have made repeated use of properties of
the observable which we referred to as being elements of a novel
condition called recursive IRC safety. Let us now assemble those
elements.

When discussing two-correlated-parton emission,
section~\ref{sec:runn-coupl-effects}, we introduced the requirement
that for two soft and/or collinear partons of similar hardness and
close in rapidity, the observable should have a value of the same
order of magnitude as in the presence of just one of those partons.
As discussed in section~\ref{sec:further-requirements}, it is
difficult in numerical codes to unambiguously check the requirement
that two values be of the same order of magnitude. It is instead
simpler (and equivalent) to check that they have the same scaling as
one varies an overall softness scale. A similar requirement, \ie that
the observable scale in the same fashion with multiple emissions as
with a single emission, also appeared in the context of our treatment
of multiple independent emission in section~\ref{sec:mult-indep-emsn}.
Both of these requirements can actually be expressed in terms of a
single condition, as follows.

We define a momentum function $\kappa_i(\zeta)$ such that
$V(\kappa_i(\zeta),\{\tilde p\}) = \zeta$, in analogy with the rescalable
momentum $k_t^{(\rho)}$ of eq.~(\ref{eq:kti_rho}), but generalised so
that it is now specified by three parameters, $\eta^0_i$, $\xi_i$, and
$\phi_i$, as well as the leg index $\ell_i$,
\begin{equation}
  \label{eq:kappaSec2Def}
  \eta_i(\zeta) = \eta^{0}_i - \frac{\xi_i \ln
    \zeta}{a+b_{\ell_i}}\,,\qquad\quad
  \kappa_{ti}(\zeta)= 
  \left( \frac{\zeta e^{b_{\ell_i}\eta_i(\zeta)}}{d_{\ell_i}
      g_{\ell_i}(\phi_i)} \right)^{\frac1a},
  \qquad\quad
  \phi_i(\zeta) \equiv \phi_i\,,
\end{equation}
corresponding to a linear path in the
$\eta$--$\ln k_t$ plane.  Asymptotically ($\zeta \to 0$), any other
functional form will either be nonsensical (\eg outside the allowed
phase-space) or else approximate a linear path.
Not all values are allowed for the parameters, in particular
$0\le\xi_i\le1$ and for certain values of the $\eta^0_i$ the emission
may be kinematically allowed only for sufficiently small $\zeta$ (or
disallowed altogether if $\xi_i = 1$).

We then require that for any $m$ such momentum functions, independently
of the $\xi_i$, $\eta^0_i$ and $\phi_i$ ($i=1\ldots m$), the following limit,
\begin{equation}
  \label{eq:Sec2rIRClimit1}
  \lim_{{\bar v} \to 0} \frac{1}{{\bar v}} 
  V(\{\tilde p\},\kappa_1({\bar v} \zeta_1), \ldots,
  \kappa_m({\bar v} \zeta_m))\,,
\end{equation}
be well-defined and non-zero, for any choice of (non-zero) values of
the $\zeta_i$. We introduce here $\bar v$, a parameter that we are
free to vary to probe the observable's properties, and distinguish it
from $v$ which, from now on, will just denote the particular value of the
observable for which we resum the distribution.

The above condition guarantees that in the limit of small $\bar v$,
any set of emissions close to the boundary $V(\{\tilde p\},k) = \bar
v$ will lead to a value for the observable of order $\bar v$ (because
otherwise, there will exist choices of $\kappa_i(\zeta)$ and $\zeta_i$
such that the limit is infinite, zero, or ill-defined).

By defining some of the $\kappa_i$ in eq.~(\ref{eq:Sec2rIRClimit1})
such that they have identical $\xi_i$, one obtains a set of momenta
that stay close in rapidity as the emissions are scaled towards the
soft-collinear limit, precisely like the correlated emissions of
section~\ref{sec:runn-coupl-effects}, thus embodying the rIRC safety
condition discussed in the first paragraph of the part of
section~\ref{sec:runn-coupl-effects} labelled `Observable's dependence
on correlated gluon emission'.
Taking different $\xi_i$ values for the various emissions, the part of
rIRC safety expressed in eq.~(\ref{eq:scalingprop}) is also embodied,
since we are free to fix the $\eta^0_i$ so as to reproduce
eq.~(\ref{eq:kti_rho}).

Eq.~(\ref{eq:Sec2rIRClimit1}) lays the basis for the next part of the
rIRC safety condition, that which actually resembles plain IRC safety.
For plain IRC safety, given some ensemble of partons, one requires
that the additional soft and/or collinear splitting of one (or more)
of those partons not change the value of the observable by more than a
positive power of the softness/collinearity of the splitting(s),
normalised to the hard scale $Q$.

In the context of resummations we effectively have two relevant scales:
$Q$ and the scale (as a function of rapidity) set by the boundary
$V(\{\tilde p\},k) = v$. To be able to carry out the resummation, in
addition to IRC safety, we needed to assume (see sections
\ref{sec:towards-all-orders}, \ref{sec:mult-indep-emsn}) that, for
sufficiently 
small $v$, there exists some $\epsilon \ll 1 $, that can be chosen
\emph{independently} of $v$, such that we can neglect any splitting
that is at a smaller scale than that defined by $\epsilon v$ (to
within an absolute correction to $V$ of order
 $\epsilon^{p}v$, with $p$ some positive power). The crucial
extension compared to plain IRC safety is in the fact that one should be
able to choose $\epsilon$ independently of $v$.

In terms of the momentum functions $\kappa_i(\zeta)$, textbook
statements of IRC safety can be expressed with requirements such as
\begin{equation}
  \label{eq:Sec2IRClimit}
  \lim_{\zeta_{m+1}\to 0} 
  V(\{\tilde p\},\kappa_1({\bar v} \zeta_1), \ldots, \kappa_m({\bar v}
  \zeta_m), \kappa_{m+1}({\bar v}
  \zeta_{m+1})) \\= 
  V(\{\tilde p\},\kappa_1({\bar v} \zeta_1), \ldots, \kappa_m({\bar v}
  \zeta_m))\,.
\end{equation}
The equivalent statement for rIRC safety exploits the scaling property
of the observable, \ie the fact that eq.~(\ref{eq:Sec2rIRClimit1}) is
well-defined and finite, to then define a double limit,
\begin{multline}
  \label{eq:Sec2rIRClimit2a}
  \lim_{\zeta_{m+1}\to 0} \lim_{{\bar v} \to 0} \frac{1}{{\bar v}} 
  V(\{\tilde p\},\kappa_1({\bar v} \zeta_1), \ldots, \kappa_m({\bar v}
  \zeta_m), \kappa_{m+1}({\bar v}
  \zeta_{m+1})) \\= 
  \lim_{{\bar v} \to 0} \frac{1}{{\bar v}} 
  V(\{\tilde p\},\kappa_1({\bar v} \zeta_1), \ldots, \kappa_m({\bar v}
  \zeta_m))\,. 
\end{multline}
The order of the limits on the left-hand-side is crucial: if the
softness/collinearity $\zeta_{m+1}$ at which the $(m+1)^\mathrm{th}$
emission becomes irrelevant depends on $\bar v$ (\eg it scales as a
power of $\bar v$), then for infinitely small $\bar v$ the
$(m+1)^\mathrm{th}$ emission will never become irrelevant and the
equality will not be satisfied.  One can also combine
eqs.~(\ref{eq:Sec2IRClimit}) and (\ref{eq:Sec2rIRClimit2a}) to obtain
an alternative statement of the rIRC safety condition in terms of the
commutator of the limits:
\begin{equation}
  \label{eq:Sec2rIRClimit2aComm}
  \left[\lim_{\zeta_{m+1}\to 0} , \lim_{{\bar v} \to 0}\right]\
  \frac{1}{{\bar v}} 
  V(\{\tilde p\},\kappa_1({\bar v} \zeta_1), \ldots, \kappa_m({\bar v}
  \zeta_m), \kappa_{m+1}({\bar v}
  \zeta_{m+1})) =  0\,.
\end{equation}

In addition to the limit where an extra emission is made
soft/collinear to the hard leg (relevant in sections
\ref{sec:towards-all-orders} and \ref{sec:mult-indep-emsn}), we also
need to consider the situation where one or more existing emissions
split softly and/or collinearly (as discussed in
sections~\ref{sec:runn-coupl-effects} and
\ref{sec:towards-all-orders}). We represent the collinear splitting of
an existing emission with the notation $\kappa_i(\zeta) \to
\{\kappa_{i_a},\kappa_{i_b}\}(\zeta, \mu)$, such that $\mu^2 =
(\kappa_{i_a} + \kappa_{i_b})^2/\kappa_{ti}^2$ and $\lim_{\mu\to0}
\left(\kappa_{i_a} + \kappa_{i_b}\right) = \kappa_i$. We then for
example require
\begin{multline}
  \label{eq:Sec2rIRClimit2b}
  \lim_{\mu\to 0} \lim_{{\bar v} \to 0} \frac{1}{{\bar v}} 
  V(\{\tilde p\},\kappa_1({\bar v} \zeta_1), \ldots, 
  \{\kappa_{i_a},\kappa_{i_b}\}({\bar v}\zeta_i, \mu),\ldots
  \kappa_m({\bar v}
  \zeta_m)) \\= 
  \lim_{{\bar v} \to 0} \frac{1}{{\bar v}} 
  V(\{\tilde p\},\kappa_1({\bar v} \zeta_1), \ldots, 
  \kappa_i({\bar v} \zeta_i),\ldots,
  \kappa_m({\bar v}
  \zeta_m))\,,
\end{multline}
regardless of how the $\mu\to0$ limit is taken --- \ie whether at
fixed relative energy fractions for $\kappa_{i_a}$, $\kappa_{i_b}$ or
with one of them simultaneously becoming softer than the other. Making
use of plain IRC safety, eq.~(\ref{eq:Sec2rIRClimit2b}) too can can be
expressed in terms of a commutator of limits, as in
eq.~(\ref{eq:Sec2rIRClimit2aComm}).

Eqs.~(\ref{eq:Sec2rIRClimit2a}) and (\ref{eq:Sec2rIRClimit2b}),
together with the condition that (\ref{eq:Sec2rIRClimit1}) be well
defined, summarise
the requirements that make up rIRC safety. Combined with the form
eq.~(\ref{eq:parametricform}) for the observable in the soft and
collinear limit, and the continuous globalness condition, rIRC safety
ensures the validity, to NLL accuracy, of the approximations that we
have made in order to arrive at eq.~(\ref{eq:Result:expRcF}) together
with eqs.~(\ref{eq:sigma1_standard}) and (\ref{eq:cF-with-vto0-limit}).

It is quite deliberately that we stated that
eqs.~(\ref{eq:Sec2rIRClimit2a}) and (\ref{eq:Sec2rIRClimit2b})
`summarise' rIRC safety. Indeed, as discussed in
appendix~\ref{sec:IRC}, widespread mathematical statements of
normal IRC safety, such as eq.~(\ref{eq:Sec2IRClimit}), are also
merely summaries of full IRC safety, for which we are not aware of a
complete mathematical formulation. Thus for IRC safety it is in some
respects more accurate to state it in words, \ie that an observable
should be insensitive to soft and/or collinear branchings. Similarly,
part 2 of rIRC safety is more fully described by the paragraph
preceding eq.~(\ref{eq:Sec2IRClimit}) than by
eqs.~(\ref{eq:Sec2rIRClimit2a}) and (\ref{eq:Sec2rIRClimit2b}), and,
for example, exceptions to eq.~(\ref{eq:Sec2rIRClimit1}) are allowed, as
long as they correspond to sets of configurations having zero measure
in the space $\prod_i (d\zeta_i d\eta^0_i d\phi_i)$.

A more general question arises as to whether rIRC safety and
continuous globalness, are the necessary and sufficient conditions for
the form eq.~(\ref{eq:vProb-general}) to be correct. Since
eq.~(\ref{eq:Result:expRcF}) is a specific case of
eq.~(\ref{eq:vProb-general}), these conditions should certainly be
sufficient.  However they are not necessary --- we know for example of
non-global observables that have a resummed distribution of the form
eq.~(\ref{eq:vProb-general}) \cite{NG1,DiscontGlobal}. Furthermore, we
believe that there exist observables that fail a subset of the rIRC
conditions, but which nevertheless also have a resummed distribution
with the structure eq.~(\ref{eq:vProb-general}) --- we suspect the
Geneva jet resolution threshold, discussed in
appendix~\ref{sec:Geneva}, of being an example of such an observable.

The specificity of observables that fail some of our applicability
conditions, but at the same time have a resummed structure as in
eq.~(\ref{eq:vProb-general}) is, we suspect, that they cannot be
resummed by just the double logarithmic resummation of virtual
contributions with corrections coming exclusively from dynamics at the
scale of the lower boundary of the excluded region. A deeper
understanding of this point would, however, require further
investigation.

\subsection{Generalisation to other Born configurations}
\label{sec:more-general-born}

The discussion till now has been limited to the case of a Born
configuration consisting of a single outgoing hard quark-antiquark
pair. Much of it carries over relatively straightforwardly
to more general cases --- in particular all the applicability
conditions that we have discussed are simply related to the general
classes of divergence that are found in multi-parton matrix elements
(soft and collinear divergences with respect to the Born partons and
also with respect to any further emitted partons). No new classes of
divergence appear when going beyond the simple 2-outgoing jet case
studied above, and therefore the applicability conditions will remain
valid quite generally.

The main non-trivial modifications relative to the results of
section~\ref{sec:allorders} concern the structure of
single-logarithmic contributions associated with incoming legs
(collinear single-logarithms) and with the colour structure of
configurations with more than two legs (soft large-angle
single-logarithms), both of which issues have been extensively
discussed in the literature.

\subsubsection{Incoming hard legs} 
\label{sec:IncomingHardLegs}

Implicit in our discussion so far is
that the probability $\vProb(v)$ multiplies the hard cross section for
the underlying Born event, \cnf eq.~(\ref{eq:Sigmacut_resummed}). In
processes with incoming legs, that hard
cross section is evaluated using a procedure which factorises
collinear divergences along each incoming leg into an associated
parton density function, $q(x,\mu_F^2)$. One generally chooses a
factorisation scale $\mu_F$ of the order of the hard scale $Q$.

The factorisation procedure for the Born cross section involves an
integration over collinear 
emissions with transverse momenta up to scale $\mu_F$, which `builds
up' the parton density function at scale $\mu_F$. However, when one
places a limit $v$ on the value of the final-state observable, one
vetoes collinear emissions with $(k_t/Q)^{a+b_\ell} \gtrsim v$.  Due
to rIRC safety, the remaining hard collinear emissions (those at lower
values of $k_t$) do not affect
the value of the observable, and can be therefore integrated over to
`build up' the parton density to a scale of the order of $Q v^{1/(a +
  b_\ell)}$. 
The probability $\vProb(v)$ therefore includes a correction factor
\begin{equation}
  \label{eq:pdf-correct}
  \frac{q(x,\mu_F^2 v^{2/(a + b_\ell)})}{q(x,\mu_F^2)}\,,
\end{equation}
so that the parton density $q(x,\mu_F^2)$ that was included in the
Born cross section is effectively replaced with a parton density at
the new, lower factorisation scale (the choice of $Q v^{1/(a +
  b_\ell)}$ or $\mu_F v^{1/(a + b_\ell)}$ being of course arbitrary,
since they differ only by NNLL corrections).  We note that above the
scale $\mu_F v^{1/(a + b_\ell)}$ there remains the virtual part of the
collinear corrections, already accounted for by the $B_\ell$ term in
eq.~(\ref{eq:sigma1_standard}).

The above result is simply a generalisation of the one for the widely
studied Drell-Yan transverse momentum resummation
\cite{CSS,KodairaTrentadue,CollinsSoper}, and has also been quite
extensively discussed for event shapes
\cite{ADS,KoutZ0,KoutDIS,AzimDIS}. Nevertheless, in
Appendix~\ref{sec:hardcollinear} we revisit the derivation of
eq.~\eqref{eq:pdf-correct}, giving special emphasise to the
requirement of rIRC safety.

\subsubsection{Three hard legs} 
\label{sec:3HardLegs}

NLL final-state resummations for
Born events consisting of a hard quark-(anti)quark pair and a hard
gluon have been discussed in \cite{eeKout,KoutZ0,KoutDIS,AzimDIS}. The
treatment of a general observable in the 3-jet case mirrors quite
closely that given above for 2 jets.  The main difference is that
$R(v)$ originates from a sum over \emph{three} dipoles, as opposed to
a single dipole: a $q q'$ dipole ($q$ and $q'$ being respectively the
quark and (anti)quark) which is associated with a colour factor $(\CF
- \CA/2)$, and the $q g$ and $q'g$ dipoles ($g$ being the gluon) each
associated with the colour factor $\CA/2$.

Schematically one can therefore write $R$ as
\begin{multline}
  \label{eq:sigma3_sumdipoles}
  R(v) = \sum_{\mathrm{dipoles}}  C_{\mathrm{dipole}} \left( \sum_{\ell \in
      \mathrm{dipole}}
    \left[r_\ell(L) + r_\ell'(L) \left(\ln {\bar d}_\ell - b_\ell
        \ln\frac{2E_\ell}{Q}\right) + 
    \right.\right.\\\left. \left. 
    + B_\ell\,
    T\!\left(\frac{L}{a+b_\ell} \right) \right]
  + 2 \, T\!\left(\frac{L}{a}\right) \ln \frac{Q_\mathrm{dipole}}{Q}\right),
\end{multline}
where $C_\mathrm{dipole}$ is the colour factor associated with the
dipole. Note that for the gluonic leg, $B_\ell$ has a different
value than in the quark case. It can be determined
from the collinear matrix elements for $g \to gg$ and $g \to q\bar q$
splitting,
\begin{subequations}
  \label{eq:gg_qg_mat_elements}
\begin{align}
  \label{eq:gg_mat_elements}
  |M^2_{\ell,g\to gg}(k)| &= \frac{\as C_A}{2\pi}\, \frac{ z^{(\ell)}
    p_{gg}(z^{(\ell)}) }{k_t^2}\,,
  \qquad p_{gg}(z) = \frac{2(1-z)}{z} + z(1-z)\,,\\
  \label{eq:qg_mat_elements}
  |M^2_{\ell,g\to q\bar q}(k)| &= \frac{\as T_R}{2\pi}\, \frac{
    z^{(\ell)} p_{qg}(z^{(\ell)}) }{k_t^2}\,,
  \qquad p_{qg}(z) = z^2 + (1-z)^2\,,
\end{align}
\end{subequations}
where we have exploited the $z\leftrightarrow 1-z$ symmetry of the
$g\to gg$ splitting to write $p_{gg}$ such that it only has a $z\to 0$
divergence (\cnf eq.~(5.41) of \cite{ESWbook}). The gluonic $B_\ell$
value is then given by 
\begin{equation}
  \label{eq:Bell_gluon}
  B_\ell^{(\mathrm{gluon})} = 
  \int_0^1 \frac{dz}{z} \left(\frac{z p_{gg}(z)}{2}  + \frac{\TR \nf\,
      zp_{qg}(z)}{2\CA} -
    1\right) = \frac{-11\CA + 4\TR\nf}{12\CA}\,,
\end{equation}
where, as in eq.~(\ref{eq:Bell}), we extract the overall colour factor
associated with the soft divergence, which will reappear below, after
explicitly summing over dipoles.

The presence of sums over dipoles and over their associated legs in
eq.~(\ref{eq:sigma3_sumdipoles}) is somewhat cumbersome (as well as
difficult to generalise subsequently). However we can invert the order
of the sums over legs and dipoles, and perform the sum over dipoles to
obtain
\begin{multline}
  \label{eq:sigma3_sumlegs}
  R(v) =   \sum_{\ell=1}^n C_\ell
    \left[r_\ell(L) + r_\ell'(L) \left(\ln {\bar d}_\ell - b_\ell
        \ln\frac{2E_\ell}{Q}\right)
    + B_\ell\,
    T\!\left(\frac{L}{a+b_\ell} \right) \right]
  \\
  - \ln S\left(T\left(\frac{L}{a}\right)\right),
\end{multline}
where $n=3$ is the number of legs, and we have exploited the fact that
for each leg, $\sum_{\mathrm{dipole} \supset \{\ell\}}
C_{\mathrm{dipole}} = C_\ell$,\footnote{With the (formal) notation
  $\mathrm{dipole} \supset \{\ell\}$
we indicate a dipole such that $\ell \in \mathrm{dipole}$.}
 with $C_\ell$ the colour factor of the
given leg, $\CF$ for the (anti)quarks and $\CA$ for the gluon. The
function $S$ collects the terms that cannot be conveniently expressed as
a sum over individual legs,
\begin{equation}
  \label{eq:S3legs}
  \ln S(t) = -t \left[{\CA}\ln \frac{Q_{qg} Q_{q'
        g}}{Q_{q q'} Q} + 2\CF \ln \frac{Q_{q q'}}{Q}\right]\,.
\end{equation}
One can verify that the $Q$ dependence of $\ln S(t)$ is reducible to
the form $t\, C_T \ln Q$, with $C_T =\sum_\ell C_\ell$, as is necessary
for $R(v)$ overall to be $Q$-independent. The remaining part of $S$
accounts for the coherent structure of large-angle radiation from the
ensemble of hard legs.

Eq.~(\ref{eq:sigma3_sumlegs}) is of course only the single-gluon
result. The full all-order result needs to be obtained by following a
procedure analogous to that given in section~\ref{sec:allorders}. As
was shown in \cite{eeKout} the decomposition into a structure of three
dipoles holds at all orders, which means that the analysis carries
through essentially unchanged, the only difference being that, for
$\cF$, the sum over two legs in
eqs.~(\ref{eq:rewrite-dk-m2})--(\ref{eq:cF-evshps}) should be
generalised to a sum over three legs and $\CF$ should be replaced with
the appropriate leg colour factor.

We finally note\footnote{We are grateful to Yuri Dokshitzer for
  bringing this to our attention.} that processes such as $gg \to
\mathrm{Higgs} + g$, which involve three gluonic legs, or equivalently
three gluon-gluon dipoles, can be treated in a similar manner, the
only difference being that each dipole is associated with a colour
factor $\CA/2$, so that in eq.~(\ref{eq:S3legs}) one needs to replace
$\CF$ with $\CA$.

\subsubsection{Four hard legs and beyond} 
\label{sec:FourHardLegs}

A crucial property of the two and three-jet cases is that there is a
unique structure of colour flow for the underlying hard process --- a
single dipole in the two-jet case, and a sum over (the $3$) dipoles
made from all pairs of hard legs in the $3$-jet case. This means that
a loop virtual correction does not change the colour structure of the
underlying hard event, and it is this property that allows us to
straightforwardly exponentiate the single-gluon term, $\int[dk] \,
|M^2_{rc}(k)|$, in the virtual corrections in
eq.~(\ref{eq:MultipleIndepEmsn}).

In processes with four or more hard jets the situation is more
complex, as was discussed originally
in~\cite{BottsSterman,KS,KOS,Oderda,KidonakisOwens}.
To 
illustrate the point concretely, let us consider the process $q\bar
q\to q\bar q$, where for example the incoming $q\bar q$ pair can form
a colour singlet or a colour octet. Both the hard matrix element and
the pattern of large-angle soft radiation (and associated virtual
corrections) depend on the overall colour of the incoming pair.
Additionally a loop correction (stretched say across the incoming and
outgoing quarks) can modify the overall colour of the $q\bar q$ pair
entering the hard scattering: loop corrections introduce mixing
between the different colour structures, and at all orders one needs
to resum the resulting mixing matrix.

As explained in detail in \cite{KOS}, one needs to keep track
separately of the colour channel of the Born amplitude and its complex
conjugate. Denoting the possible colour channels by an index $I$,
one has (in the notation of the above papers) a matrix $H_{II'}$
for the product of the Born amplitude and its complex conjugate,
respectively in colour channels $I$, $I'$, modulo the
normalisation associated with the colour algebra, contained in a
matrix $M_{II'}$ (which, for the orthogonal, but non-normal choice
of colour basis in \cite{KOS,Oderda} is a diagonal matrix). The
differential Born cross section (modulo an overall kinematic
normalisation) is given by $\Tr(HM) \equiv H_{II'}
M_{I'I}$.

In the two and three-jet cases we had been able to write the
(double-logarithmic) virtual corrections as the exponential of an
integral over the single-gluon matrix element, $\exp\left(-R(v)\right)
= \exp\left(-\int_v [dk] |M^2_{rc}(k)|\right)$.  As already indicated,
in the multi-jet case, the virtual corrections are now matrices in
colour space, which act separately on the amplitude and its complex
conjugate. Given the form of these matrices as presented in
\cite{BottsSterman,KS,KOS,Oderda,KidonakisOwens} (specifically for
$2\to2$ scattering) and \cite{Bonciani:2003nt} (in generic form),
$\exp\left(-R(v)\right)$ can be written
\begin{multline}
  \label{eq:matrixsuppression}
  e^{-R(v)} \equiv \frac{
    (e^{-\frac12\int_v [dk] \gamma(k) - \frac{i}{2}  C
      T\left(\frac{L}{a}\right)})_{JI}
    (e^{-\frac12\int_v [dk] \gamma(k) + \frac{i}{2}  C
      T\left(\frac{L}{a}\right)})_{J'I'}
    H_{II'} M_{J'J}} 
    {\Tr(HM)} 
    =\\=
  \frac{
    \Tr\left(e^{-\frac12\int_v [dk] \gamma(k)- \frac{i}{2} C 
        T\left(\frac{L}{a}\right)} 
    H e^{-\frac12\int_v [dk] \gamma^\dagger(k)+ \frac{i}{2} C^\dagger
      T\left(\frac{L}{a}\right)} M\right) } 
    {\Tr(HM)} \,.
\end{multline}
where both $\gamma_{JI}(k)$ and $C_{JI}$ are real matrices.  Each
entry of $\gamma_{JI}(k)$ can be written as a linear combination of
dipole emission matrix elements. Furthermore in the limit of $k$ being
collinear to a hard leg $\ell$, $\gamma_{JI}(k) = |M^2_{\ell,rc}(k)|
(\delta_{JI} + \order{\smash{e^{-\eta^{(\ell)}}}})$, where
$|M^2_{\ell,rc}(k)|$ is the matrix element for emission collinear to
leg $\ell$, eq.~(\ref{eq:matrixelement_ell}), with the addition of a
running coupling\footnote{For the exact formula in the case of
  incoming legs, we refer the reader to
  appendix~\ref{sec:hardcollinear}.} --- this reduction, in the
collinear limit, of $\gamma_{JI}(k)$ to $|M^2_{\ell,rc}(k)| \delta_{JI}$
is a manifestation of coherence, \ie the independence of the dynamics
at small angular scales from that at large angular scales.
In the cases studied so
far (and we believe more generally), the entries of the $C_{JI}$
matrix are simply combinations of colour factors (\ie without any
dependence on the Born kinematics), whose structure is closely related
to that of the $\gamma_{JI}(k)$. They give rise to the non-cancelling
parts of Coulomb phases.\footnote{There can be additional Coulomb
  phases proportional to the unit matrix, however these cancel between
  the amplitude and its complex conjugate.  Indeed, Coulomb phases are
  also present in cases with fewer than four jets, but there cancel
  completely and so can be ignored. For a further discussion of
  Coulomb phases in the context of QCD
  resummations, we refer the reader to \cite{Bonciani:2003nt}.} %
One notes that $C_{JI}$ is multiplied by
$T\left(\frac{L}{a}\right)$, a signal of its large-angle origin --- quite
generally, it is only the (single-logarithmic) virtual corrections
that are the counterpart of large-angle emission that lead to a mixing
of colour states.

Let us now trace the structure of the calculation of $\vProb(v)$ in
section~\ref{sec:mult-indep-emsn} and examine how to modify it in
light of the above complications.  We introduce a function ${\overline
  M}_{rc}^2(k) \delta_{JI}$ which coincides with $\gamma_{JI}(k)$ in
all collinear regions, but with some simple behaviour at large angles
(for concreteness we use $q\bar q \to q \bar q$ as an example case,
and there take ${\overline M}_{rc}^2(k)$ to be the incoherent sum of
emission from a $12$ dipole and from a $34$ dipole).

We can still express $R(v)$ as in eq.~(\ref{eq:sigma3_sumlegs})
(where, we recall, $n$ is the number of hard legs).  It is convenient
to write
\begin{equation}
  \label{eq:SSbar}
  \ln S(t) = \ln \overline S(t) + \ln \Delta(t)\,,
\end{equation}
where $\ln \overline S(t)$ accounts for the large-angle single
logarithms that arise with our approximate ${\overline M}_{rc}^2(k)
\delta_{JI}$ matrix element (in our $q\bar q \to q \bar q$
example, $\ln \overline S(t) = -t \cdot 4\CF \ln Q_{12}/Q$). One then
has, to NLL accuracy,
\begin{equation}
  \label{eq:Delta}
  \Delta(t) = \frac{\mathrm{Tr} (H
       e^{-t\Gamma^\dagger/2} M
       e^{{-t}\Gamma/2})}{\mathrm{Tr} (H M)}\,,
\end{equation}
where the `anomalous dimension' 
coefficient matrix $\Gamma$ \cite{BottsSterman} is defined through
\begin{equation}
  \label{eq:Gamma}
  \frac{\as(k_t^2)}{\pi} \Gamma_{JI} = \int [dk']
  \left(\gamma_{JI}(k') - 
    {\overline M}_{rc}^2(k') \delta_{JI}\right) 
  \delta\left(\ln \frac{{k_t'}^2}{k_t^2}\right) + 
  iC_{JI}  \frac{\as(k_t^2)}{\pi} + 
  \order{\as^2}\,,
\end{equation}
and is independent of $k_t$.  We recall that at large angles the
definition of $k_t^2$ is arbitrary to within a factor of order one ---
this arbitrariness is related to NNLL $\order{\as^2}$ corrections to
the above equation.

In addition to treating the virtual corrections that lead to the
factor $e^{-R(v)}$ in eq.~(\ref{eq:Result:expRcF}), we also need to
carry out the sum over real emissions (and corresponding virtual
corrections) at scales $\epsilon v < V(\{\tilde p\},k) \lesssim v$ in
order to obtain the single-logarithmic `multiple-emission' correction
factor $\cF$. Accounting for the full colour-matrix structure of
large-angle real-emission would complicate this task quite
substantially. However, at NLL accuracy, $\cF$ only receives
contributions from the region of soft \emph{and} collinear emissions
and we are free to carry out the calculation with the approximate
${\overline M}_{rc}^2(k)$ matrix element --- the error in this
approximation comes from a region of large angles and transverse
momentum scales $(\epsilon v)^{1/a} \lesssim k_t/Q \lesssim v^{1/a}$,
and so is of order $\as \ln 1/\epsilon$, which by virtue of coherence
is a multiplicative correction, \ie truly NNLL.

Thus the resummed result with for processes with four or more hard
legs is of the form given in eq.~(\ref{eq:Result:expRcF}), with an
$R(v)$ defined as in eq.~(\ref{eq:sigma3_sumlegs}), $S$ as in
eq.~(\ref{eq:SSbar}), and with $\cF$
calculated with an approximate real-emission matrix element that is
required to be correct only in the collinear limit. The colour-matrix
structure of the problem appears only in the large-angle soft
resummation function $S(t)$.\footnote{As we have seen, because of this
  multi-channel nature of the problem, $R(v)$ loses its direct
  interpretation as an integral over the single-gluon emission
  probability. }

The only case with four or more jets in which the mixing has been
explicitly calculated in the literature is that of dijet production in
hadron-hadron scattering; $\ln S(t)$ can be written as a sum of two
terms, as discussed above:
\begin{equation}
  \label{eq:S-hadronic-dijet}
  \ln S(t) = -t \sum_\ell C_\ell \ln \frac{Q_{12}}{Q}
       + \ln \frac{\mathrm{Tr} (H
       e^{-t\Gamma^\dagger/2} M
       e^{{-t}\Gamma/2})}{\mathrm{Tr} (H M)} \,.
\end{equation}
The first term is obtained by performing the resummed calculation as
if we were dealing with four `independent' hard legs each carrying a
momentum of half of a dipole of invariant mass $Q_{12}$. It contains
all the dependence on our arbitrary hard scale $Q$ (ensuring again the
independence of $R(v)$ on the hard scale $Q$).  The second term gives
the correction needed to properly account for the colour mixing of
large-angle radiation, as derived by the Stony Brook group
in~\cite{BottsSterman,KS,KOS,Oderda,KidonakisOwens}, largely in the
context of threshold resummations (though the results apply here too).
The full details, and the explicit forms for the matrices are
reproduced in appendix~\ref{sec:softlargeangle}.

As yet, analogous results for other processes do not exist in detail.
A general solution of the problem (for factorised observables), in
terms of an exponentiated matrix in the space of the colours of each
hard leg (as opposed to the smaller, minimal basis of \cite{KOS} for
the 4-jet case), has however been given in~\cite{Bonciani:2003nt}.
From that formulation one could envisage extracting, for an arbitrary
process, the function $S(t)$ that provides the large-angle single
logarithmic resummation contribution to our result,
eqs.~(\ref{eq:Result:expRcF}), (\ref{eq:sigma3_sumlegs}).

\section{Presentation and discussion of master formula}
\label{sec:master}

The different elements and applicability conditions of the resummed
prediction for a general observable are somewhat spread out across the
previous section. It is therefore convenient to summarise them all in
one location. This is the purpose of
section~\ref{sec:summary-master}. It is also of use to illustrate them
with some examples, notably (\cnf section~\ref{sec:thrust}) a case 
where the analytical resummation is well known, but also one
where the applicability conditions fail to hold (see
section~\ref{sec:rIRC}). Finally section~\ref{sec:div} discusses
issues related to the convergence of the function $\cF$.

\subsection{Master formula and applicability conditions}
\label{sec:summary-master}

Let us start by summarising the applicability conditions, including
some brief reminders of their physical origins. For the details of the
notation we refer the reader to the previous section.
\begin{itemize}
\item For a resummation that is to be carried out in the $n$-jet ($n$-leg)
  limit, the observable should vanish smoothly as a single extra
  ($n$+$1$)$^\mathrm{th}$ parton of momentum $k$ is made
  asymptotically soft and collinear to any leg $\ell$, the functional
  dependence being of the form (\cnf eq.~(\ref{eq:parametricform})): 
  \begin{equation}
    \label{eq:simple-second-time}
    V(\{{\tilde p}\}, k)=
    d_{\ell}\left(\frac{k_t^{(\ell)}}{Q}\right)^{a_\ell}
    e^{-b_\ell\eta^{(\ell)}}\, 
    g_\ell(\phi^{(\ell)})\>.
  \end{equation}
  As we have seen, the restriction to this (near universal) form makes
  it possible to carry out the LL part of the resummation entirely
  analytically.  IRC safety implies $a_\ell > 0$ and $b_\ell >
  -a_\ell$ (see also \cite{BKS03}). It is also necessary for the
  observable to be positive definite --- this is essential in order to
  retain the connection between an upper limit on the value of the
  observable, and an upper limit on the momenta of any emissions.  We
  recall that the $\{\tilde p\}$ are the Born momenta after recoil
  from the emission $k$. The functional dependence on $k$ of the
  relation between the original Born momenta $\{p\}$ and the $\{\tilde
  p\}$ is discussed in appendix~\ref{sec:recoil}.
   
\item The observable should be global \cite{NG1}, meaning that it
  departs from zero for any emission of an ($n$+$1$)$^\mathrm{th}$
  parton that is not infinitely soft or collinear. Furthermore it
  should be \emph{continuously} global \cite{DiscontGlobal}. Roughly,
  this means that the power of $k_t$ should be the same everywhere,
  implying $a_1 = \cdots = a_n \equiv a$. More formally the
  condition can be expressed as 
  \begin{equation}
    \label{eq:DiscontGlobal}
    \left. \frac {\partial \ln V(\{\tilde p\}, k)}{\partial \ln k_t^{(\ell)}} 
    \right|_{\mathrm{fixed}\;\eta^{(\ell)},\,\phi^{(\ell)}} \!\!\!\! = a\,,
    \qquad\quad
    \left. \frac {\partial \ln V(\{\tilde p\}, k)}{\partial \ln k_t^{(\ell)}} 
    \right|_{\mathrm{fixed}\;z^{(\ell)},\,\phi^{(\ell)}} \!\!\!\!=
    \;\;a + b_\ell\,,
  \end{equation}
  where $z^{(\ell)}$ is the longitudinal momentum fraction (or
  normalised Sudakov component) of emission $k$ along the direction of
  leg $\ell$. The two forms in (\ref{eq:DiscontGlobal}) should be
  valid respectively in the soft (and optionally collinear) region and
  in the collinear (and optionally soft) region.  The reason for the
  different formulations in the soft and in the collinear regions is
  that different forms of deviations from
  eq.~(\ref{eq:simple-second-time}) are permissible (\ie associated at
  most with NNLL contributions) in the soft large-angle and in the
  hard-collinear regions.
  
  We recall, from section~\ref{sec:singlegluonemission}, that the
  continuous globalness condition is useful because without it, any
  general resummation result would need to encode information about
  potential boundaries between regions with different $k_t$
  dependences of the observable and such boundaries could be
  arbitrarily complex.
\end{itemize}
The above two conditions are required in order to obtain the
(analytical) single-gluon result for the probability $\vProb(v)$ that
the observable is smaller than some value $v$. For a given emission
angle, condition~1 determines the maximum allowable transverse
momentum scale; condition~2 guarantees that small changes in angle do
not drastically change that scale.\footnote{A drastic change in scale
  being one involving a power of $v$.} In order to straightforwardly
resum the result we also need to ensure that the addition of extra
emissions does not drastically change this scale. This need appeared
in various contexts in section~\ref{sec:allorders}, and we express it
here through the following novel condition.
\begin{itemize}
\item The observable should be \emph{recursively} IRC (rIRC) safe ---
  given an ensemble of arbitrarily soft and collinear emissions, the
  addition of further emissions of similar softness or collinearity
  should not change the value of the observable by more than a factor
  of order one (\ie without any powers of $v$). The addition of
  relatively much softer or more collinear emissions (whether with
  respect to the hard leg or one of the other emissions) should not
  change the value of the observable by more than some power of the
  relative extra softness or collinearity.
  
  One can also express these conditions more mathematically in terms
  of limits.\footnote{The mathematical expression of rIRC safety that
    follows may seem more precise than the somewhat vague description
    that precedes it. But as discussed in appendix~\ref{sec:IRC} in
    the context of normal IRC safety, it turns out to be non-trivial
    to embody the full generality of IRC or rIRC safety using such 
    (seemingly precise) mathematical statements.} %
  We use momentum functions $\kappa_i(\zeta_i)$ (a concrete class of which was
  introduced in section~\ref{sec:rIRC-deriv}), that depend
  on the parameters $\zeta_i$ such that,
  \begin{equation}
    \label{eq:kidef}
    V(\{\tilde p\},\kappa_i(\zeta_i)) = \zeta_i \,,
  \end{equation}
  with the condition that in the soft and/or collinear limits,
  $\zeta_i \to 0$, the azimuthal angle $\phi_i$ of
  $\kappa_i(\zeta_i)$ should be fixed.  Each of the momentum
  functions $\kappa_1(\zeta)$, $\kappa_2(\zeta)$, etc.  may be
  different as long as they all satisfy eq.~(\ref{eq:kidef}) --- for
  example $\kappa_1(\zeta)$ might involve a scaling of
  $\kappa_{t1}\sim \zeta^{1/a} $ at fixed rapidity, while
  $\kappa_2(\zeta)$ might involve a scaling of $\kappa_{t2} \sim
  \zeta^{1/(a+b_\ell)}$ at fixed longitudinal momentum fraction. We
  also denote the collinear splitting of an existing emission by
  $\kappa_i(\zeta) \to
  \{\kappa_{i_a},\kappa_{i_b}\}(\zeta, \mu)$, such that $\mu^2 =
  (\kappa_{i_a} + \kappa_{i_b})^2/\kappa_{ti}^2$ and $\lim_{\mu\to0}
  \left(\kappa_{i_a} + \kappa_{i_b}\right) = \kappa_i$.
  
  The conditions for rIRC safety are then that
  \begin{itemize}
  \item[1.] the limit 
    \begin{equation}
      \label{eq:rIRClimit1}
      \lim_{{\bar v} \to 0} \frac{1}{{\bar v}} 
      V(\{\tilde p\},\kappa_1({\bar v} \zeta_1), \ldots,
      \kappa_m({\bar v} \zeta_m))
    \end{equation}
    should be well-defined and non-zero (except possibly in a region
    of phase-space of zero measure). This 
    expresses the
    requirement that the soft and collinear
    scaling properties of the observable should be the same regardless
    of whether there is just one, or many emissions.
  \item[2a.] the following two limits should be well-defined and identical,
    \begin{subequations}
      \label{eq:rIRClimit2}
    \begin{multline}
      \label{eq:rIRClimit2a}
      \lim_{\zeta_{m+1}\to 0} \lim_{{\bar v} \to 0} \frac{1}{{\bar v}} 
      V(\{\tilde p\},\kappa_1({\bar v} \zeta_1), \ldots, \kappa_m({\bar v}
      \zeta_m), \kappa_{m+1}({\bar v}
      \zeta_{m+1})) \\= 
      \lim_{{\bar v} \to 0} \frac{1}{{\bar v}} 
      V(\{\tilde p\},\kappa_1({\bar v} \zeta_1), \ldots, \kappa_m({\bar v}
      \zeta_m))\,, 
    \end{multline}
    \ie having taken the limit eq.~(\ref{eq:rIRClimit1}), the addition
    of an extra much softer and/or more collinear emission should not
    affect the value of the observable.
  \item[2b.] The analogue of eq.~(\ref{eq:rIRClimit2a}) should hold
    also for the collinear splitting of an existing emission
    \begin{multline}
      \label{eq:rIRClimit2b}
      \lim_{\mu\to 0} \lim_{{\bar v} \to 0} \frac{1}{{\bar v}} 
      V(\{\tilde p\},\kappa_1({\bar v} \zeta_1), \ldots, 
      \{\kappa_{i_a},\kappa_{i_b}\}({\bar v}\zeta_i, \mu),\ldots
      \kappa_m({\bar v}
      \zeta_m)) \\= 
      \lim_{{\bar v} \to 0} \frac{1}{{\bar v}} 
      V(\{\tilde p\},\kappa_1({\bar v} \zeta_1), \ldots, 
      \kappa_i({\bar v} \zeta_i),\ldots,
      \kappa_m({\bar v}
      \zeta_m))\,, 
    \end{multline}
    \end{subequations}
    this, regardless of how precisely the collinear limit is taken (it
    can for example involve a simultaneous soft limit of one of the
    daughters from the collinear splitting). Such equalities should
    hold also for the case of multiple extra emissions and/or
    collinear splittings.  
  \end{itemize}
  We note that at first sight eqs.~(\ref{eq:rIRClimit2}) closely
  resemble normal IRC safety --- however they actually differ
  critically, because of the order of the limits on the left-hand
  sides.  The novelty of the recursive IRC conditions is such that
  they deserve to be studied and explained with the aid of some
  concrete examples.  This will be done in
  section~\ref{sec:rIRC} and appendix~\ref{sec:further-rIRC-unsafe}.
  
\end{itemize}
Given the above conditions, the resummed probability $\vProb(v)$ that
an observable has a value less than $v$ can be written to
NLL accuracy as follows:
\begin{equation}
\begin{split}
  \label{eq:Master}
  \ln \vProb(v) &= -\sum_{\ell=1}^n C_\ell \left[r_\ell(L) + 
    r_\ell'(L) \left(\ln {\bar d}_\ell - b_\ell \ln
      \frac{2E_\ell}{Q}\right) 
    + B_\ell \, T\!\left(\frac{L}{a+b_\ell}\right)
  \right] \\
  &+ \sum_{\ell=1}^{n_i} \ln \frac{q^{(\ell)}(x_\ell,e^{-\frac{2L}{a+b_\ell}}
    \muf^2)}{q^{(\ell)}(x_\ell, \muf^2)}  
  + \ln S\left(T(L/a\right)) 
  + \ln \cF(C_1,\ldots,C_n;\lambda)
  \,,
\end{split}
\end{equation}
where ${\bar d}_\ell$ was defined in eq.~(\ref{eq:dbarell}), while $r_\ell$,
$r'_\ell$, and $T$ were given in eqs.~(\ref{eq:rell})--(\ref{eq:rpell})
and are evaluated in appendix~\ref{sec:rad}; $L = \ln 1/v$,
$\lambda = \beta_0\as L$; $C_\ell$ is the colour factor associated
with leg $\ell$; and the hard collinear correction term $B_\ell$ is
given by
\begin{equation}
  \label{eq:BellGen}
  B_\ell = \left\{
    \begin{array}{cl}
      {\displaystyle -\frac34}                          
        & \qquad \mathrm{quarks}\,,\medskip\\
      {\displaystyle -\frac{11\CA - 4\TR \nf}{12\CA}}   
        & \qquad \mathrm{gluons}\,.
    \end{array}
    \right.
\end{equation}
The number of incoming hadronic legs is denoted by $n_i$, and each of
them is associated with a parton distribution
$q^{(\ell)}(x_\ell,\muf^2)$ at Bjorken momentum fraction $x_\ell$ and,
in the Born cross section, at a hard factorisation scale $\muf^2 \sim
Q^2$. To guarantee the NLL accuracy of $\vProb(v)$ it is
sufficient to use just LL DGLAP evolution \cite{DGLAP} to resum the
collinear
(single) logarithms in the ratio
$q^{(\ell)}(x_\ell,e^{-\frac{2L}{a+b_\ell}} \muf^2)/q^{(\ell)}(x_\ell,
\muf^2)$.

Process dependence enters also through the large-angle soft
single-logarithms $S(T(L/a))$, discussed in
section~\ref{sec:more-general-born}, which can be summarised as
follows
\begin{subequations}
  \label{eq:S-summary}
\begin{align}
  n = 2:\!\! &\quad \ln S(t) = -t\cdot 2C_F \,\ln \frac{Q_{qq'}}{Q}\,,\\
  n = 3:\!\! &\quad \ln S(t) = -t \left[{\CA}\ln \frac{Q_{qg} Q_{q'
        g}}{Q_{q q'} Q}
    + 2\CF \ln \frac{Q_{q q'}}{Q}\right] \,,\\
  n = 4:\!\! &\quad \ln S(t) = -t \sum_\ell C_\ell \ln
  \frac{Q_{12}}{Q}
  + \ln \frac{\mathrm{Tr} (H e^{-t\Gamma^\dagger/2} M
    e^{{-t}\Gamma/2})}{\mathrm{Tr} (H M)} \,,
\end{align}
\end{subequations}
For the cases with $n=2,3$, there are also purely gluonic processes
(notably Higgs production), for which one simply replaces $\CF$ with
$\CA$ (and $q,q',g$ with $g_1, g_2, g_3$); the matrices $H$, $M$ and
$\Gamma$ in the $n=4$ case have currently been calculated only for
hadronic dijet production \cite{BottsSterman,KS,KOS,Oderda,KidonakisOwens} --- they are collected
in appendix~\ref{sec:softlargeangle}.

The last part of the general result~(\ref{eq:Master}) is the
single-logarithmic function $\cF$, discussed in
section~\ref{sec:mult-indep-emsn}. Since it is closely connected with
the third of our applicability conditions, it is convenient to adopt a
similar notation in its definition, giving
\begin{multline}
  \label{eq:cF-with-vto0-limit-anyn}
  \cF (C_1,\ldots,C_n;\lambda) = \\\lim_{\epsilon\to0}
  \frac{\epsilon^{R'}}{R'}
    \sum_{m=0}^{\infty} \frac{1}{m!}
  \left( \prod_{i=1}^{m+1} \sum_{\ell_i=1}^n
    \int_{\epsilon}^{1} \frac{d\zeta_i}{\zeta_i} 
    \frac{C_{\ell_i} r_{\ell_i}'}{\cN_{\ell_i}(\lambda/\beta_0)}
    \int_0^1 \frac{d\xi_i}{1 + \frac{a + (1-\xi_i)b_{\ell_i}}{a(a+b_{\ell})}
    2\lambda} 
  \int_0^{2\pi} \frac{d\phi_i}{2\pi}
  \right) 
  \times \\ \times 
  \delta\!\left(\ln \zeta_1\right)
  \exp\left(-R'\ln \lim_{\bar v\to0} \frac{V(\{\tilde
      p\},\kappa_1(\zeta_1 \bar v), \ldots,
      \kappa_{m+1}(\zeta_{m+1} \bar v))}{\bar v}\right),
\end{multline}
where $R' = \sum_\ell C_\ell r_\ell'$, $\cN_\ell$ is simply a
normalisation, defined in eq.~(\ref{eq:xi-and-norm-defs}), and the
$\kappa_{i}(\bar v)$ are a shorthand for
\begin{equation}
  \label{eq:kappa_i}
  \kappa_i(\bar v) \equiv \kappa (\bar v; \ell_i, \phi_i, \xi_i)\,,
\end{equation}
where $\kappa(\bar v; \ell, \phi, \xi)$ is the momentum collinear to
leg $\ell$ with azimuthal angle $\phi$, and rapidity $\eta =
\frac{\xi}{a+b_{\ell}} \ln \frac1{\bar v}$, such that $V(\{\tilde
p\},\kappa(\bar v; \ell, \phi, \xi)) = \bar v$.

Note that in eq.~(\ref{eq:cF-with-vto0-limit-anyn}), relative to
eq.~(\ref{eq:cF-with-vto0-limit}), we have explicitly introduced the
limit $\epsilon\to0$. One thus clearly sees the reason for the rIRC
requirements --- condition (a) ensures that the $\bar v \to 0$ limit
is well defined, while condition (b), specifically
eq.~(\ref{eq:rIRClimit2a}), ensures that one can also safely take the
$\epsilon \to 0$ limit, the particular order of the limits being
dictated by the condition $\ln \frac1{\bar v} \gg \ln \frac1\epsilon
\gg 1$ that is crucial in making the approximations that ensure that
eq.~(\ref{eq:cF-with-vto0-limit-anyn}) is truly single logarithmic.
In many cases of rIRC unsafe observables, the result for $\cF$
diverges as one takes the limits $\bar v\to 0$ and $\epsilon\to0$.

A further point concerns the arguments of $\cF$. As can be seen from
eq.~(\ref{eq:cF-with-vto0-limit-anyn}), $\cF$ depends on the
$r'_\ell$, which were defined, eq.~(\ref{eq:rpell}), as functions of
$L$ (and implicitly, $\as$). At NLL accuracy, $r'_\ell$ actually
depends only on the combination $\lambda=\as \beta_0 L$. However it
was natural to write it as a function separately of $L$ and $\as$,
eqs.~(\ref{eq:T}), (\ref{eq:rpell}), since one might wish to compute
the relevant integrals beyond NLL accuracy. While a simple such
extension might make sense for $r'_\ell$, it would make much less
sense for $\cF$, because of the various sources of NLL approximation
that entered its derivation. We emphasise this by explicitly writing
$\cF$ in terms of the NLL parts of the $r'_\ell$, making it a function
solely of $\lambda$ and the colour factors.
In some contexts we will use a more
compact notation, $\cF(R')$. This is motivated by the fact that, for
many observables, $\cF$ depends principally (or even exclusively) on
the overall value of $R'$ rather than on the separate $C_\ell$ and
$\lambda$ values.

Finally, we quote also a simplified form for $\cF$, corresponding to
eq.~(\ref{eq:cF-evshps}).  This is valid (and of considerable
practical importance) for the
many observables that have the property that they do not depend on the
values of the $\xi_i$, \ie for which (for sufficiently small $\bar v$)
one can exchange any given set of $\xi_i$ values with a new set
$\xi_i'$ without changing the value of the observable:
\begin{multline}
  \label{eq:exchange_xi}
    V(\{\tilde p\} \,,\, \kappa(\zeta_1\bar
    v;\ell_1,\phi_1,\xi_1 )\,,\,\ldots \,,\, \kappa(\zeta_n\bar
    v;\ell_n,\phi_n,\xi_n ))
    =\\ =
    V(\{\tilde p\} \,,\, \kappa(\zeta_1\bar
    v;\ell_1,\phi_1,\xi_1' ) \,,\, \ldots \,,\, \kappa(\zeta_n\bar
    v;\ell_n,\phi_n,\xi_n' ))
    \,.
\end{multline}
In this situation, in which we refer to the observable as
`event-shape like', the $\xi_i$ integrations can be carried out
trivially, giving 
\begin{multline}
  \label{eq:cF-evshps-anyn}
  \cF(C_1,\ldots,C_n;\lambda) = \lim_{\epsilon\to0}
  \frac{\epsilon^{R'}}{R'}
    \sum_{m=0}^{\infty} \frac{1}{m!}
  \left( \prod_{i=1}^{m+1} \sum_{\ell_i=1}^n  C_{\ell_i} r_{\ell_i}'
    \int_{\epsilon}^{1} \frac{d\zeta_i}{\zeta_i} 
    \int_0^{2\pi} \frac{d\phi_i}{2\pi}
  \right)   \delta\!\left(\ln \zeta_1\right)
  \times \\ \times
  \exp\left(-R'\ln \lim_{{\bar v}\to0} \frac{V(\{\tilde p\},\kappa_1(\zeta_1
      {\bar v}) , \ldots,
      \kappa_{m+1}(\zeta_{m+1}
      {\bar v}))}{\bar v}\right),
  \qquad \xi_i = \mathrm{any}\,.
\end{multline}
This form tends to be numerically more convenient than
eq.~(\ref{eq:cF-with-vto0-limit-anyn}).

\subsection{A worked example: the thrust}
\label{sec:thrust}

The thrust is one of the most widely known and studied event shape
observables. It is therefore an appropriate choice to illustrate the
various elements of our general approach.
While the analysis of the thrust here presented can be obtained in a
fully automatically way, we chose to give here a manual, step-by-step
derivation of all elements needed for an NLL resummation.

The thrust is defined for $\ee$ events as~\cite{ThrustDef},
\begin{equation}
  \label{eq:thrust}
  T = \max_{\vec n} \frac{\sum_i |{\vec q}_i \cdot {\vec n}|}{\sum_i
    |{\vec q}_i|} \,,
\end{equation}
where the sum runs over all particles in the final state and the
maximisation is carried out over all unit vectors $\vec n$. Physical
observable definitions do not distinguish between Born partons (denoted
by $p_\ell$ up to now) and soft/collinear partons ($k_i$) and to
reflect this we have used the notation $q_i$ in eq.~(\ref{eq:thrust})
to refer to a general parton.

In the $2$-jet limit, $T=1$, so it is $\tau = 1-T$ that measures the
departure from the $2$-jet limit.  It can be written as 
\begin{equation}
  \label{eq:taucostheta}
  \tau = \min_{\vec n} \frac{\sum_i |{\vec q}_i| (1 - |\cos
    \theta_{i\vec n}|)}{\sum_i |{\vec q}_i|} \,,
\end{equation}
where $\theta_{i\vec n}$ is the angle between particle $i$ and the
thrust axis $\vec n$. 

Let us now work through the applicability conditions. We first need to
establish whether, for a soft and collinear emission, one can write
the observable in the form eq.~(\ref{eq:simple-second-time}). Let us
define $Q$ as the centre-of-mass energy. Then it is straightforward to
show that
\begin{equation}
  \label{eq:tau_kt_eta_n}
  \tau = \min_{\vec n} \sum_i \frac{ q_{ti}^{(n)}}{Q} e^{-|\eta^{(n)}| } \,,
\end{equation}
where $q_{ti}^{(n)}$ and $\eta^{(n)}$ are transverse momenta and
rapidities defined with respect to the thrust axis $\vec n$. This
already looks somewhat similar to eq.~(\ref{eq:simple-second-time}),
except that we still have a minimisation over the direction of the
thrust axis, the transverse momenta and rapidities are defined with
respect to that thrust axis, and the sum runs over all partons,
including the Born partons.

In the case of just soft and/or collinear emissions the minimisation over
the thrust is straightforward as a result of the following fact
\cite{CTTW}: dividing the event into two hemispheres by a plane
perpendicular to the thrust axis, then in each hemisphere the vector
sum of transverse momenta ${\vec q}_{ti}^{\,(n)}$ is zero. Thus,
relating the (recoiled) Born momentum $\ptilde_\ell$ to the emissions
in the associated hemisphere $\cH_\ell$, we have
\begin{equation}
  \label{eq:ktcons}
  \vec{\ptilde}_{t\ell}^{\,(n)} = - \sum_{i \in \cH_\ell} {\vec
    k}_{ti}^{(n)}\,,\qquad\quad
  \ptilde_{z\ell} \simeq \frac{Q}{2}\left(1 - \sum_{i \in \cH_\ell}
    z_i\right)\,,  
\end{equation}
where $z_i$ is the longitudinal momentum fraction $2|k_{iz}|/Q$ of
emission $i$ and the departure of $\ptilde_{z\ell}$ from $Q/2$ is
accurate (as well as relevant) only when the sum over $z_i$ is
dominated by collinear partons. This allows us to write
\begin{equation}
  \label{eq:tau_of_ks}
  \tau \simeq \sum_i \frac{ k_{ti}^{(n)}}{Q} e^{-|\eta^{(n)}| } + 
  \sum_{\ell = 1,2} \frac{1}{Q^2}\frac{|\sum_{i\in\cH_\ell} {\vec
      k}_{ti}^{\,(n)}|^2}{(1-\sum_{i\in\cH_\ell} z_i)}\,.
\end{equation}
To reach a form similar to eq.~(\ref{eq:simple-second-time}) we need
to exploit two further observations. Firstly the angle of the
recoiling Born partons to the thrust axis, $2
{\ptilde}_{t\ell}^{\,(n)}/\ptilde_{z\ell}$ is much smaller than that
of all but hard collinear emissions, allowing one to replace
$k_{ti}^{(n)}$ and $|\eta^{(n)}|$ with $k_{ti}^{(\ell)}$ and
$\eta^{(\ell)}$ respectively. Secondly, again for all but hard
collinear emissions, $k_{ti}^{(\ell)} e^{-\eta^{(\ell)}} \gg
(k_{ti}^{(\ell)}/Q)^2$, allowing one to neglect the second term of
eq.~(\ref{eq:tau_of_ks}). Thus for soft (and optionally collinear)
emissions we can write
\begin{equation}
  \label{eq:tau_of_ks_Soft}
  \tau \simeq \sum_{\ell=1,2} \sum_{i\in \cH_\ell} \frac{
    k_{ti}^{(\ell)}}{Q} e^{-\eta^{(\ell)} } \,,
\end{equation}
which for a single emission is precisely of the form
eq.~(\ref{eq:simple-second-time}) with 
\begin{equation}
  \label{eq:abdg_thrust}
  a_\ell = b_\ell = d_\ell = g_\ell(\phi) = 1, \qquad \ell = 1, 2\,,
\end{equation}
as anticipated at the beginning of section~\ref{sec:masterderiv}.

Next we need to check the (continuous) globalness conditions. Firstly
one notes that the thrust receives contributions for emissions in all
directions and that $a_1 = a_2 \equiv a$. Furthermore in the soft (and
optionally collinear) region, using eq.~(\ref{eq:tau_of_ks_Soft}), it
is straightforward to see that
\begin{align}
  \left. \frac {\partial \ln \tau(\{\tilde p\}, k)}{\partial \ln k_t^{(\ell)}} 
  \right|_{\mathrm{fixed}\;\eta^{(\ell)},\,\phi^{(\ell)}} \!\!\!\! &=
  1 = a\,.
\\\intertext{%
In the region of collinear (and optionally soft) emissions, we need to
revert to eq.~(\ref{eq:tau_of_ks}). Noting that at fixed $z_i$,
$d\eta_i/ d\ln k_{ti} = -1$, and that $k_t^{(\ell)}$ and $k_t^{(n)}$
are proportional to one another, we see that both terms in
eq.~(\ref{eq:tau_of_ks}) scale as $k_t^2$,
}
  \left. \frac {\partial \ln \tau(\{\tilde p\}, k)}{\partial \ln k_t^{(\ell)}} 
  \right|_{\mathrm{fixed}\;z^{(\ell)},\,\phi^{(\ell)}} \!\!\!\! &=
  2 = a + b_\ell\,,
\end{align}
as required. 

The final condition to be verified is that of recursive IRC safety.
Let us first deal with the situation in which the momentum functions
$\kappa_i(\bar v)$ are such that, as $\bar v\to0$, all emissions remain
in the soft and collinear region. In that case we are entitled to use
eq.~(\ref{eq:tau_of_ks_Soft}) for $\tau$ and we have that the
observable is \emph{additive}:
\begin{equation}
  \label{eq:tau_additive}
  \tau (\{\tilde p\},\kappa_1({\bar v} \zeta_1), \ldots,
      \kappa_m({\bar v} \zeta_m)) = \sum_{i=1}^m \tau (\{\tilde
      p\},\kappa_i({\bar v} \zeta_i)) = {\bar v} \sum_{i=1}^m \zeta_i\,.
\end{equation}
Using this result, it is trivial to demonstrate the validity of
eqs.~(\ref{eq:rIRClimit1}) and (\ref{eq:rIRClimit2}).

We also need to examine what happens if some of the momentum functions
$\kappa_i(\bar v)$ are such that asymptotically their corresponding
emissions are collinear and hard.  This is possible only if their
rapidities satisfy $d\eta_i(\bar v)/d\ln \bar v = -1/(a + b_\ell)$ ---
a smaller value would mean that for $\bar v\to 0$ an emission would
become soft, while larger values are kinematically unallowed. The
corresponding scaling of the transverse momentum is $d\ln k_{ti}(\bar
v)/d\ln \bar v = 1/(a + b_\ell)$. Let us now examine how the two terms
of eq.~(\ref{eq:tau_of_ks}) behave with respect to the first of the
rIRC conditions, eq.~(\ref{eq:rIRClimit1}). The first term clearly
satisfies the condition, as was the case with just soft emissions. The
second term involves a non-linear dependence on combinations of
momenta. However, asymptotically, as $\bar v\to0$ both the numerator
and the denominator come to be dominated entirely by the emissions
with $d\eta_i(\bar v)/d\ln \bar v = -1/(a + b_\ell)$ (for other
emissions $z_i\to0$ and $d\ln k_{ti}(\bar v)/d\ln \bar v > 1/(a +
b_\ell)$). Since all these emissions scale in the same fashion,
$k_t(\bar v) \sim \bar v^{1/(a+b_\ell)}$ (in our specific case,
$k_t(\bar v) \sim \sqrt{\bar v}$) the second term of
eq.~(\ref{eq:tau_of_ks}), like the first term, scales as $\bar v$,
ensuring the validity of the first rIRC condition,
eq.~(\ref{eq:rIRClimit1}). Based on eq.~(\ref{eq:tau_of_ks}) it is
straightforward to show also the validity of the remaining parts of
the rIRC condition, eqs.~(\ref{eq:rIRClimit2}).

Having established that the applicability conditions are satisfied by
the thrust (!) we have nearly all the elements needed for the NLL
resummation. What remains is the function $\cF$. 
We have seen that for soft and collinear emissions the thrust is
additive, eq.~(\ref{eq:tau_additive}). This immediately allows one to
integrate analytically over the $\xi_i$ in
eq.~(\ref{eq:cF-with-vto0-limit-anyn}). Some caution is needed
however, because for hard collinear emissions we have to account for
the second term of eq.~(\ref{eq:tau_of_ks}) which breaks the
additivity. Fortunately, since as $\bar v \to 0$ this is relevant in
an ever smaller region of $\xi$, $1-\xi \lesssim \frac{1}{\ln 1/{\bar
    v}}$, it is associated with a NNLL correction and can be ignored.
Thus we can take eqs.~(\ref{eq:cF-evshps-anyn}) and
(\ref{eq:tau_additive}) and write\footnote{The following treatment can
  be somewhat simplified using the form
  eq.~(\ref{eq:cFalt-evshps-anyn}) for $\cF$ in
  appendix~\ref{sec:anl_convenient_F}, rather than
  eq.~(\ref{eq:cF-evshps-anyn}). We nevertheless choose to illustrate
  the determination of $\cF$ using eq.~(\ref{eq:cF-evshps-anyn}), since
  it is this form that will be used numerically for the automated
  resummation.}
\begin{equation}
  \label{eq:cF_tau}
  \cF = \lim_{\epsilon\to0}
  \frac{\epsilon^{R'}}{R'}
  \sum_{m=0}^{\infty} \frac{1}{m!}
  \left( \prod_{i=1}^{m+1}  R'
    \int_{\epsilon}^{1} \frac{d\zeta_i}{\zeta_i} 
  \right)
  \delta\!\left(\ln \zeta_1\right)
  e^{-R' \ln \sum_{j=1}^{m+1} \zeta_j}\,,
\end{equation}
where we have summed over legs for each emission. To evaluate this
integral, we essentially follow the now standard method of
\cite{CTTW}, introducing a Mellin transform representation,
\begin{equation}
  \label{eq:mellin}
  e^{-R' \ln \sum_{j=1}^{m+1} \zeta_j} 
  = R' \int \frac{dZ}{Z} \, e^{-R' \ln Z} \int \frac{d\nu}{2\pi
    i\nu} e^{\nu Z} \prod_{j=1}^{m+1} e^{-\nu \zeta_j}\,,
\end{equation}
and performing the sum over $m$ to give
\begin{equation}
  \label{eq:cF_tau_mellin}
  \cF = R' \int \frac{dZ}{Z} \, e^{-R' \ln Z} \int \frac{d\nu}{2\pi
    i\nu} e^{\nu (Z - 1)} \exp\left({R' \int^1_0 \frac{d\zeta}{\zeta}
    \left(e^{-\nu\zeta} - 1\right)}\right)\,.
\end{equation}
After some manipulation this can be reduced to the form
\begin{equation}
  \label{eq:cF_tau_final}
  \cF = \int \frac{d\nu}{2\pi i \nu}\, e^{\nu - R'\ln \nu - R' \gae} =
  \frac{e^{-\gae R'}}{\Gamma(1 + R')}\,,
\end{equation}
where $\gae$ is the Euler constant.  Inserting this expression for
$\cF$ into eq.~(\ref{eq:Master}), one can then verify that the
resulting resummed distribution coincides at NLL accuracy with that
originally calculated in \cite{CTTW}.

\subsection{Example of rIRC unsafety: combinations of event shapes}
\label{sec:rIRC}

The condition of recursive infrared and collinear safety is one of the
main novel developments in this article. At first sight, certain parts
of it bear a strong resemblance to normal IRC safety, so we devote
some attention to understanding how precisely they differ. This is
most easily accomplished by studying observables that are IRC safe but
not rIRC safe. We give here one simple example, and refer the reader
to appendix~\ref{sec:further-rIRC-unsafe} for further cases that
illustrate each of the rIRC subconditions.

A rather simple class of observables that has not to our knowledge
previously been considered, consists of products and ratios of normal
$\ee$ event shapes. The example that we shall consider here is $V =
(1-T) B_T$, \ie the product of (one minus) the thrust,
eq.~(\ref{eq:thrust}), and the total jet broadening,
\begin{equation}
  \label{eq:bt}
  B_T = \frac{\sum_i |\vec q_i \times {\vec n}_T|}{2\sum_i |\vec q_i|}\,,
\end{equation}
where ${\vec n}_T$ is the thrust axis. Its dependence on a single emission
is associated with the following coefficients
\begin{equation}
  \label{eq:tauBcoeffs}
  a = 2\,,\quad b_\ell = d_\ell = g_\ell(\phi) = 1\,,\qquad\quad
  \ell = 1, 2\,.
\end{equation}
Introducing ${\bar v}$ and $\xi_i$ to parametrise an emission
$\kappa_i({\bar v})$, as in section~\ref{sec:summary-master}, we
have\footnote{Note that while this parametrisation embodies sufficient
  degrees of freedom for the purpose of our discussion here, it is not
  sufficient for a fully general test of rIRC safety, where one should
  maintain the freedom of adding an arbitrary constant to each of the
  $\eta_i(\bar v)$, as well as considering the azimuthal angles
  $\phi_i$.}
\begin{equation}
  \label{eq:ketadep}
  \begin{split}
    \ln \frac{\kappa_{ti}({\bar v})}{Q} & = 
    \left(\frac{1}{a} - \frac{ b_\ell\, \xi_i}{a(a+b_\ell)}\right)\ln {\bar v}
    = \left(\frac12 - \frac{\xi_i}6\right) \ln {\bar v}\,, \\
    \eta_i({\bar v}) &= -\frac{\xi_i}{a+b_\ell}\ln {\bar v} 
    =  -\frac{\xi_i}3\ln {\bar v}\,,    
  \end{split}
\end{equation}
and for such an emission the corresponding values of $\tau=1-T$ and
$B_T$ are
\begin{equation}
  \label{eq:tauB1emsn}
  \ln \tau(\{\tilde p\},\kappa_i({\bar v})) = \left(\frac12 +
    \frac{\xi_i}6\right) \ln {\bar v}\,, \qquad 
  \ln B_T(\{\tilde p\},\kappa_i({\bar v})) = \left(\frac12 -
    \frac{\xi_i}6\right) \ln {\bar v}\,,
\end{equation}
reproducing $\tau B_T = {\bar v}$.

Now let us consider the value of the observable with two emissions,
$\kappa_1({\bar v})$ and $\kappa_2({\bar v})$ (for
  simplicity, we choose $\zeta_1 = \zeta_2=1$).
As long as $|(\xi_1 - \xi_2) \ln {\bar v}|$ is large then, separately
$\tau$ and $B_T$ are dominated by just one of the emissions.
However because the $\xi_i$ appear with different signs in the thrust
and the broadening, the emission that dominates the broadening (that
with the larger $\xi_i$) is not
the same as that dominating in the thrust (that with the smaller
$\xi_i$), and the product of the two 
observables behaves as follows
\begin{equation}
  \label{eq:tauB2emsnsRes}
  \ln (\tau B_T)(\{\tilde p\},\kappa_1({\bar v}),\kappa_2({\bar v})) \simeq
  \left(1 - \frac{|\xi_1 - \xi_2|}6\right) \ln {\bar v}\,.
\end{equation}
Thus the limit eq.~(\ref{eq:rIRClimit1}) is infinite and the first
rIRC safety condition is not satisfied. Though we have chosen the
$\zeta_1$ and  $\zeta_2$ of eq.~(\ref{eq:rIRClimit1}) both equal to $1$, the argument can be extended
more generally, and one finds that there is a double-logarithmic
region in which eq.~(\ref{eq:rIRClimit1}) is infinite, corresponding
to `corrections' to the resummation at order $\as^2 L^4$, \ie a
breakdown of exponentiation. Similar conclusions hold for a range of
other products and ratios of `standard' event shapes --- essentially
any (IRC safe) product or ratio of event shapes with different values
for the ratio $b_\ell/a$.

Further examples of observables that violate the rIRC conditions are given in appendix~\ref{sec:further-rIRC-unsafe}.

\subsection{Convergence issues for $\cF$}
\label{sec:div}

We have discussed, using the concept of rIRC safety, the
conditions that are necessary for the limits and individual elements
of eq.~(\ref{eq:cF-with-vto0-limit-anyn}) for $\cF$ to be
well-defined. This alone however does not guarantee that the resulting
integrals are all finite. In particular it is known
\cite{BSZ,DSBroad,RakowWebber} that for certain observables, the
resummed distribution defined in terms of exponentiated leading and
next-to-leading logarithmic functions can have a divergence at a
finite value of $\as L$.

To see the origin of potential problems, let us introduce the
probability $d\cP(y)/dy$,
\begin{multline}
  \label{eq:ProbRatio}
  \frac{d\cP(y)}{dy} \!=\! \lim_{\epsilon\to0}
  \frac{\epsilon^{R'}}{R'}
    \sum_{m=0}^{\infty} \frac{1}{m!}
  \left( \prod_{i=1}^{m+1} \sum_{\ell_i=1}^n
    \int_{\epsilon}^{1}\!\! \frac{d\zeta_i}{\zeta_i} 
    \frac{C_{\ell_i} r_{\ell_i}'}{\cN_{\ell_i}(\lambda/\beta_0)}
    \int_0^1 \!\!\frac{d\xi_i}{1 + \frac{a + (1-\xi_i)b_{\ell_i}}{a(a+b_{\ell})}
    2\lambda} 
  \int_0^{2\pi} \!\!\frac{d\phi_i}{2\pi}
  \right)   
  \times \\ \times 
  \delta\!\left(\ln \zeta_1\right)
  \delta\left(y -  \lim_{\bar v\to0} \frac{V(\{\tilde
      p\},\kappa_1(\zeta_1 \bar v), \ldots,
      \kappa_{m+1}(\zeta_{m+1} \bar v))}{\bar v}\right),
\end{multline}
for a given set of $\{C_{\ell} r'_\ell\}$, of having a configuration of
momenta such that the ratio between the full observable and $\bar v$
is equal to some given value $y$ (in the limit $\bar v \to 0$).  We
can then rewrite eq.~(\ref{eq:cF-with-vto0-limit-anyn}) as
\begin{equation}
  \label{eq:cF-rewritten}
  \cF = \int_0^\infty dy \,\frac{d\cP(y)}{dy}\, 
  y^{-R'}
  \,.
\end{equation}
Without the $y^{-R'}$ factor, the integral is by definition convergent
both at large $y$ 
and small $y$, since the total probability is $1$. The inclusion of
the $y^{-R'}$ factor
improves the convergence at large $y$, but worsens it at small $y$.
For many observables this does not pose a problem because the value of
the observable in the presence of multiple emissions is systematically
larger than in the presence of any single one of the emissions, \ie\
$\cP(y) \equiv  \int_0^y dy' d\cP(y')/dy' = 0$ for $y<1$.

There are however observables for which there can be a cancellation
between the contributions from different emissions. The classic
example is the transverse momentum of a Drell-Yan pair --- since the
pair transverse momentum is given by the recoil from all emissions,
cancellations \cite{CollinsSoper,ParPet} in the vector sum of emitted
transverse
momenta imply that $y$ can have values down to $0$. Other examples of
observables where $y$ can approach zero include the $\ee$ oblateness
\cite{BSZ} and the broadening (with respect to the photon axis) in DIS
\cite{DSBroad}, as well as the indirectly-global hadronic dijet
observables defined in \cite{BSZhh}.

The consequences of this for $\cF$ depend on the analytical behaviour
of $\cP(y)$ in the neighbourhood of $y=0$.  Let us assume that
$\cP(y)$ vanishes as a power of $y$ for $y\to0$, $\cP(y) \sim y^{p}$
(as is usually the case in this kind of problem). Then the integral
eq.~(\ref{eq:cF-rewritten}) is finite only for $R'< R'_c \equiv p$,
\begin{equation}
  \label{eq:cF-divergence}
  \cF \sim \frac{1}{R'_c - R'} \,,\qquad \mathrm{for}\;\;0<R'_c -R' \ll 1\,.
\end{equation}
For the case of the Drell-Yan $p_t$ distribution, the result of the
vector sum can be loosely identified with the result of a random walk,
which has a uniform distribution in $\vec p_t$ for $p_t$ close to
zero. This corresponds to $\cP(y) \sim y^2$ for small $y$, so the
integral for $\cF$ diverges for $R' \ge 2$.

Physically the origin of this divergence is as follows. Normally the
requirement that the observable be small is satisfied by forbidding
radiation --- it is this that leads to the appearance of the double
logarithmic Sudakov form factor in the resummed distribution. So a
reduction in the maximum allowed value of the observable from, say,
$v$ to a moderately smaller value $v'$, leads to an extra suppression
in $\vProb$, eq.~(\ref{eq:Master}), of the form
\begin{equation}
  \label{eq:vProb-extra-suppression}
  \vProb(v') \simeq \vProb(v) \left(\frac{v'}{v}\right)^{R'}\,.
\end{equation}
The appearance here of $R'$ comes from the expansion of the LL Sudakov
structure and is not modified by the NLL function $\cF$.

For observables with cancellations, $\cP(y)\sim y^p$ ($y \ll 1$),
there is an alternative mechanism for reducing $v$ to $v'$, \ie by
choosing the configurations that have the strongest
cancellations. This corresponds to paying 
a price of $(v'/v)^{p}$. As long as $R' < p$, the cancellation
mechanism simply gives a NLL correction to the Sudakov suppression,
which is taken into account in the function $\cF$. Instead, for
sufficiently small values of the observable ($R' > p$), it is the
cancellation mechanism that dominates,
\begin{equation}
  \label{eq:vProb-canelling-suppression}
  \vProb(v') \simeq \vProb(v) \left(\frac{v'}{v}\right)^{p}\,.
\end{equation}
Since it is impossible for an NLL $\cF$ function to transform the
behaviour of eq.~(\ref{eq:vProb-extra-suppression}) into that of
eq.~(\ref{eq:vProb-canelling-suppression}), the master formula
eq.~(\ref{eq:Master}) can no longer be used to represent the full
resummed prediction. This is reflected in a divergence of $\cF$,
eq.~(\ref{eq:cF-divergence}). Were one able to calculate the analogous
function at NNLL one would expect to see an even stronger divergence.

The divergence is not a specificity of our semi-numerical approach to
the resummation, but appears also in purely analytical resummed
calculations, \eg \cite{DSBroad,RakowWebber}.  In such situations,
current techniques for obtaining a full resummed answer usually
require that one carry out the resummation in some appropriate
transform space (\eg $b$-space resummation for the Drell-Yan $p_t$
distribution). Within the context of a semi-numerical approach such as
ours, the divergence could be
eliminated by including in eq.~(\ref{eq:splitVirtual}), and elsewhere
in section~\ref{sec:mult-indep-emsn}, the $R''$ (and possible higher)
terms of the expansion of $R(v)$.

Even when $\cF$ has a divergence, it may still be possible to make use
of eq.~(\ref{eq:Master}) for phenomenological applications. For
observables without divergences, for $R'$ of order $1$, the N$^n$LL
term is suppressed relative to the LL term by a power $\as^n$. Since
the LL term is of order $\as L^2\sim 1/\as$ in this region, the
neglected NNLL terms give corrections in the exponent of order $\as$.
For observables with a divergence in $\cF$, it seems \cite{DSBroad}
that the N$^n$LL term is suppressed relative to the LL by
$(\as/(R'_c-R'))^n$. As long as one stays sufficiently far from the
divergence, \ie in a region where $R'_c-R'\gtrsim 1$, the neglected
NNLL corrections remain small, of order $\as$. When $R'_c-R'\sim
\sqrt{\as}$ there is still a hierarchy in the series of N$^n$LL terms,
N$^n$LL$\;\sim \as^{n/2} \times $LL, however the neglected NNLL contribution
becomes significant since it amounts to a correction of order $1$ in the
exponent. Finally when $R'_c-R'\sim {\as}$, the N$^n$LL hierarchy
breaks down completely, since all terms are of the same order.

The critical question therefore is whether the region where problems
start to appear, $R'_c-R'\sim \sqrt{\as}$, is relevant
phenomenologically. If $R'_c$ is sufficiently large, then the
divergence of $\cF$ affects the resummed distribution only in a region
far into the Sudakov-suppressed tail of the distribution. One can show
that the maximum of the distribution of the observable, $df(v)/dv$, is
situated at $R'\simeq1$ and beyond this point, Sudakov suppression
sets in very rapidly. Accordingly if $R'_c$ is somewhat larger than
this (in our experience, if $R'_c \gtrsim 3$), then the divergence
will be sufficiently strongly suppressed that it can be ignored.
Normal one and two-dimensional cancellations usually lead to $R'_c =
1$ and $2$ respectively. The question of how higher values of $R'_c$
arise and a variety of related issues are discussed in the context of
a more general treatment of divergences of $\cF$ in
appendix~\ref{sec:divergences-ff}.

\section{Computer automated expert semi-analytical resummation}
\label{sec:caesar}

In the previous sections we have outlined a well-defined procedure
for obtaining resummed predictions for a given observable. Its
strength is that it is a closed procedure --- to carry out the
resummation, it is sufficient to know how to evaluate the observable
for arbitrary configurations of partons.

Nevertheless, even using the results of section~\ref{sec:master}, a
certain amount of straightforward, though tedious analysis of the
observable is required in order to obtain a resummed prediction.
Furthermore one needs to implement some form of numerical integration
for the determination of the function $\cF$. Given that the approach
is well-defined it is therefore natural to investigate, instead, the
possibility
of implementing a computer program to follow it through.

One possible tactic would be to attempt to code the procedure for use
in a symbolic manipulation program such as Form, Mathematica or
Maple.  However, even with the simplest of the observables one
would quickly encounter difficulties. For example, the definition of
the thrust, eq.~(\ref{eq:thrust}), involves a maximisation over the
direction of a projection axis. Such a maximisation is a highly
non-trivial operation if it is to be carried out entirely
analytically, in closed form, by a symbolic manipulation program.

We choose instead an approach inspired by the field of Experimental
Mathematics \cite{ExpMath}, and that incorporates also some
characteristics of expert systems~\cite{ExpertSyst}. The observable is
coded as a computer subroutine,\footnote{It is to be kept in mind that
  there are observables for which this requires some thought!} which
is then called with a range of partonic configurations. By taking the
soft and collinear limits for the emissions it is possible to obtain
the information required for the resummation, \cnf
section~\ref{sec:summary-master}. This analysis is carried out for a
single Born configuration. The general approach and certain specific
details are discussed in section~\ref{sec:analysis}. Issues associated
with the subsequent integration over Born configurations are then
considered in section~\ref{sec:Born-integrate}, finally in
\ref{sec:res} we discuss applications of \caesar.

\subsection{The analysis}
\label{sec:analysis}

The study of the observable for a given Born configuration follows the
sequence outlined in the flowchart of figure~\ref{fig:flowchart}. The
overall structure should be self-explanatory, so rather than
proceeding with a step-by-step explanation of each entry of the
flowchart, we will discuss
(section~\ref{sec:Caesar-General-considerations}) issues that are
common to many parts of the analysis, and then concentrate on points
that require more detailed attention, that is tests of rIRC safety
(section~\ref{sec:rIRCtests}), the general determination of $\cF$
(section~\ref{sec:efficiency_for_cF}).

\FIGURE{
  \scalebox{0.7}{{\begin{picture}(0,0)%
\includegraphics{flowchart.pstex}%
\end{picture}%
\setlength{\unitlength}{4144sp}%
\begingroup\makeatletter\ifx\SetFigFont\undefined%
\gdef\SetFigFont#1#2#3#4#5{%
  \reset@font\fontsize{#1}{#2pt}%
  \fontfamily{#3}\fontseries{#4}\fontshape{#5}%
  \selectfont}%
\fi\endgroup%
\begin{picture}(8027,12894)(941,-13573)
\put(4501,-916){\makebox(0,0)[b]{\smash{{\SetFigFont{12}{14.4}{\familydefault}{\mddefault}{\updefault}User supplies observable}}}}
\put(4501,-1141){\makebox(0,0)[b]{\smash{{\SetFigFont{12}{14.4}{\familydefault}{\mddefault}{\updefault}and Born momenta}}}}
\put(4501,-3256){\makebox(0,0)[b]{\smash{{\SetFigFont{12}{14.4}{\familydefault}{\mddefault}{\updefault}Determination of}}}}
\put(4501,-3481){\makebox(0,0)[b]{\smash{{\SetFigFont{12}{14.4}{\familydefault}{\mddefault}{\updefault}sufficiently soft and}}}}
\put(4501,-3706){\makebox(0,0)[b]{\smash{{\SetFigFont{12}{14.4}{\familydefault}{\mddefault}{\updefault}collinear region for}}}}
\put(4501,-3931){\makebox(0,0)[b]{\smash{{\SetFigFont{12}{14.4}{\familydefault}{\mddefault}{\updefault}subsequent steps}}}}
\put(4501,-6361){\makebox(0,0)[b]{\smash{{\SetFigFont{12}{14.4}{\familydefault}{\mddefault}{\updefault}[eqs.(\ref{eq:rIRClimit1},\ref{eq:rIRClimit2})]}}}}
\put(4501,-6136){\makebox(0,0)[b]{\smash{{\SetFigFont{12}{14.4}{\familydefault}{\mddefault}{\updefault}rIRC safe?}}}}
\put(8326,-6046){\makebox(0,0)[b]{\smash{{\SetFigFont{12}{14.4}{\familydefault}{\mddefault}{\updefault}Declare}}}}
\put(8326,-6271){\makebox(0,0)[b]{\smash{{\SetFigFont{12}{14.4}{\familydefault}{\mddefault}{\updefault}failure of}}}}
\put(8326,-6496){\makebox(0,0)[b]{\smash{{\SetFigFont{12}{14.4}{\familydefault}{\mddefault}{\updefault}resummation}}}}
\put(4501,-4966){\makebox(0,0)[b]{\smash{{\SetFigFont{12}{14.4}{\familydefault}{\mddefault}{\updefault}global? [$a_1 = \cdots = a_n$}}}}
\put(4501,-4741){\makebox(0,0)[b]{\smash{{\SetFigFont{12}{14.4}{\familydefault}{\mddefault}{\updefault}Continuously}}}}
\put(4501,-5191){\makebox(0,0)[b]{\smash{{\SetFigFont{12}{14.4}{\familydefault}{\mddefault}{\updefault}and eqs.(\ref{eq:DiscontGlobal})]}}}}
\put(4510,-7306){\makebox(0,0)[b]{\smash{{\SetFigFont{12}{14.4}{\familydefault}{\mddefault}{\updefault}Additive?}}}}
\put(4636,-7846){\makebox(0,0)[lb]{\smash{{\SetFigFont{12}{14.4}{\familydefault}{\mddefault}{\updefault}no}}}}
\put(3196,-7261){\makebox(0,0)[lb]{\smash{{\SetFigFont{12}{14.4}{\familydefault}{\mddefault}{\updefault}yes}}}}
\put(4510,-7531){\makebox(0,0)[b]{\smash{{\SetFigFont{12}{14.4}{\familydefault}{\mddefault}{\updefault}[eq.\ref{eq:V_additive}]}}}}
\put(4501,-9285){\makebox(0,0)[b]{\smash{{\SetFigFont{12}{14.4}{\familydefault}{\mddefault}{\updefault}Determine zeroes and }}}}
\put(4501,-9510){\makebox(0,0)[b]{\smash{{\SetFigFont{12}{14.4}{\familydefault}{\mddefault}{\updefault}study their properties}}}}
\put(4501,-9735){\makebox(0,0)[b]{\smash{{\SetFigFont{12}{14.4}{\familydefault}{\mddefault}{\updefault}(used in computation of $\cF$)}}}}
\put(5761,-11806){\makebox(0,0)[lb]{\smash{{\SetFigFont{12}{14.4}{\familydefault}{\mddefault}{\updefault}no}}}}
\put(4636,-12481){\makebox(0,0)[lb]{\smash{{\SetFigFont{12}{14.4}{\familydefault}{\mddefault}{\updefault}yes}}}}
\put(4510,-11838){\makebox(0,0)[b]{\smash{{\SetFigFont{12}{14.4}{\familydefault}{\mddefault}{\updefault}$\cF$ calculable in}}}}
\put(4510,-12042){\makebox(0,0)[b]{\smash{{\SetFigFont{12}{14.4}{\familydefault}{\mddefault}{\updefault}double precision?}}}}
\put(1943,-12976){\makebox(0,0)[b]{\smash{{\SetFigFont{12}{14.4}{\familydefault}{\mddefault}{\updefault}$\cF$ and $\cF_2$ known}}}}
\put(1943,-13201){\makebox(0,0)[b]{\smash{{\SetFigFont{12}{14.4}{\familydefault}{\mddefault}{\updefault}analytically}}}}
\put(1943,-13426){\makebox(0,0)[b]{\smash{{\SetFigFont{12}{14.4}{\familydefault}{\mddefault}{\updefault}[eqs.(\ref{eq:cF_tau_final},\ref{eq:F2_additive})]}}}}
\put(7028,-12976){\makebox(0,0)[b]{\smash{{\SetFigFont{12}{14.4}{\familydefault}{\mddefault}{\updefault}Calculate $\cF$ and $\cF_2$}}}}
\put(7028,-13201){\makebox(0,0)[b]{\smash{{\SetFigFont{12}{14.4}{\familydefault}{\mddefault}{\updefault}in multiple precision}}}}
\put(7028,-13426){\makebox(0,0)[b]{\smash{{\SetFigFont{12}{14.4}{\familydefault}{\mddefault}{\updefault}[eqs.(\ref{eq:cF-with-vto0-limit-anyn},\ref{eq:F2})]}}}}
\put(4501,-12976){\makebox(0,0)[b]{\smash{{\SetFigFont{12}{14.4}{\familydefault}{\mddefault}{\updefault}Calculate $\cF$ and $\cF_2$}}}}
\put(4501,-13201){\makebox(0,0)[b]{\smash{{\SetFigFont{12}{14.4}{\familydefault}{\mddefault}{\updefault}in double precision}}}}
\put(4501,-13426){\makebox(0,0)[b]{\smash{{\SetFigFont{12}{14.4}{\familydefault}{\mddefault}{\updefault}[eqs.(\ref{eq:cF-evshps-anyn},\ref{eq:F2})]}}}}
\put(4500,-1877){\makebox(0,0)[b]{\smash{{\SetFigFont{12}{14.4}{\familydefault}{\mddefault}{\updefault}Determination of}}}}
\put(4500,-2102){\makebox(0,0)[b]{\smash{{\SetFigFont{12}{14.4}{\familydefault}{\mddefault}{\updefault}leg properties}}}}
\put(4500,-2327){\makebox(0,0)[b]{\smash{{\SetFigFont{12}{14.4}{\familydefault}{\mddefault}{\updefault}$a_\ell$, $b_\ell$, $d_\ell$, $g_\ell(\phi)$}}}}
\put(4500,-2552){\makebox(0,0)[b]{\smash{{\SetFigFont{12}{14.4}{\familydefault}{\mddefault}{\updefault}[eq.(\ref{eq:simple-second-time})]}}}}
\put(5941,-1996){\makebox(0,0)[lb]{\smash{{\SetFigFont{12}{14.4}{\familydefault}{\mddefault}{\updefault}failure}}}}
\put(4636,-2896){\makebox(0,0)[lb]{\smash{{\SetFigFont{12}{14.4}{\familydefault}{\mddefault}{\updefault}success}}}}
\put(5941,-4876){\makebox(0,0)[lb]{\smash{{\SetFigFont{12}{14.4}{\familydefault}{\mddefault}{\updefault}no}}}}
\put(5941,-6136){\makebox(0,0)[lb]{\smash{{\SetFigFont{12}{14.4}{\familydefault}{\mddefault}{\updefault}no}}}}
\put(4636,-6766){\makebox(0,0)[lb]{\smash{{\SetFigFont{12}{14.4}{\familydefault}{\mddefault}{\updefault}yes}}}}
\put(4636,-5596){\makebox(0,0)[lb]{\smash{{\SetFigFont{12}{14.4}{\familydefault}{\mddefault}{\updefault}yes}}}}
\put(4501,-8341){\makebox(0,0)[b]{\smash{{\SetFigFont{12}{14.4}{\familydefault}{\mddefault}{\updefault}Establish integration}}}}
\put(4501,-8566){\makebox(0,0)[b]{\smash{{\SetFigFont{12}{14.4}{\familydefault}{\mddefault}{\updefault}range ($\epsilon$) for $\cF_2$ and $\cF$}}}}
\put(5761,-10546){\makebox(0,0)[lb]{\smash{{\SetFigFont{12}{14.4}{\familydefault}{\mddefault}{\updefault}no}}}}
\put(4636,-11221){\makebox(0,0)[lb]{\smash{{\SetFigFont{12}{14.4}{\familydefault}{\mddefault}{\updefault}yes}}}}
\put(4510,-10636){\makebox(0,0)[b]{\smash{{\SetFigFont{12}{14.4}{\familydefault}{\mddefault}{\updefault}Event-shape like?}}}}
\put(4510,-10861){\makebox(0,0)[b]{\smash{{\SetFigFont{12}{14.4}{\familydefault}{\mddefault}{\updefault}[eq.(\ref{eq:exchange_xi})]}}}}
\end{picture}%
}}
  \caption{Flowchart of analysis for automated resummation. See main text
    for details.}
  \label{fig:flowchart}
}

\subsubsection{General considerations}
\label{sec:Caesar-General-considerations}

Many of the limits that arise in section~\ref{sec:summary-master} are
approached accurately only for extremely soft and collinear emissions.
Rounding errors often make it impossible to correctly calculate the
value of an observable in such limits using standard double precision
arithmetic. Therefore an essential tool in the numerical analysis of
the observable is multiple-precision (MP) arithmetic. 

We have chosen to use the MP arithmetic package by David Bailey
\cite{MP}. It exploits Fortran~90's operator overloading abilities
to provide transparent access to nearly all operations (including
special functions) on MP quantities, so that one can write normal
Fortran~90 code, with only minimal changes needed for it to work in
multiple precision.\footnote{We have added functionality to this
  package, extending its operator overloading to many common array
  operations that were not supported, and introducing a basic template
  mechanism analogous to that of C++, making it possible to write
  routines in unspecified precision and have them converted to
  explicit double-precision and multiple-precision versions.}

The user is expected to provide a subroutine for the observable, and
to specify a configuration of Born momenta, $\{p\}$, for which the
resummation is to be carried out. This is the starting point for the
flowchart of fig.~\ref{fig:flowchart}. Given these inputs it is
possible for the program to check the applicability conditions of
section~\ref{sec:summary-master}, to determine the various leg
coefficients in eq.~(\ref{eq:simple-second-time}) and to calculate
$\cF$ (as well its expansion coefficients, needed for matching).

We find it convenient to exploit a combination of deterministic and
Monte Carlo procedures. The former are used to help formulate
hypotheses, the latter to test them. For example, for the coefficients
in eq.~(\ref{eq:simple-second-time}) the program uses a restricted set
of momentum configurations to establish the probable values of the
$a_\ell$, $b_\ell$, $d_\ell$. It then verifies that those values hold
for a large number of further (randomly generated)
configurations.\footnote{This Monte Carlo check simultaneously finds a
  region of the $\eta$--$\ln k_t$ plane that is sufficiently asymptotic for
  the rest of the analysis (including a determination of the $\bar
  v$ used in various equations of
  section~\ref{sec:summary-master}).}

Sometimes the hypotheses that are formulated concern functions rather
than just numbers. This is the case for $g_\ell(\phi)$ and $\cF(R')$.
For certain observables these functions have simple analytical forms,
and that information can be of value.

For example, quite often $g_\ell(\phi)$ is just an integer power of
$\sin(\phi)$ or $\cos(\phi)$, and this can easily be established. In
the remaining cases $g_\ell(\phi)$ is tabulated over a large number of
points so as to have an accurate representation for it.\footnote{One
  current technical restriction concerns possible zeroes of
  $g_\ell(\phi)$.  Recall, eq.~(\ref{eq:kidef}), that we define
  momenta $\kappa(v)$ by the requirement that $V(\{\tilde
  p\},\kappa(v)) = v$. If $g_\ell(\phi)$ has a zero at some $\phi =
  \phi_0$, then in the limit $\phi \to \phi_0$, the transverse
  momentum of $\kappa(v)$ can grow large (\ie no longer soft and
  collinear). Cuts on $\phi$ can be used to circumvent such problems,
  but only given good knowledge about $\phi_0$ and the value of $d\ln
  g_\ell(\phi)/d\ln (\phi-\phi_0)$ in the neighbourhood of $\phi_0$.
  To simplify the determination of this information, we currently
  require that if $g_\ell(\phi)$ has zeroes, they be either at
  $\phi=0,\pi$ or $\phi=\pi/2,3\pi/2$.} One could of course use the
methods of experimental mathematics \cite{ExpMath} to expand the range
of functions that one tests for.

In the case of $\cF$, fully analytical results can be obtained for
observables that are additive, like the thrust. Such observables
satisfy the condition (\cnf
eq.~(\ref{eq:tau_additive}))\footnote{Recall that the momenta $\{\tilde
p\}$ are defined as the recoiling Born momenta after all emissions, so
they differ in the l.h.s. and r.h.s. of eq.~\eqref{eq:V_additive}. See
Appendix~\ref{sec:recoil} for more details.}
\begin{equation}
  \label{eq:V_additive}
  V (\{\tilde p\},k_1, \ldots,
      k_m) = \sum_{i=1}^m V(\{\tilde p\},k_i).
\end{equation}
Given this property, the derivation of $\cF$ closely follows that for
the thrust, and one has the general result
eq.~(\ref{eq:cF_tau_final}), alleviating the need for $\cF$ to be
calculated numerically.

As mentioned above, once a hypothesis has been formulated with the aid
of deterministic methods, it is checked using Monte Carlo methods. In
such a check, two parameters should be supplied by the user: the
number of random tests and the accuracy to which they should be
satisfied. As is always the case in any `experimental' verification of
a hypothesis, it suffices to have a single negative test result to
falsify the hypothesis, whereas formally an infinite number of
positive tests is needed in order to verify it. In practice, for the
various observables that we have studied (about 50), we find that on
the occasions when a hypothesis is falsified this occurs after at most
a few hundred test events, and usually after just a few test events.

Concerning the accuracy ($\varepsilon$) of the tests, again formal
certainty regarding the tests can only be achieved in the limit of
arbitrarily high accuracy, $\varepsilon \to 0$. For small but finite
$\varepsilon$ we believe that an undetected violation of a condition at a
level below the accuracy $\varepsilon$ will translate to a relative
incorrectness of the logarithmic structure of the resummation that is
bounded by a positive power of $\varepsilon$. 

There are certain tests where good accuracy is critical. For example
it is important that the coefficients $a$ and $b_\ell$ be well
determined, because any uncertainties in the $a$ and $b_\ell$
will be magnified by the values of $\ln Q/k_t$ and $\eta$, which can
be large. In such cases we typically insist on having close to the
full accuracy that can be represented in double-precision (used to
store the values of the coefficients). In many other situations high
accuracy is less critical and leads to an unnecessary slowing down of
the program. For example for hypotheses involving multiple emissions
(such as the tests of rIRC safety, or exploration of the structure of
any divergences of $\cF$) we find that an absolute accuracy
requirement of $\varepsilon=10^{-3}$ (on $\ln V$), reliably
establishes the veracity of the hypotheses.

Given that we are using MP arithmetic it may seem surprising that we
should have such a `poor' accuracy requirement.  Schematically this
can be understood by noting that there can be effects that manifest
themselves through corrections that scale as $1/\ln V$. 
The number of digits of internal arithmetic precision that is needed
then scales as $1/\varepsilon$. We note though that there is room
for going
to higher accuracies than are currently used, since run times for the
full analysis (except the computation of $\cF$) are of the order of a
few minutes.

\subsubsection{Tests of rIRC safety}
\label{sec:rIRCtests}

The rIRC tests are among the least trivial in \caesar, essentially
because of the double limits in
eqs.~(\ref{eq:rIRClimit1},\ref{eq:rIRClimit2}). 

We use a randomly generated sample of events and require that the
conditions hold for each event. An event is built up first by choosing
the number, $m$, of emissions (currently we take $2\le m\le4$). Then
for each emission $i$ one specifies the leg, $\ell$, to which it is closest,
its azimuthal angle, a value for the $\zeta_i$ and the form of the
function $\kappa_i(\zeta)$.  The latter is chosen according to
eq.~(\ref{eq:kappaSec2Def}), so that a
variation of $\zeta$ corresponds to following a linear path in the
$\eta$--$\ln k_t$ plane.  

The first of the rIRC conditions, eq.~(\ref{eq:rIRClimit1}), is tested
by examining the value of the ratio 
\begin{equation}
  \label{eq:fratioV_vbar}
  y(\bar v; \zeta_1, \ldots, \zeta_m) = \frac{V(\{\tilde
  p\},\kappa_1(\zeta_1 \bar v),\ldots,\kappa_m(\zeta_m \bar v))}{\bar
  v}\,,
\end{equation}
for two widely separated values of $\ln 1/\bar v$. If the difference
between the two results for $\ln y$ is larger than the accuracy
requirement $\varepsilon$, then the two $\ln 1/\bar v$ values are
increased further to establish whether a limit is being reached for
$y$. This procedure is continued until the available (multiple)
precision is insufficient to correctly calculate the observable. If at
this point $y$ has still not reached a limit, the observable is deemed
to fail the first rIRC condition.

Having found a region that is asymptotic with respect to rescalings of
$\bar v$, one establishes the change to $\ln y$ on removing emission
$m$. If the effect is larger than $\varepsilon$, then one determines
the threshold value $\zeta_{m,\mathrm{crit}}$ such that if $\zeta_m >
\zeta_{m,\mathrm{crit}}$, then we have
$|\ln y(\bar v; \zeta_1,\ldots,\zeta_m) - \ln
y(\bar v; \zeta_1,\ldots,\zeta_{m-1})| > \varepsilon$ and if $\zeta_m
< \zeta_{m,\mathrm{crit}}$, then $|\ln y(\bar v;
\zeta_1,\ldots,\zeta_m) - \ln 
y(\bar v; \zeta_1,\ldots,\zeta_{m-1})| < \varepsilon$.\footnote{This
  particular formulation is necessary because $y$ may be discontinuous
  with respect to variations of $\zeta_m$. Currently no explicit check
  is carried out for the existence of multiple solutions for
  $\zeta_{m,\mathrm{crit}}$, it being assumed that if any one of these multiple
  solutions is `dangerous', it will be found as a result of the Monte
  Carlo sampling of $\zeta_i$ values and functional forms for $\kappa_i$.} %
If no value can be found for $\zeta_{m,\mathrm{crit}}$, then this is
usually an indication that the observable is IRC unsafe (this should
not however be considered as a complete test of IRC safety). If a
$\zeta_{m,\mathrm{crit}}$ is found, then a second value of $\ln 1/\bar
v$ is taken and $\zeta_{m,\mathrm{crit}}$ is redetermined.  As for the
first rIRC condition, the two $\ln 1/\bar v$ values are both increased
until $\zeta_{m,\mathrm{crit}}$ becomes independent of $\bar v$. If
this does not occur within the accessible range of $\ln 1/\bar v$, the
observable is deemed to fail on eq.~(\ref{eq:rIRClimit2a}) of the rIRC
condition.

A similar procedure is used to check eq.~(\ref{eq:rIRClimit2b}), it
being a critical value of $\mu$ that is searched for.

\subsubsection{Efficiency considerations for calculating $\cF$}
\label{sec:efficiency_for_cF}

The slowest part of our automated resummation approach is the
calculation of $\cF$. 
This is because it is necessary to carry out a separate Monte Carlo
integration for $\cF$ for each of a range of values of $R'$. The issue
of speed becomes particularly relevant if one has to use high-accuracy
multiple-precision arithmetic in the evaluation of the limits in
eqs.~(\ref{eq:cF-with-vto0-limit-anyn}), (\ref{eq:cF-evshps-anyn}).

There is of course a trade-off between speed and the accuracy of the
final result. The determining factors for the accuracy are the number
of Monte Carlo events used, the non-asymptoticity of the result due to
the use of finite $\epsilon$ and $\bar v$, and rounding errors in the
calculation of $V$.

The first thing to be established in the numerical calculation of
$\cF$ is a suitable value for $\epsilon$ in
eqs.~(\ref{eq:cF-with-vto0-limit-anyn}) and (\ref{eq:cF-evshps-anyn}),
which formally should be taken to zero. One specifies some target
accuracy $\varepsilon$ (note that $\epsilon$ and $\varepsilon$ are
different quantities). Schematically one sets the value of $\epsilon$
such that for most configurations, eliminating those emissions with
$\zeta_i < \epsilon$ changes the value of the observable by less than
some fraction of $\varepsilon$. In practice rather than explicitly
probing the observable to determine $\epsilon$ for a given
$\varepsilon$, one determines the integer power $q$ such that for
(almost) all double-emission configurations
\begin{equation}
  \label{eq:q-range-det}
    \bar v (1-\zeta^{1/q})^q <  V(\{\tilde p\}, \kappa_1(\bar
        v), \kappa_2(\zeta
        \bar v)) < \bar v(1+\zeta^{1/q})^q\,,
\end{equation}
and then uses this to set $\epsilon$ as a function of $\varepsilon$,
$\epsilon \lesssim \varepsilon^q/q^2$.

Next one should choose a value of $\bar v$.  The easiest situation is
that for `event-shape-like' observables, for which the integration in
the $\xi_i$ can performed analytically and one can use
eq.~(\ref{eq:cF-evshps-anyn}) for $\cF$. Typically in this situation
the $\bar v \to 0$ limit converges rapidly --- errors due to the use
of a finite value of $\bar v$ are essentially associated with
corrections to eq.~(\ref{eq:simple-second-time}), which usually vanish
as a power of the softness/collinearity of the emissions. It is
therefore possible to evaluate $\cF$ with reasonable accuracy without
going to extremely small values of $\bar v$. 

Depending on the details of the algorithm used to calculate the
observable, there may even exist a range of $\bar v$ in which one can
use (fast) double precision arithmetic to evaluate the observable while
maintaining small errors both from numerical rounding and
non-asymptoticity.  The freedom to choose an arbitrary rapidity for
each emission is useful in this respect. The simplest choice would be
to take some arbitrary fixed rapidity fraction $\xi$. However one
finds that both rounding errors and the degree of non-asymptoticity
can depend substantially on a non-trivial combination of rapidity
fraction, azimuthal angle and value of $\zeta \bar v$. Thus to
minimise the combination of rounding and non-asymptoticity errors it
is convenient, for each emission, to choose the rapidity fraction most
appropriate to the specific $\phi$ and $\zeta \bar v$, as stored in a
lookup table calculated once for each observable (at the stage `$\cF$
calculable in double precision?' in the flowchart,
figure~\ref{fig:flowchart}).

There are also (event-shape) observables for which there is no
range of $\bar v$ in which both (double-precision) rounding and
non-asymptoticity errors
are simultaneously small enough. In such cases it is necessary to
resort to multiple precision, though usually a fairly moderate number
of digits is sufficient to keep the rounding error $\ll \varepsilon$
in a region where the non-asymptoticity error is also smaller than
$\varepsilon$.

Observables for which one cannot integrate analytically over the
$\xi_i$ tend to be more challenging. This is because for finite $\bar
v$ there can be corrections to $\cF$ (associated physically with NNLL
contributions) originating from regions where two values of $\xi$ are
close, $|\xi_i - \xi_j| \ln \bar v \lesssim 1$. After integration over
the $\xi_i$,
such corrections scale as $(\ln 1/\bar v)^{-1}$, \ie much larger
non-asymptoticity errors than in the case of event-shape-like
observables. Accordingly to obtain an accuracy $\varepsilon$ one
should choose $\ln 1/\bar v \sim \varepsilon^{-1}$, with a
correspondingly large number of digits being needed to avoid rounding
errors. In such situations, reasonable results for $\cF$ can require
up to a hundred days of CPU time on a modern processor (though on
today's large computing clusters this typically corresponds to a few
days' real time).  The procedure can be rendered more efficient by using
correlated events with different values of $\bar v$, from which one
can estimate the small corrections to $\cF$ due to non-asymptoticity
with far fewer events than are needed to evaluate $\cF$ itself.

\subsection{Integration over Born configurations}
\label{sec:Born-integrate}

The discussion so far has been based on the study of a single Born
configuration with a given structure of flavour indices and associated
colour factors. For a Born process such as $\ee \to 2$~jets, DIS 1+1
jet, or Drell-Yan production, the Born kinematics (normalised to the
one dimensionful scale) and associated colour factors are unique, so
the result as given so far is sufficient to obtain a full resummed
prediction.

In general, however, this is not the case, notably when the Born
process involves three or more ($n$) hard legs. In such a situation
one has to select a subset of events such that there are always at
least $n$ hard jets, using some cut, such as the function
$\cH(q_1,q_2,\ldots)$ introduced in section~\ref{sec:problem}.  Then,
eq.~(\ref{eq:Sigmacut_resummed}), one has to integrate over all Born
configurations $\momConf$ that satisfy the cut, summing over hard
scattering channels $\subProc$, and evaluating the resummation
individually $\vProb_{\momConf,\subProc}$ individually for each Born
configuration and scattering channel.

In principle, for each $\momConf$ and $\subProc$, one should
redetermine all the inputs to the master formula for
$\vProb_{\momConf,\subProc}$, \ie the 
$a$, $b_\ell$, $d_\ell$, $g_\ell(\phi)$ and $\cF$ (and the
applicability conditions). This would be rather slow, especially the
redetermination of $\cF$. Fortunately, for most of the cases that we
have examined it is only the $d_\ell$ that have any dependence on
$\momConf$  (modulo permutations of the indices $\ell$, to be
discussed shortly).\footnote{One can of course design observables for which
  this is not the case, for example $D^{y_3}$ in the three-jet limit,
  where $D$ is the $D$-parameter \cite{ERT} and $y_3$ is the Durham
  three-jet
  resolution.} %
This means that the analysis of the observable can be carried out in
full for just a single momentum configuration $\momConf_\mathrm{ref}$
and then for each new momentum configuration one redetermines only the
$d_\ell$, which is a straightforward procedure.

The exact property that is required is that for each Born
configuration $\momConf$, there should exist a permutation
function
$P_\momConf: \{\ell\} \to \{\ell'\}$ such that
\begin{subequations}
  \label{eq:Born-conf-props}
\begin{align}
  a_{\ell,\momConf} &= a_{\ell',\momConf_\mathrm{ref}}\,,\\
  b_{\ell,\momConf} &= b_{\ell',\momConf_\mathrm{ref}}\,,\\
  g_{\ell,\momConf} (\phi) &= g_{\ell',\momConf_\mathrm{ref}}(\phi)\,,
\end{align}
and furthermore
\begin{multline}
  \label{eq:mult-emsn-independence}
  V(\{p\}_{\momConf},\kappa(\zeta_1\bar
  v;\ell_1,\phi_1,\xi_1),\ldots, \kappa(\zeta_m\bar
  v;\ell_m,\phi_m,\xi_m))
  = \\ = 
  V(\{p\}_{\momConf_\mathrm{ref}},\kappa(\zeta_1\bar
  v;\ell_1',\phi_1,\xi_1),\ldots, \kappa(\zeta_m\bar
  v;\ell_m',\phi_m,\xi_m)) \,.
\end{multline}
\end{subequations}
Given these conditions it is straightforward to show that the function
$\cF$ to be used is
\begin{equation}
  \label{eq:F-perm-col-facts}
  \cF_{\momConf}(C_{1},\ldots,C_{n}; \lambda) \equiv
  \cF_{\momConf_\mathrm{ref}}(C_{P_\momConf^{-1}(1)},\ldots,C_{P_\momConf^{-1}(n)}; 
  \lambda)\,,  
\end{equation}
where the $\{C_\ell\}$ are the colour factors for the legs of
$\momConf$. Accordingly the problem of evaluating $\cF$ for an
arbitrary Born configuration reduces to that of evaluating it for a
single reference Born configuration, but for all permutations of
colour factors. One should of course also consider that different sets
of colour factors may arise for different Born subprocesses.

With this approach, the calculation of the integral over Born
configurations, eq.~(\ref{eq:Sigmacut_resummed}), now involves the
following steps: for each configuration $\momConf$, one should find, 
if it exists, a permutation such that
eqs.~(\ref{eq:Born-conf-props}) hold, determine the $d_\ell$, and then
compute the resulting distribution $f_{\momConf,\subProc}(v)$. This is
still a moderately slow procedure, because establishing the existence
of a suitable permutation involves probing the observable with a
number of test configurations of soft and collinear emissions for each
$\momConf$.

So, as a further simplification, we make the additional assumption that
a common rule holds for determining the permutation for a wide range
of observables. This rule differs according to the process:
\begin{itemize}
\item For $\ee\to3$~jets we choose the permutation that ensures
  $E_{P_\momConf^{-1}(1)} > E_{P_\momConf^{-1}(2)}> E_{P_\momConf^{-1}(3)}$.
\item For DIS $2+1$ jet events we permute only outgoing legs,
  such that $p_{P_\momConf^{-1}(3)}.p_1 < p_{P_\momConf^{-1}(2)}.p_1$.
\item For hadronic dijet events we permute only outgoing legs, such
  that $p_{P_\momConf^{-1}(3)}.p_1 < p_{P_\momConf^{-1}(4)}.p_1$.
\end{itemize}
While this is not a general solution to the problem of determining
$P_\momConf$, we find it to be adequate for the whole of range of
observables that are normally studied. It is of course mandatory that
one tests its validity. This is done for a random (sub)sample of Born
configurations during the (Monte Carlo) evaluation of the integral in
eq.~(\ref{eq:Sigmacut_resummed}).

\subsection{(Meta-)Results}
\label{sec:res}

It would be natural at this point, having given an extensive
discussion of the basis and implementation of \caesar, to illustrate
its capabilities with some example resummations.

One of the main potential applications of \caesar is the resummation
of event-shapes and jet-rates in hadronic-dijet production.  With the
aid of resummed predictions (and recent progress also in fixed-order
calculations \cite{NLOJET,Nagy03,TriRad}), event-shape and jet-rates
studies at hadronic experiments should allow studies of a number of
interesting issues, related for instance to the underlying event, or,
from a purely perturbative point of view, to the non-trivial structure
of interference between large-angle soft emissions from different
dipoles, a characteristic of events with four or more jets.

A first resummed result for hadronic dijet events was given in
\cite{BSZ03}, for a global variant of a transverse thrust.  We are
aware of only one experimental measurement of a hadronic dijet event
shape distribution, \cite{D0Thrust}, also a transverse thrust, but
with a non-global definition --- it is therefore beyond the current
scope of our approach.

Compared to $\ee$ environments, the issue of globalness for
observables in had\-ronic collisions turns out to be particularly
critical, because limited detector reach at forward rapidities
restricts the measurement of the properties of the beam remnant jets
(which form an integral part of any global measurement).
Nevertheless, it turns out to be possible to define various types of
observables, specifically designed to be global 
but hopefully still measurable at hadronic colliders.

The presentation of a systematic definition of classes of hadronic
dijet event-shapes, together with sample output results for the
analysis and the resummation from \caesar, is naturally accompanied by
a discussion on how complementary properties of various observables
can be tuned to address various aspects of the physics of hadron
colliders. Accordingly, rather than present example resummations here,
we have chosen to devote a second, companion article to the subject
\cite{BSZhh}.

Additionally some illustrative examples are given in
appendix~\ref{sec:ee-observables} for the BKS (or angularity)
continuous class of $\ee$ observables \cite{BKS03,BergerMagnea,BS03}
and also for a new alternative class that is better behaved with
respect to variation of the continuous parameter that defines
individual elements of the class. Many more examples, in a range of
hard processes, are available from \cite{qcd-caesar.org}.

\section{Conclusions and outlook}
\label{sec:conclu}

In this article we have presented a detailed derivation of a master
formula for NLL final-state resummations, and discussed the properties
that an observable has to fulfil in order for the approach to be
valid --- principally continuous globalness and a novel property,
recursive infrared and collinear safety. We have also outlined the
elements that were needed to construct a computer program, \caesar,
that can determine, given a subroutine for the observable, all the
observable-dependent inputs to the master formula. It
will be made public in the near future.

The breadth of results already obtained with \caesar, presented
elsewhere \cite{BSZhh,qcd-caesar.org} (and in
appendix~\ref{sec:ee-observables}), testifies as to the power of the
approach. Therefore, rather than review, once again, the achievements
of the method, we discuss here briefly the scope for future work.

The most immediate direction for future work is that of
phenomenological applications,
including the study
of hadronic dijet event shapes discussed in \cite{BSZhh}. All such
studies require matching to fixed order predictions and in processes
with three or more jets, certain new conceptual issues
arise~\cite{BSZPrep} compared to the well understood two-jet
case~\cite{CTTW}, related to the identification of separate
hard-scattering channels in the fixed-order calculation.

More generally, it would  of course be of interest to extend the
approach to non-global 
observables, which are often easier to measure than global
observables, especially in processes with incoming hadrons.  Partially
analytical resummations exist for a range of non-global final-state
observables \cite{NG1,DiscontGlobal,BMS,ApplebySeymour,BanfiDasgupta}
in the large-$\NC$ limit and advances are also being made beyond
leading $\NC$ \cite{WeigertNG}.  Relative to the global case, the
additional complication within an automated approach comes from the
need to treat boundaries that separate regions with different
sensitivities to the transverse momenta of the emissions.

Yet another possible extension includes the case of final-state
observables in processes with heavy quarks, for which few resummed
results \cite{KraussRodrigo} exist as yet.

Further progress could also be made for observables that involve
cancellations between different emissions (for example due to vector
transverse momentum sums), for which the resummation applies only up
to some finite value of $\as L$. For a number of observables (\eg
\cite{DSBroad}) the breakdown is in a sufficiently suppressed region
that it can be ignored, however this is not always the case.  Beyond,
one must currently resort to standard analytical methods, based on
appropriate integral transforms, as in \cite{Bonciani:2003nt}. The
methods developed here already make it possible to identify many of
the most common cases of such observables. A full solution to the
problem might conceivably make use of that information to actually
carry out the resummation.

Some final comments relate to recursive infrared and collinear
safety. For many years it has been known that there are observables
for which double logarithms do not exponentiate, \ie the resummed
series cannot be expressed in the form eq.~(\ref{eq:vProb-general}).
One of the significant developments made here is the formulation of a
\emph{sufficient} condition for exponentiation, namely rIRC safety.
There are a number of analogies between rIRC safety and normal IRC
safety: for example, just as IRC unsafe observables can lead to NLO
predictions that diverge as an infrared regulator is taken to zero,
rIRC observables often have an NLL $\cF$ function that diverges as
infrared regulator is taken to zero. However, while the general consequences
of IRC safety are well understood --- it is the necessary and
sufficient condition for an observable to be calculable at all fixed
orders in perturbation theory --- rIRC safety remains somewhat more
nebulous.  One reason for this is that it is not yet clear how to
formulate an approach like ours at all logarithmically resummed
orders. Only within the framework of such a systematic approach would
it then have any sense to make an analogous statement about rIRC
safety.

\paragraph{Acknowledgements}

We have benefited from conversations on issues related to automated
resummation with a number of people, including Stefano Catani, Mrinal
Dasgupta, Yuri Dokshitzer, Jeff Forshaw, Stefan Gieseke, Eric Laenen,
Pino Marchesini and Mike Seymour. We also thank Mrinal Dasgupta for
his comments on the manuscript and Zoltan Nagy for having provided us
with the latest version of NLOJET++ (and help in running it), of use in
a number of cross checks. Each of us has greatly appreciated the
hospitality of the others' institutions both current and past.
Finally, we are grateful to CERN, the IPPP, Durham and Milano Bicocca
University for the continued use of computing facilities.

\appendix

\section{Analytical ingredients}
\label{sec:analytics}

In this section we collect the analytical formulae needed to evaluate
\eqref{eq:Master} and its order-by-order expansion.  The knowledge of
the latter is needed when matching resummed predictions to fixed order
calculations, a step which we leave for a future work.  We also
present some alternative representations of the function $\cF$, which
are more convenient for analytical evaluation of the function.

\subsection{The radiators}
\label{sec:rad}
It has become standard to give all resummed quantities in terms of
$\lam=\as\be_0 L$, with $\as=\alpha_{s,\MSbar}$.
We consider first the function $r_\ell$ defined in \eqref{eq:rell},
and split it into a pure LL term and a pure NLL term as follows,
$r_{\ell}(L)=Lr_{1,\ell}(\as L)+r_{2,\ell}(\as L)$. 
In terms of $\lambda$ one then has, for $b_\ell\ne 0$,
\begin{equation}
\label{rad-lam}
\begin{split}
r_{1,\ell}(\as L)&=\frac{1}{2\pi\be_0 \lam b_{\ell}}
\left((a-2 \lambda)
\ln\left(1-\frac{2\lam}{a}\right)
-(a+b_{\ell}-2\lam)
\ln\left(1-\frac{2\lam}{a+b_{\ell}}\right)
\right)\>,\\
r_2(\as L)&=\frac{1}{b_{\ell}}\left[\frac{K}{4\pi^2\be_0^2}
\left((a+b_{\ell})\ln\left(1-\frac{2\lam}{a+b_{\ell}}\right)
-a\ln\left(1-\frac{2\lam}{a}\right)
\right)\right.\\
&\left.+\frac{\be_1}{2\pi\be_0^3}\left(
\frac{a}{2}\ln^2\left(1-\frac{2\lam}{a}\right)
-\frac{a+b_{\ell}}{2}\ln^2\left(1-\frac{2\lam}{a+b_{\ell}}\right)
\right.\right.\\&\left.\left.
+a\ln\left(1-\frac{2\lam}{a}\right)
-(a+b_{\ell})\ln\left(1-\frac{2\lam}{a+b_{\ell}}\right)\right)\right]\>,
\end{split}
\end{equation}
where the first two coefficients of the beta function are 
\begin{equation}
  \label{eq:b0b1}
  \beta_0=\frac{11 C_A - 4T_R n_f}{12 \pi}\>,\qquad
  \beta_1=\frac{17 C_A^2-5C_A n_f -3 C_F n_f}{24\pi^2}\>,
\end{equation}
and $K$ is the constant that relates the physical scheme of
ref.~\cite{CMW} to the $\MSbar$ scheme:
\begin{equation}
  \label{eq:K}
  K = \CA\left(\frac{67}{18}-\frac{\pi^2}{6}\right)-\frac{5}{9}\nf\,.
\end{equation}
These expressions have a finite limit for $b_{\ell}=0$:
\begin{equation}
\label{rad-lam-0}
\begin{split}
r_1(\as L)&=-\frac{1}{2\pi\be_0\lam}
\left(
\frac{2\lam}{a}+\ln\left(1-\frac{2\lam}{a}\right)
\right)\>,\\
r_2(\as L)&=\frac{K}{4\pi^2\be_0^2}
\left(
\ln\left(1-\frac{2\lam}{a}\right)
+\frac{2}{a}\frac{\lam}{1-\frac{2}{a}\lam}
\right)
\\
&-\frac{\be_1}{2\pi\be_0^3}
\left(
\frac12 \ln^2\left(1-\frac{2\lam}{a}\right)
+\frac{\ln\left(1-\frac{2\lam}{a}\right)
+\frac{2}{a}\lam}{1-\frac{2\lam}{a}}
\right)\>.
\end{split}
\end{equation}
The function $T(L)$ in \eqref{eq:T} is given by
\begin{equation}
  \label{eq:sl-lam}
  T(L)=-\frac{1}{\pi\be_0}\ln\left(1-2\lam\right)\>,
\end{equation}
and the function $r_\ell'(L)$ can be obtained from \eqref{eq:rpell}.
We give here only its (finite) limit for $b_\ell = 0$:
\begin{equation}
  \label{eq:rprime-lam-0}
  r'(L)=\frac{2}{a^2}\frac{1}{\pi\be_0}
  \frac{\lam}{1-\frac{2\lam}{a}}\>,\qquad b_\ell = 0\>.
\end{equation}
Finally, a change in renormalisation scale from $Q$ to $\mu$ results in a
change in $r_{2,\ell}$ as follows
\begin{equation}
  \label{eq:ren-change}
  r_{2,\ell}\to r_{2,\ell}+\lam\ln\frac{\mu^2}{Q^2}(r'_{\ell}-r_{1,\ell})\>.
\end{equation}

\subsection{The expansion of $\cF$ to order  $R'^2$}
\label{sec:F2}

One can consider $\cF$ as an expansion in powers of $R'$,
\begin{equation}
  \label{eq:cFexpansion}
  \cF(R') = 1 + \sum_{p=2}^\infty \cF_2 R'^p\,.
\end{equation}
The first term in the expansion, $\cF_2$ is relatively simple, and can
be written as follows
\begin{multline}
  \label{eq:F2}
  \cF_2 =  -\left(\sum_{\ell=1}^n \frac{C_\ell}{a(a + b_\ell)} \right)^{-2}
\left(\prod_{i=1}^{2} \sum_{\ell_i=1}^n
   \int_{0}^{1} \frac{d\zeta_i}{\zeta_i} 
   \frac{C_{\ell_i}}{a(a + b_{\ell_i})}
   \int_0^1 d\xi_i\int_0^{2\pi} \frac{d\phi_i}{2\pi}\right)
 \times \\ \times
   \delta(\ln \zeta_1)
   \ln \lim_{\bar v\to0} \frac{V(\{\tilde
      p\},\kappa_1(\zeta_1 \bar v),\kappa_2(\zeta_2 \bar v))}{\bar v}\,.
\end{multline}
For observables that belong to the event-shapes class, the integrals
over the $\xi_i$ can be evaluated analytically and just give $1$. For
additive observables,
\begin{equation}
  \label{eq:F2_additive}
    \cF_2 = -\int_0^1 \frac{d\zeta_2}{\zeta_2}\ln(1+\zeta_2) = 
    -\frac{\pi^2}{12}\,.
\end{equation}
In arriving at eq.~(\ref{eq:F2}), various manipulations have been
carried out assuming rIRC safety, as was the case also for
eqs.~(\ref{eq:cF-with-vto0-limit-anyn}) and (\ref{eq:cF-evshps-anyn}).
If one wishes to give a quantitative interpretation to $\cF_2$ when
investigating non rIRC observables (specifically for double
logarithmic violations of exponentiation), one is not entitled to
carry out those manipulations and rather one should explicitly derive
$\cF_2$ from eq.~(\ref{eq:MultipleIndepReg}), as a function of the
base scale $\bar v$ at which one evaluates $R$,
\begin{multline}
  \label{eq:F2nonrIRC}
  \cF_2(\bar v) = \frac{1}{2!}
  \left(\sum_{\ell=1}^n \frac{C_\ell}{a(a + b_\ell)} \right)^{-2}\!\!
  \left(\prod_{i=1}^{2} \sum_{\ell_i=1}^n
    \int_{0}^{\infty} \frac{d\zeta_i}{\zeta_i} 
    \frac{C_{\ell_i}}{a(a + b_{\ell_i})}
    \frac{\ln \zeta_i \bar v}{\ln \bar v}
    \int_0^1 \!\!d\xi_i\int_0^{2\pi} \frac{d\phi_i}{2\pi}\right)
  \times \\ \times
  \left(\Theta(\bar v - V(\{\tilde
    p\},\kappa_1(\zeta_1 \bar v),\kappa_2(\zeta_2 \bar v)))
    - \Theta(1 - \zeta_1) \Theta(1 - \zeta_2)
  \right)\,.
\end{multline}
It is simple to verify that for rIRC safe observables it coincides
with eq.~(\ref{eq:F2}) in the limit $\bar v \to 0$.

\subsection{The fixed order expansions}
\label{sec:fixedord}

In order to compare resummed results with fixed order calculations, it
is useful to know the fixed order expansion for $\vProb(v)$ in
eq.~\eqref{eq:Master}.
We choose to write $\vProb(v)$
in the form\footnote{This differs from the convention adopted
  in~\cite{CTTW} in which the whole probability $\vProb(v)$ is written as an
exponential.}
\begin{equation}
  \begin{split}
    \label{eq:Sigma-exp}
    \vProb(v)&=\cF(R')S(T(L/a))
    \prod_{\ell=1}^{n_i}\frac{q^{(\ell)}(x_\ell,e^{-\frac{2L}{a+b_\ell}}\muf^2)} 
    {q^{(\ell)}(x_\ell,\muf^2)}
    \exp\left\{\sum_{n=1}^{\infty}\sum_{m=0}^{n+1} G_{nm} \asb^n
      L^m\right\} \\
    &=\sum_{n=0}^{\infty}\sum_{m=0}^{2n} H_{nm} \asb^n
    L^m
    \>,    
  \end{split}
\end{equation}
with, as usual, $L=\ln 1/v$ and $\asb=\alpha_{s,\MSbar}/(2\pi)$. In
the first line of eq.~\eqref{eq:Sigma-exp} we isolate the contributions
to $\vProb(v)$ that can be straightforwardly written as an exponential,
while the second line defines $H_{mn}$,
the coefficients of the expansion of $\vProb(v)$ in powers of
$\asb$ and of $L$.

In order to derive the explicit form for $G_{nm}$ we need the
expansions of $r_{1,\ell}(L)$, $r_{2,\ell}(L)$, $r'_\ell(L)$ and
$T(L)$:
\begin{equation}
  \label{eq:r-exp}
  \begin{split}
    r_{1,\ell}(L)&=\sum_{n=1}^{\infty}\frac{4(4\pi\be_0)^{n-1} 
      \bar \as^n L^n}{n(n+1)\, b_\ell}
    \left(\frac{1}{a^n} -\frac{1}{(a+b_\ell)^n}\right)\>,\\
    r_{2,\ell}(L)&=K\sum_{n=2}^{\infty}\frac{4(4\pi\be_0)^{n-2} 
      \bar \as^n L^n  }{ n\, b_{\ell}}
      \left(\frac{1}{a^{n-1}} -\frac{1}{(a+b_\ell)^{n-1}}\right)\\
    &+ 32 \pi^2\be_1\sum_{n=3}^{\infty}\frac{(4\pi\be_0)^{n-3}
      \bar \as^n L^n
    }{n\, b_{\ell}}(\gamma_E + \psi(n)-1)
      \left(\frac{1}{a^{n-1}} 
        -\frac{1}{(a+b_\ell)^{n-1}}\right)\>,\\
    T(L)&=\sum_{n=1}^{\infty}\frac{4(4\pi\be_0)^{n-1} }{n}\asb^n L^n
    \>,\\
    r'_{\ell}(L)&=\sum_{n=1}^{\infty}\frac{4(4\pi\be_0)^{n-1}\asb^n L^n}
    {n\, b_\ell}  \left(\frac{1}{a^n} -\frac{1}{(a+b_\ell)^n}\right)\>. 
  \end{split}
\end{equation}
Notice that the above expressions have well defined $b_\ell \to 0$
limits, which read:
\begin{equation}
  \label{eq:r-exp-b=0}
  \begin{split}
    r_{1,\ell}(L)&=\sum_{n=1}^{\infty}\frac{4(4\pi\be_0)^{n-1} 
      \bar \as^n L^n}{(n+1)\, a^{n+1}}\>,\qquad 
    r'_{\ell}(L)=\sum_{n=1}^{\infty}\frac{4(4\pi\be_0)^{n-1}\asb^n L^n}
    { a^{n+1}} \>,\\
    r_{2,\ell}(L)&=K\sum_{n=2}^{\infty}\frac{4(4\pi\be_0)^{n-2} 
      \bar \as^n L^n (n-1) }{ n \,a^n }\\
    &+ 32 \pi^2\be_1\sum_{n=3}^{\infty}\frac{(4\pi\be_0)^{n-3}
      \asb^n L^n (n-1)
      }{n\, a^n}(\gamma_E + \psi(n)-1)\>. 
  \end{split}
\end{equation}

Substituting \eqref{eq:r-exp} (or \eqref{eq:r-exp-b=0})
in~\eqref{eq:Master} we are able to extract the coefficients $G_{nm}$.
Here we report only the terms that we are able to control at NLL
accuracy up to second order in $\as$, which correspond to the current
accuracy of fixed order calculations (the expansion to higher
orders being just a trivial exercise):
\begin{equation}
  \label{eq:Gnm}
  \begin{split}
 G_{12}&=-\frac{2}{a}\sum_\ell \frac{C_\ell}{a+b_\ell}\>, \\
 G_{11}&=-\sum_\ell C_\ell\left(\frac{4B_\ell}{a+b_\ell}+
    \frac{4}{a(a+b_\ell)}\left(\ln\bar
      d_{\ell}-b_\ell\ln\frac{2E_\ell}{Q}\right)\right)\>,\\
 G_{23}&=-\frac{8\pi\be_0}{3a^2}\sum_\ell 
  C_\ell\frac{2a+b_\ell}{(a+b_\ell)^2}\>,\\
 G_{22}&=-\frac{8\pi\be_0}{a^2}\sum_\ell
  C_\ell\frac{2a+b_\ell}{(a+b_\ell)^2}
  \left(\ln\bar d_{\ell}-b_\ell\ln\frac{2E_\ell}{Q}\right)\\
  &-8\pi\be_0\sum_\ell \frac{C_\ell B_\ell}{(a+b_\ell)^2}
  -\frac{2K}{a}\sum_\ell \frac{C_\ell}{a+b_\ell}\>. 
  \end{split}
\end{equation}

The last step is to expand also $\cF$, $S(T(L/a))$ and
$q^{(\ell)}(e^{-\frac{2L}{a+b_\ell}}\muf^2)$. The expansion for $\cF$ can be
found in eq.~(\ref{eq:cFexpansion}), while
\begin{equation}
  \label{eq:Sexpansion}
  S(t)=1+\sum_{p=1}^{\infty} S_p \,t^p\>, 
\end{equation}
where the coefficients $S_p$ can be easily extracted from
eq.~\eqref{eq:S-summary}. In particular, for $n<4$ we have simply 
$S(t)=\exp\{S_1 \, t\}$, with
\begin{align}
  \label{eq:S1-nlt4}
  n = 2:\!\! &\quad S_1 = - 2C_F \,\ln \frac{Q_{qq'}}{Q}\,,\\
  n = 3:\!\! &\quad S_1 = -\left[{\CA}\ln \frac{Q_{qg} Q_{q'
        g}}{Q_{q q'} Q}  
+ 2\CF \ln \frac{Q_{q q'}}{Q}\right] \,,
\end{align}
while for $n = 4$ the situation becomes more complicated and we have
\begin{equation}
  \label{eq:S1-n=4}
  \begin{split}
    S_1 &= -\sum_\ell C_\ell \ln \frac{Q_{12}}{Q} -\frac12
    \frac{\mathrm{Tr} (H\Gamma^\dagger M+HM\Gamma)}{\mathrm{Tr} (H M)} \,,\\
    S_2 &= 
    \frac12\left(\sum_\ell C_\ell \ln \frac{Q_{12}}{Q}\right)^2
    +\frac12\left(\sum_\ell C_\ell \ln
      \frac{Q_{12}}{Q}\right)\frac{\mathrm{Tr} (H\Gamma^\dagger
      M+HM\Gamma)}{\mathrm{Tr} (H M)} \\
    &+\frac18 \frac{\mathrm{Tr} (H(\Gamma^\dagger)^2 M+2 (H
      \Gamma^\dagger M \Gamma) +
      HM\Gamma^2)}{\mathrm{Tr} (H M)} \,,
  \end{split}
\end{equation}
with the matrices $H$, $\Gamma$ and $M$ reproduced in
appendix~\ref{sec:softlargeangle}.

In order to compute the expansion for
$q^{(\ell)}(x_\ell,e^{-\frac{2L}{a+b_\ell}} \muf^2)$
we use the following
notation,
\begin{equation}
  \label{eq:f-matrix}
  {\bf q}(x,\muf^2)=\left(
    \begin{array}[c]{c}
      q_u(x,\muf^2) \\
      q_{\bar u}(x,\muf^2)\\
      \vdots \\
      g(x,\muf^2)
    \end{array}
  \right)
\,,\qquad
{\bf P}(x)=\left(
  \begin{array}[c]{cccc}
    P^{(0)}_{qq}(x) & 0 & \cdots & P^{(0)}_{qg}(x) \\
    0 & P^{(0)}_{qq}(x) &  &  \\
    \vdots & & \ddots & \\
    P^{(0)}_{gq}(x) & & & P^{(0)}_{gg}(x)
  \end{array}
  \right)\,,
\end{equation}
where $P^{(0)}_{ij}(x)$ are the leading order Altarelli-Parisi splitting
functions, taken from ref.~\cite{ESWbook}, which we reproduce here for
completeness:
\begin{equation}
  \label{eq:Pij}
  \begin{split}
  P^{(0)}_{qq}(x) =\>& \CF\left[
    \frac{1+x^2}{(1-x)_+}+\frac{3}{2}\delta(1-x)
  \right]\,,\\     
  P^{(0)}_{qg}(x) =\>& \TR\left[
    x^2+(1-x)^2
  \right]\,,\\    
P^{(0)}_{gq}(x)  =\>& \CF\left[
    \frac{1+(1-x)^2}{x}
  \right]\,,\\    
P^{(0)}_{gg}(x)  =\>& 2\CA\left[
  \frac{x}{(1-x)_+}+
  \frac{1-x}{x}+x(1-x)
  \right]\\
  &+\delta(1-x)\frac{(11\CA-4n_f\TR)}{6}
  \,.    
  \end{split}
\end{equation}
We also make the identification $q^{(\ell)}(x_\ell,\muf^2)={\bf
  q}^{(\ell)}_i(x_\ell,\muf^2)$, with $i$ the flavour of hard
parton $p_\ell$.
To NLL accuracy, one can express
$q^{(\ell)}(e^{-\frac{2L}{a+b_\ell}}\muf^2)$ in terms of the 
function $T(L/(a+b_\ell))$ as follows
\begin{equation}
  \label{eq:q-evolve}
  q^{(\ell)}(x_\ell,v^{\frac{2}{a+b_\ell}}\muf^2)=
  \left[e^{-\left\{\frac{t}{2}({\bf P} \>\otimes)\right\}}\>{\bf q}^{(\ell)}\right]_i (x_\ell,\muf^2)\> \>, \qquad t = 
  T\left(\frac{L}{a+b_\ell}\right)\>,
\end{equation}
where we have used the notation
\begin{equation}
  \label{eq:otimes}
  [{\bf P} \otimes {\bf q}]_{i} (x,\muf^2) = 
  \int_{x}^1 \!\frac{dz}{z}
  {\bf P}_{ij}\left(\frac{x}{z}\right)
  \>{\bf q}_j(z,\muf^2)\,
\end{equation}
to indicate both matrix multiplication and convolution in $x$ space.
The expansion of $q^{(\ell)}(e^{-\frac{2L}{a+b_\ell}}\muf^2)$ in powers
of $T(L/(a+b_\ell))$ can be then trivially obtained from
eq.~(\ref{eq:q-evolve}).

If one wants to compute $q^{(\ell)}(e^{-\frac{2L}{a+b_\ell}}\muf^2)$
with a NLL DGLAP evolution, one should modify the expansion
accordingly.  Of course the differences in the two treatments of the
evolution would appear only at NNLL level in $\vProb(v)$.

All these ingredients can be merged together to obtain the
coefficients $H_{nm}$ defined in~\eqref{eq:Sigma-exp}, given here again
only to second order in $\bar \as$:
\begin{equation}
  \label{eq:Hij}
  \begin{split}
    H_{12}&= G_{12} \>, \quad H_{11}=
    G_{11}+\frac{4}{a}S_1-\sum_{\ell=1}^{n_i}\frac{2}{a+b_\ell}
    \frac{[{\bf P}\otimes {\bf q}^{(\ell)}]_i}{q^{(\ell)}}\>,\\
    H_{24}&=\frac12  G_{12}^2 \>, \qquad H_{23}= 
    G_{23}+ G_{12}\left( G_{11}+\frac{4}{a}S_1-\frac{2}{a+b_\ell}
      \frac{[{\bf P}\otimes  {\bf q}^{(\ell)}]_{i}}{q^{(\ell)}}\right)\>,\\
    H_{22}&= \frac12  G_{11}^2 + C_1  G_{12} + 
    G_{22}
    +\frac{8\pi\be_0}{a^2}S_1+\frac{16}{a^2} S_2+\frac{4}{a}S_1  G_{11}\\
    &+\frac{16 \cF_2}{a^2}\left(\sum_\ell
      \frac{C_\ell}{a+b_\ell}\right)^2 -\left(
      G_{11}+\frac{4}{a}S_1\right) \sum_{\ell=1}^{n_i}
    \frac{2}{a+b_\ell}\frac{[{\bf P}\otimes {\bf
        q}^{(\ell)}]_{i}}{q^{(\ell)}} 
    \\&-\sum_{\ell=1}^{n_i}
    \frac{4\pi\be_0}{(a+b_\ell)^2}\frac{[{\bf P}\otimes {\bf
        q}^{(\ell)}]_{i}}{q^{(\ell)}}
    \!+\!\sum_{\ell_1}^{n_i}\sum_{\ell_2 < \ell_1}^{n_i}\!\!
    \left(\frac{2}{a+b_{\ell_1}}\frac{[{\bf P}\otimes {\bf
          q}^{(\ell_1)}]_{i}}{q^{(\ell_1)}}\right)
    \left(\frac{2}{a+b_{\ell_2}}\frac{[{\bf P}\otimes {\bf
          q}^{(\ell_2)}]_{i}}{q^{(\ell_2)}}\right) 
    \\&+\frac
    12\sum_{\ell=1}^{n_i}\left(\frac{2}{a+b_\ell}\right)^2 \frac{[{\bf
        P}\otimes {\bf P}\otimes {\bf q}^{(\ell)}]_{i}}{q^{(\ell)}}\>.
  \end{split}
\end{equation}
Analogous expressions can be obtained for the higher order coefficients.

\subsection{More analytically convenient forms for $\cF$}
\label{sec:anl_convenient_F}

From the point of view of analytical evaluations of $\cF$ (not that
this should really be necessary!) it can be convenient, rather than
using eqs.~(\ref{eq:cF-with-vto0-limit-anyn}) and
(\ref{eq:cF-evshps-anyn}), to resort to the following equivalent
forms. Retaining explicit explicit $\xi_i$ integrations, we have
\begin{multline}
  \label{eq:cFalt-with-vto0-limit-anyn}
  \cF = \lim_{\epsilon\to0}
  \epsilon^{R'}
    \sum_{m=0}^{\infty} \frac{1}{m!}
  \left( \prod_{i=1}^{m} \sum_{\ell_i=1}^n
    \int_{\epsilon}^{\infty} \frac{d\zeta_i}{\zeta_i} 
    \frac{C_\ell r_{\ell_i}'}{\cN_{\ell_i}(\lambda/\beta_0)}
    \int_0^1 \frac{d\xi_i}{1 + \frac{a + (1-\xi_i)b_{\ell_i}}{a(a+b_{\ell})}
    \lambda} 
  \int_0^{2\pi} \frac{d\phi_i}{2\pi}
  \right)   
  \times \\ \times 
  \Theta\left(1 - \lim_{\bar v\to0} \frac{V(\{\tilde
      p\},\kappa_1(\zeta_1 \bar v), \ldots,
      \kappa_{m}(\zeta_{m} \bar v))}{\bar v}\right),
\end{multline}
while for observables where $\xi_i$ can be integrated out
analytically, the result is
\begin{multline}
  \label{eq:cFalt-evshps-anyn}
  \cF = \lim_{\epsilon\to0}
  \epsilon^{R'}
    \sum_{m=0}^{\infty} \frac{1}{m!}
  \left( \prod_{i=1}^{m} \sum_{\ell_i=1}^n  C_\ell r_{\ell_i}'
    \int_{\epsilon}^{\infty} \frac{d\zeta_i}{\zeta_i} 
    \int_0^{2\pi} \frac{d\phi_i}{2\pi}
  \right) 
  \times \\ \times
  \Theta\left(1 - \lim_{v\to0} \frac{V(\{\tilde p\},\kappa_1(\zeta_1
      {\bar v}) , \ldots,
      \kappa_{m}(\zeta_{m}
      {\bar v}))}{\bar v}\right),
  \qquad \xi_i = \mathrm{any}\,.
\end{multline}
It is to be kept in mind however that for numerical evaluations these
forms are considerably less efficient than
eqs.~(\ref{eq:cF-with-vto0-limit-anyn}) and (\ref{eq:cF-evshps-anyn}).

\section{Soft large angle contributions for $n=4$}
\label{sec:softlargeangle}
We reproduce here the explicit expressions for the matrices $\Gamma$,
$H$ and $M$ needed to compute the function $S$ which accounts for soft
large-angle emission for processes which involve two incoming and two
outgoing hard partons at Born level.

All the matrices are taken from \cite{BottsSterman,KS,KOS,Oderda,KidonakisOwens}, with
slightly changed conventions.  First our definition of the
$\Gamma$-matrix differs from the one in \cite{KOS} in that we
extract a factor $\as/\pi$.
Furthermore, the normalisation of $H$ and $M$ is fixed here in such a
way that the Born partonic cross section for a given (partonic)
subprocess $\subProc$ (with flavour content $ij\to kl$) is given by:
\begin{equation}
  \label{eq:dsigma0}
  \frac{d\sigma_\subProc}{d \that} =
  \frac{\pi\as^2}{\shat}\mathrm{Tr}(HM)
  \frac{1}{1+\delta_{kl}}
\>. 
\end{equation}
Here $1/(1+\delta_{kl})$ represents the needed symmetry factor for
producing two identical particles, $\shat\equiv(p_1 + p_2)^2$, and
$\that\equiv(p_1 - p_3)^2$.  A comment here is in order concerning the
labelling of parton momenta.  In the whole section all hard parton
momenta are labelled according to the flavour. For instance for the
subprocess $qg \to qg$, $p_1$ and $p_2$ will denote the momenta of the
{\em incoming} quark and gluon respectively, while $p_3$ and $p_4$
will denote respectively the momenta of the {\em outgoing} quark and
gluon.  In cases such as $qq\to qq$, in which such a labelling does
not lead to a unique parton identification, an arbitrary choice will
be performed, which of course will have no influence on all physical
quantities, such as the cross section~\eqref{eq:dsigma0} or the soft
function~(\ref{eq:S-hadronic-dijet}).

All the matrices $\Gamma$, $H$ and $M$ can be expressed in terms of
the Mandelstam invariants $\shat$, $\that$ and
$\hat u\equiv (p_1-p_4)^2$. It is also convenient to introduce 
\begin{equation}
  \label{eq:defTandU}
    T\equiv \ln(\frac{-\hat t}{\hat s})+i\pi\>, \qquad
    U\equiv \ln(\frac{-\hat u}{\hat s})+i\pi\>.
\end{equation}
For an extensive discussion on the physical meaning of all these
matrices, the reader is referred to \cite{BottsSterman,KS,KOS,Oderda}.
Here we collect only explicit results for all possible partonic subprocesses. 

The results we present here correspond to a particular choice of the
colour bases for each subprocess, which we will explicitly
indicate, denoting with $r_i$ the colour of parton $p_i$.  
\begin{itemize}

\item $q\qbar \to q\qbar$
  
  For this subprocess we choose the $t$-channel singlet-octet basis
  \begin{equation}
    \label{eq:basis-qq2qq}
    \begin{split}
      c_1 & = \delta_{r_1 r_3}\delta_{r_2 r_4}\,,\\
      c_2 & = -\frac{1}{2\NC}\delta_{r_1r_3}\delta_{r_2r_4}+\frac12\delta_{r_1r_2}\delta_{r_3r_4}\,.\\
    \end{split}
  \end{equation}

With this basis the expression for $H$ reads
\begin{equation}
\label{eq:H-qqbar}
H=\frac{2}{\NC^2} \, \left( 
\begin{array}{cc}
\frac{C_F^2}{\NC^2} \chi_1  & \frac{C_F}{\NC^2} \chi_2 \vspace{2mm} \\
\frac{C_F}{\NC^2} \chi_2 & \chi_3
\end{array}
\right) \,,
\end{equation}
where, in the case $q\bar{q} \to q\bar{q}$, $\chi_1$, $\chi_2$ and
$\chi_3$ are defined by
\begin{equation}
\label{eq:chi-qqbar}
  \begin{split}
    \chi_1&=\frac{\hat{t}^2+\hat{u}^2}{\hat{s}^2}\>,      \\
    \chi_2&=\NC\frac{\hat{u}^2}{\hat{s}\hat{t}}-
    \frac{\hat{t}^2+\hat{u}^2}{\hat{s}^2}\>,    \\
    \chi_3&=\frac{\hat{s}^2+\hat{u}^2}{\hat{t}^2}+
    \frac{1}{\NC^2}\frac{\hat{t}^2+\hat{u}^2}{\hat{s}^2}
    -\frac{2}{\NC}\frac{\hat{u}^2}{\hat{s}\hat{t}}\,.    
  \end{split}
\end{equation}
This result can be also exploited to describe the subprocesses $q\bar q \to
q' \bar q'$ and $q \bar q' \to q\bar q'$. 
For  $q\bar q \to q' \bar q'$ one has to keep in
\eqref{eq:chi-qqbar} only the
$s$-channel contributions, i.e. drop all terms containing
$\hat t$ in the denominator, while for $q \bar q' \to q\bar q'$ one
needs only the $t$-channel terms.

The matrix $\Gamma$ is given by
\begin{equation}
\label{eq:Gamma-qqbar}
  \Gamma=\left(
    \begin{array}{cc}
      2{C_F}T  &   -\frac{C_F}{\NC} U  \vspace{2mm} \\
      -2U    &-\frac{1}{\NC}(T-2U)
    \end{array} \right)\,,
  \end{equation}
and the matrix $M$ is 
\begin{equation}
\label{eq:M-qq}
  M=\left(
    \begin{array}{cc}
      \NC^2  & 0  \\
      0 &  \frac{1}{4}(\NC^2-1)
    \end{array}
    \right) \> .
\end{equation}

\item $qq\to qq$

The $t$-channel singlet-octet basis for this process is
  \begin{equation}
    \label{eq:basis-qq}
    \begin{split}
      c_1 & = \delta_{r_1 r_3}\delta_{r_2 r_4}\,,\\
      c_2 & = -\frac{1}{2\NC}\delta_{r_1r_3}\delta_{r_2r_4}+\frac12\delta_{r_1r_4}\delta_{r_2r_3}\,.\\
    \end{split}
  \end{equation}

  Since this subprocess is related to $q\bar{q} \rightarrow q\bar{q}$
  by the crossing transformation $\hat{s} \leftrightarrow \hat{u}$,
  the matrix $H$ has the same form as in eq.~\eqref{eq:H-qqbar}, with the
  functions $\chi_1$, $\chi_2$ and $\chi_3$ given by
\begin{equation}
\label{eq:chi-qq}
  \begin{split}
    \chi_1&=\frac{\hat{t}^2+\hat{s}^2}{\hat{u}^2}\>,       \\
    \chi_2&=\NC \frac{\hat{s}^2}{\hat{t}\hat{u}}-
    \frac{\hat{s}^2+\hat{t}^2}{\hat{u}^2}\>,   \\
    \chi_3&=\frac{\hat{s}^2+\hat{u}^2}{\hat{t}^2}+
    \frac{1}{\NC^2}\frac{\hat{s}^2+\hat{t}^2}{\hat{u}^2}-
    \frac{2}{\NC}\frac{\hat{s}^2}{\hat{t}\hat{u}}\,.
  \end{split}
\end{equation}

The unequal-flavour case $q q' \to q q'$ can be obtained from equation
\eqref{eq:chi-qq} by keeping only the $t$-channel terms.

The matrix $\Gamma$ for this subprocess reads
  \begin{equation}
    \Gamma=\left(
      \begin{array}{cc}
        2{C_F}T & \frac{C_F}{\NC} U  \vspace{2mm} \\
        2U & -\frac{1}{\NC}(T+U)+2C_F U  
      \end{array} \right)\,,
  \end{equation}
while the matrix $M$ is given in eq.~\eqref{eq:M-qq}.

\item $qg\to qg$

We use here the $t$-channel basis
\begin{equation}
  \label{eq:basis-qg}
  \begin{split}
    c_1 &= \delta_{r_1 r_3}\delta_{r_4 r_2}\,,\\
    c_2 &= d_{r_2 r_4 c}\,(t^{c})_{r_3 r_1}\,,\\
    c_3 &= i f_{r_2 r_4 c}\,(t^{c})_{r_3 r_1}\,.
  \end{split}
\end{equation}

The matrix $H$ is then given by: 
\begin{equation}
  \label{eq:H-qg}
H=\frac{1}{2 \phi(\NC)}
  \left( 
    \begin{array}{ccc}
      \frac{1}{\NC^2}\chi_1    &
      \frac{1}{\NC}\chi_1   &
      \frac{1}{\NC}\chi_2  \vspace{2mm} \\
      \frac{1}{\NC}\chi_1  & \chi_1 & \chi_2 \vspace{2mm} \\
      \frac{1}{\NC} \chi_2 & \chi_2 & \chi_3 
    \end{array}
\right) \> ,
\end{equation}
where the factor $\phi(\NC)$ represents the average over incoming
colours, that is $\phi(\NC)=\NC(\NC^2-1)$, and 
the functions $\chi_1$, $\chi_2$ and $\chi_3$ are
given by
\begin{equation}
  \label{eq:chi-qg}
  \begin{split}
    \chi_1&=- \frac{\shat^2+\hat u^2}{\hat{s}\hat{u}}\>,     \\
    \chi_2&=\left(1+\frac{2\hat{u}}{\hat{t}}\right) \, \chi_1\>,          \\
\chi_3&= \left(1-4\frac{\shat \hat u}{\hat t ^2}\right)\chi_1\,.
  \end{split}
\end{equation}

The matrix $\Gamma$ for this subprocess is given by
\begin{equation}
  \label{eq:gamqg2qg}
  \Gamma =\left(
\begin{array}{ccc}
\vspace{2mm}
 \left( C_F+C_A \right) T  &   0  & U  \\ \vspace{2mm}
 0  &   C_F T+ \frac{C_A}{2} U     & \frac{C_A}{2} U \\ \vspace{2mm}
 2U  & \frac{\NC^2-4}{2\NC}U  &  C_F T+ \frac{C_A}{2}U
\end{array} \right),
\end{equation}
and the soft matrix $M$ reads
\begin{equation}
  \label{eq:M-qg}
  M= \CF
  \left( 
    \begin{array}{ccc}
      2 \NC^2   & 0  & 0 \\
      0 & \NC^2-4 & 0 \\
      0 & 0 & \NC^2
    \end{array}
\right) \>.
\end{equation}

\item $q\qbar \to gg$ and $gg \to q\qbar$

The subprocess $q\qbar \to gg$ is better described with the $s$-channel basis
  \begin{equation}
    \label{eq:basis-qq2gg}
    \begin{split}
    c_1 &= \delta_{r_1 r_2}\delta_{r_3 r_4}\,,\\
    c_2 &= d_{r_3 r_4 c}\,(t^{c})_{r_2 r_1}\,,\\
    c_3 &= i f_{r_3 r_4 c}\,(t^{c})_{r_2 r_1}\,,
    \end{split}
  \end{equation}
  while the basis for $gg \to q\qbar$ can be obtained from 
  eq.~\eqref{eq:basis-qq2gg} by exchanging incoming and outgoing indices.
  
  In this case $H$ has the same form as in eq.~(\ref{eq:H-qg}), with
  the appropriate flux factor, $\phi(\NC)=\NC^2$ for $q\qbar \to gg$
  and $\phi(\NC)=(\NC^2-1)^2$ for $gg \to q\qbar$. The functions
  $\chi_1$, $\chi_2$ and $\chi_3$ can be obtained from those in
  eq.~(\ref{eq:chi-qg}) by performing the crossing transformation
  $\hat s \leftrightarrow \hat{t}$ and multiplying the answer by
  $(-1)$, since one fermion is involved in the crossing. The explicit
  result then reads
\begin{equation}
  \label{eq:chi-qqgg}
  \begin{split}
    \chi_1 &= \frac{\hat{t}^2 + \hat{u}^2}{\hat{t}\hat{u}}\>,   \\
    \chi_2 &= \left(1+\frac{2\hat{u}}{\hat{s}}\right) \, \chi_1\>, \\
    \chi_3 &= \left(1-\frac{4\hat{t}\hat{u}}{\hat{s}^2}\right) \,
    \chi_1 \, .
  \end{split}
\end{equation}

The expression for $\Gamma$ in this case is
\begin{equation}
  \label{eq:softqq2gg}
  \Gamma=\left(
                \begin{array}{ccc}
                  \vspace{2mm}
                 0  &   0  & U-T  \\ \vspace{2mm}
                 0  &   \frac{C_A}{2}\left(T+U \right) 
                 & \frac{C_A}{2}\left(U-T \right) \\ \vspace{2mm}
                 2\left(U-T \right)  
                 & \frac{\NC^2-4}{2\NC}\left(U-T \right)  
                 & \frac{C_A}{2}\left(T+U \right)
                \end{array} \right)\,,
\end{equation}
while the soft matrix $M$ is given in eq.~(\ref{eq:M-qg}).

\item $gg \to gg$
  
  Considering all possible colour structures for this subprocess would
  lead to $9 \times 9$ matrices, which can be written in a block
  diagonal form, involving $3\times 3$ and $6\times 6$
  submatrices~\cite{KOS}. For $\NC=3$ however, the basis
  vectors which give rise to the $6\times 6$ submatrix become linearly
  dependent, so that this matrix can be reduced to $5 \times 5$. We
  will therefore reproduce here all results only for $\NC=3$.  
  
  The basis we choose can be expressed partly in terms of $t$-channel
  $SU(3)$ projectors as follows:
  \begin{equation}
    \label{eq:basis-gggg}
    \begin{split}
    c_1 &= \frac{i}{4}[f_{r_1 r_2 c}d_{r_3 r_4 c}-
    d_{r_1 r_2 c}f_{r_3 r_4 c}]\>, \\
    c_2 &= \frac{i}{4}[f_{r_1 r_2 c}d_{r_3 r_4 c}+
    d_{r_1 r_2 c}f_{r_3 r_4 c}]\>, \\
    c_3 &= \frac{i}{4}[f_{r_1 r_3 c}d_{r_2 r_4 c}+
    d_{r_1 r_3 c}f_{r_2 r_4 c}]\>, \\
    P_1    &= \frac{1}{8}\delta_{r_1 r_3}\delta_{r_2 r_4}\>, \\
    P_{8_S} &= \frac{3}{5}d_{r_1 r_3 c}d_{r_2 r_4 c}\>,\\      
    P_{8_A} &= \frac{1}{3}f_{r_1 r_3 c}f_{r_2 r_4 c}\>,\\      
    P_{10\oplus \bar{10}} &= \frac12(\delta_{r_1 r_2}\delta_{r_3 r_4}-
    \delta_{r_1 r_4}\delta_{r_2 r_3})
    -\frac{1}{3}f_{r_1 r_3 c}f_{r_2 r_4 c}\>,\\      
    P_{27} &= \frac12(\delta_{r_1 r_2}\delta_{r_3 r_4}+
    \delta_{r_1 r_4}\delta_{r_2 r_3})-
    \frac{1}{8}\delta_{r_1 r_3}\delta_{r_2 r_4}-
    \frac{3}{5}d_{r_1 r_3 c}d_{r_2 r_4 c}\>.
    \end{split}
  \end{equation}

The matrix $H$ can be written in the form
\begin{equation}
 H=\left(\begin{array}{cc}
             0_{3 \times 3}        & 0_{3 \times 5} \\
              0_{5 \times 3}      &  H_{5 \times 5}
\end{array} \right)\, ,  
\end{equation}
where the $5 \times 5$ submatrix $H_{5 \times 5}$ is given by
\begin{equation}
H_{5 \times 5}=\frac{1}{16} \, 
\left(\begin{array}{ccccc}
9\chi_1 & \frac{9}{2}\chi_1 & \frac{9}{2}\chi_2 & 0 & -3\chi_1 \vspace{2mm}\\
\frac{9}{2}\chi_1 & \frac{9}{4}\chi_1 & \frac{9}{4}\chi_2 & 0 & -\frac{3}{2}\chi_1 \vspace{2mm}\\
\frac{9}{2}\chi_2 & \frac{9}{4}\chi_2 & \chi_3 & 0 & -\frac{3}{2}\chi_2 \vspace{2mm}\\
0 & 0 & 0 & 0 & 0 \vspace{2mm}\\
-3\chi_1 & -\frac{3}{2}\chi_1 & -\frac{3}{2}\chi_2 &0 &\chi_1 
\end{array}\right) \, ,
\end{equation}
and the functions $\chi_1$, $\chi_2$ and $\chi_3$ are defined as follows:
\begin{equation}
  \begin{split}
    \chi_1&=1-\frac{\hat{t}\hat{u}}{\hat{s}^2}-\frac{\hat{s}\hat{t}}{\hat{u}^2}+
    \frac{\hat{t}^2}{\hat{s}\hat{u}} \>,      \\
    \chi_2&=\frac{\hat{s}\hat{t}}{\hat{u}^2}-\frac{\hat{t}\hat{u}}{\hat{s}^2}+
    \frac{\hat{u}^2}{\hat{s}\hat{t}}-
    \frac{\hat{s}^2}{\hat{t}\hat{u}}\>,         \\
    \chi_3&=\frac{27}{4}-9\left(\frac{\hat{s}\hat{u}}{\hat{t}^2}+\frac{1}{4}
      \frac{\hat{t}\hat{u}}{\hat{s}^2}+\frac{1}{4}\frac{\hat{s}\hat{t}}{\hat{u}^2}\right)
+\frac{9}{2}\left(\frac{\hat{u}^2}{\hat{s}\hat{t}}+\frac{\hat{s}^2}{\hat{t}\hat{u}}-
\frac{1}{2}\frac{\hat{t}^2}{\hat{s}\hat{u}}\right) \,.    
  \end{split}
\end{equation}

The same can be done for the hard matrix $\Gamma$:
\begin{equation}
  \label{eq:softgg2gg}
  \Gamma=\left(\begin{array}{cc}
            \Gamma_{3 \times 3} & 0_{3 \times 5} \\
            0_{5 \times 3}      & \Gamma_{5 \times 5}
          \end{array} \right),  
\end{equation}
with $\Gamma_{3 \times 3}$ given by
\begin{equation}
  \Gamma_{3 \times 3}= \left(
                \begin{array}{ccc}
                  \NC T  &   0  & 0  \\ 
                  0  &  \NC U & 0    \\
                  0  &  0  &  \NC\left(T+U \right) 
                   \end{array} \right),
\end{equation}
and $\Gamma_{5 \times 5}$ given by
\begin{equation}
  \Gamma_{5 \times 5}=\left(\begin{array}{ccccc} \vspace{2mm}
      6T & 0 & -6U & 0 & 0 \\ \vspace{4mm} 
      0  & 3T+\frac{3U}{2} & -\frac{3U}{2} & -3U & 0 \\ \vspace{4mm}
      -\frac{3U}{4} & -\frac{3 U}{2} &3T+\frac{3U}{2} & 0 & -\frac{9U}{4} \\ 
\vspace{4mm}
0 & -\frac{6U}{5} & 0 & 3U & -\frac{9U}{5} \\ \vspace{2mm} 
0 & 0 &-\frac{2U}{3} &-\frac{4U}{3} & -2T+4U 
\end{array} \right).
\end{equation}

Finally, we give the expression of the matrix $M$:
\begin{equation}
M=\left( 
\begin{array}{cccccccc}
-5 & 0  & 0 & 0 & 0 & 0 & 0 & 0   \\
0  & -5 &0 & 0 & 0 &0 & 0 & 0    \\
0  & 0  &-5 & 0 & 0 &0 & 0 & 0     \\
0  & 0  & 0 & 1 & 0 &0 & 0 & 0      \\  
0  & 0  & 0 & 0 & 8 &0 & 0 & 0      \\
0  & 0  & 0 & 0 & 0 &8 &0 &0      \\
0  & 0  & 0 & 0 & 0 &0 &20 &0     \\
0  & 0  & 0 & 0 & 0 &0 &0 &27     
\end{array}
\right) \, .    
\end{equation}
\end{itemize}

\section{Treatment of recoil}
\label{sec:recoil}

A subtlety that we have largely ignored in the main text concerns the
technicalities of the insertion of multiple soft and collinear
momenta. In our discussion in section~\ref{sec:masterderiv} we
referred to the transverse momentum $k_t$ with respect to an original
`parent' dipole, as defined in eqs.~(\ref{eq:SudakovForK}),
(\ref{eq:ktinv}).  Strictly speaking however, the soft and collinear
divergences in QCD amplitudes are of the form
\begin{equation}
  \label{eq:QCDdiv}
  \frac{d{\tilde k}_t^2}{{\tilde k}_t^2} \frac{dz}{z}\,,
\end{equation}
where ${\tilde k}_t$ is measured not with respect to the original Born
momenta (\ie the event without emissions), but with respect to the
actual `final' Born momenta after the inclusion of all recoils,
\begin{equation}
  \label{eq:kttilde}
  {\tilde k}_t^2 = \frac{(2k.{\tilde p}_i)(2k.{\tilde p}_j)}{(2{\tilde
      p}_i.{\tilde p}_j)}\,,
\end{equation}
where we consider the transverse momentum with respect to an arbitrary
Born dipole $ij$, as opposed to just the dipole $12$ used in
section~\ref{sec:single-emission}.

To understand the relationship between $k_t$ and ${\tilde k}_t$ it is
necessary to relate the $p_{i,j}$ to the ${\tilde p}_{i,j}$.
When both $p_i$ and $p_j$ are outgoing momenta, the most general way
of writing $\tilde p_{i,j}$ in terms of $p_{i,j}$ and $k$, such that
energy-momentum is conserved, is
\begin{subequations}
  \label{eq:recoilFormulae}
\begin{align}
  {\tilde p}_i &= Yp_i - f k + (1-X) p_j\,,\\
  {\tilde p}_j &= Xp_j - (1-f) k + (1-Y) p_i\,,
\end{align}
\end{subequations}
where $X$, $Y$ and $f$ are free parameters. Requiring all the momenta
to be massless leads to two further non trivial conditions relating
the $X$, $Y$ and $f$. There is therefore one degree of freedom (let us
choose it to be $f$) left in how one distributes the recoil between
$\tilde p_i$ and $\tilde p_j$. Physically, when $k$ is collinear to
one or other of the legs, then it is natural that this leg should
absorb the dominant (longitudinal and transverse) part of the recoil
--- this is simply because collinear emission occurs on long time
scales relative to the Born interaction and in such a limit the two
legs become independent. This corresponds to choosing $f = 1$ when $k$
is collinear to $p_i$, giving
\begin{equation}
  \label{eq:X}
  X  = 1 - \frac{p_i.k}{p_j.(p_i-k)}\,, \qquad Y=1\,.
\end{equation}
Analogous formulae hold for the case of $k$ collinear to $p_j$ (taking
$f=0$). Eq.~(\ref{eq:X}) is essentially that given in the work of
Catani and Seymour \cite{CataniSeymour} in terms of spectators and
emitters.

Note that regardless of these `naturalness' arguments, 
when considering a single emission, we are free to make any choice for
$f$. This translates into an ambiguity in the relationship between
$k_t$ and $\tilde k_t$. For an emission with Sudakov components $z^{(i)}$
and $z^{(j)}$ with respect to $p_{i,j}$ (as in
eq.~(\ref{eq:SudakovForK})), in the limit where the emission is
collinear to $i$ ($z^{(i)} \gg z^{(j)}$), we have
\begin{equation}
  \label{eq:kttilde-v-kt}
  f=1: \quad {\tilde k}_t = \frac{k_t}{1-z^{(i)}}\,,\qquad\qquad
  f=0: \quad {\tilde k}_t = k_t\,.
\end{equation}
In the soft limit there is therefore no difference between the various
definitions of transverse momentum. For hard collinear emissions there
is a difference, however it is irrelevant from the point of view of
the NLL structure of the matrix element and phase space, $[dk]|M^2(k)|$,
since one always has $d{\tilde k}_t^2/{\tilde k}_t^2 =dk_t^2/k_t^2$.
Sensitivity to the differences in definition arises when considering
the exact integration limits and the scale of the coupling, however
both of these issues are of relevance starting only from NNLL
accuracy.

One should be aware that there is dependence on the recoil
prescription also in the relation between $k_t$ and the value of the
observable, since the observable is defined in terms of $k$ and the
final Born momenta.  Again there should be no sensitivity in the soft
limit, while for hard collinear emissions any sensitivity will amount
to an ambiguity of a factor of order $1$, which corresponds to a NNLL
correction.

An equivalent analysis can be carried out for a dipole consisting of
one incoming ($i$) and one outgoing leg ($j$). Here there is no
ambiguity, because the incoming leg must remain collinear to the beam
direction, giving
\begin{subequations}
  \label{eq:recoilFormulaeIncoming}
\begin{align}
  {\tilde p}_i &= Yp_i\,,\\
  {\tilde p}_j &= p_j -  k + (Y-1) p_i\,,
\end{align}
with 
\begin{equation}
  \label{eq:Y-for-incoming-recoil}
  Y = 1 + \frac{p_j.k}{p_i.(p_j-k)}\,.
\end{equation}
\end{subequations}

So far our discussion has assumed that we were able to uniquely
identify a dipole $ij$ from which the gluon is emitted. However for
processes with more than two legs this identification is not unique
--- the leg to which the emission is collinear is well identified (let
us call it $i$), however the other leg, $j$, can be any of the legs in
the process (one can even have a combination of legs), with the
restriction that if it is incoming, it should not take any transverse
recoil. It is straightforward to show that the freedom in identifying
leg $j$ has no effect on the resummation at NLL accuracy.

While, as we have seen, there are no major subtleties for the recoil
from a single emission, the situation with multiple emissions is more
complex. Let us examine the successive insertion of two emissions,
$k_{t1}$ and $k_{t2}$, into a dipole $ij$. We take the situation where
both Born momenta are outgoing, the emissions are collinear to parton
$i$, and use the $f=1$ recoil prescription.  We use ${\tilde p_{i,j}}$
to denote the Born momenta after the insertion of $k_1$, and ${\tilde
  {\tilde p}}_{i,j}$ after the insertion of $k_{2}$. The notation with
multiples tildes is specific to this section, elsewhere a single tilde
being used to denote the Born momenta after recoil from all emissions.

We write $k_2$ in terms of Sudakov components with respect to ${\tilde
  p_{i,j}}$,
\begin{equation}
  \label{eq:SudakovFork2}
  k_2 = z^{(i)}_2 \tilde p_i + z^{(j)}_2 \tilde p_j + 
  k_{t2}
  \cos \phi_2\, \tilde n_\mathrm{in} + 
  k_{t2} \sin 
  \phi_2\, \tilde n_\mathrm{out}\,.
\end{equation}
Note that the transverse unit vectors $\tilde n_\mathrm{in}$ and
$\tilde n_\mathrm{out}$  differ between insertions $1$ and $2$,
though that difference can in practice be neglected.

In the limit where both insertions are soft, it follows from the
reasoning above, eq.~(\ref{eq:kttilde-v-kt}), that ${\tilde {\tilde
    k}}_{t2}$, the transverse momentum of $k_2$ as measured with
respect to the final ${\tilde {\tilde p}}_{i,j}$ dipole, is equal to
${k}_{t2}$. However ${\tilde {\tilde k}}_{1}$, also defined with
relative to the ${\tilde {\tilde p}}_{i,j}$, is given by
\begin{equation}
  \label{eq:kt1-tilde-tilde}
  {\tilde {\tilde k}}_{t1} \simeq k_{t1} + z^{(i)}_1 
  {k}_{t2}\,.
\end{equation}
Even when both $k_1$ and $k_2$ are soft, the recoil from the
second emission can have the effect of substantially modifying the
transverse momentum $k_{1}$ with respect to the Born dipole,
specifically if $z^{(i)}_1 {k}_{t2} \gg k_{t1}$. One thus loses
the correspondence between the generated transverse momentum,
$k_{t1}$, and the transverse momentum relative to the final Born
momenta, ${\tilde {\tilde k}}_{t1}$, the divergence in the matrix
element being with respect to the latter.

One way of avoiding this problem, \ie of ensuring a correspondence
between the `intended' transverse momentum and the actual transverse
momentum relative to the final Born particles, is by making an
appropriate choice of insertion order. For example, in our example
above, if one first inserts the emission with the larger transverse
momentum, and then that with the smaller transverse momentum, the
transverse momenta with respect to the final Born momenta will, in the
soft limit, be identical to the inserted transverse momenta. The
statement is equally true if one inserts first the emission with
largest angle.

The above analysis can be generalised to any number of emissions.
Specifically, we use the following procedure to ensure the
correspondence between the $k_{t}$ and $\eta$ values that are
`specified', and the actual resulting $k_{t}$ and $\eta$ values with
respect to the final Born momenta. It is to be kept in mind that it is
in no way unique, but rather one of many possible solutions to the
problem.
\begin{itemize}
\item Emissions are ordered such that, first, one inserts those on leg
  $1$, then those on leg $2$, and so forth (we recall our convention
  that incoming legs come first in the numbering sequence). The order
  of emissions on a same incoming leg is irrelevant, while on a same
  outgoing leg, emissions should be ordered in increasing $\eta$ (or
  alternatively decreasing $k_t$).
  
\item Each emission is inserted such that its $k_t$ and $\eta$ are
  correct with respect to Born momenta that include the recoil from
  all previous emissions.
  
\item For an emission on an incoming leg $i$, the other leg that takes
  the recoil is chosen freely among any of the outgoing legs. For an
  emission on an outgoing leg $i$, one either takes the secondary
  recoil from a freely-chosen incoming leg $j$
  (eq.~(\ref{eq:recoilFormulaeIncoming}) with $i$ and $j$ exchanged)
  or from a freely-chosen outgoing leg $j$, taking
  $f=1$.\footnote{Note that there also exist valid insertion
    procedures using $f=0$ for certain legs. One is however then more
    restricted in the choice of recoil legs.}
\end{itemize}
The results for the observable should be independent of the details of
the procedure, for example whether one takes transverse momentum or
angular ordering. The combination of matrix element and phase space,
expressed in terms of the `intended' $k_t$ and $\eta$ values, is also
independent of the details of the procedure.  The only exception is in
the case of collinear emissions, where both the transverse momentum
with respect to the final Born momenta and the value of the observable
may depend on the details of the insertion procedure, any differences
being a factor of order $1$, which translates to a NNLL ambiguity.

\section{Higher-order corrections to independent multiple-emission}
\label{sec:highorder-multemsn}

Here, we illustrate how the corrections to the picture of multiple-independent
emission can be shown, at all orders, to lead to a factorised NNLL
correction for rIRC safe observables, thus generalising the
fixed-order discussion of section~\ref{sec:runn-coupl-effects}, along
the lines of section~\ref{sec:towards-all-orders}. First we shall
discuss the decomposition of the matrix element into clusters of
emissions, and then we shall make use of this decomposition to derive
the correction. We recall that when considering multiple emission, it
is important to be aware of the recoil-related issues discussed
in appendix~\ref{sec:recoil}.

\subsection{Decomposition of $m$-parton matrix element into clusters}
\label{sec:mpartonDecomposition}
Schematically, rather than associating any given event with some
unique classification into clusters, we allow \emph{all} possible
classifications into clusters, and associate each one with a weight
calculated from an appropriate decomposition of the matrix element. In
the case of a two-gluon tree-level event, the two-cluster and
one-cluster weights are defined to be proportional respectively to
$M^2(k_1) M^2(k_2)$ and ${\widetilde M}^2(k_1,k_2)$. So, for example,
cases where the two gluons are at very different angles have weight
close to one for the two-cluster hypothesis and a weight close to zero
for the single-cluster hypothesis, in accord with one's intuition.

To make more general use of such a classification scheme, one needs to
extend the decomposition of the matrix element to $m$-parton
ensembles. We shall proceed here only for the real part of the matrix
element, though it is possible to extend the discussion to include
virtual corrections as well.\footnote{Of course it is only for the
  integration over the real part of the matrix element that our
  treatment extends what exists in the literature, since it is only in
  this part that the generality of the observable is of any relevance
  --- for the virtual corrections in contrast, since they are
  independent of the observable, existing discussions for any specific
  observable contain all the generality that is needed.} %
We introduce the concept of a `cluster'
matrix element for
$m$ partons ${\widetilde M}^2(k_1,\ldots,k_m)$, which is suppressed
unless all partons are close in rapidity. It is defined iteratively,
by rewriting the full matrix element for $m$-particle ensembles,
${M}^2(k_1,\ldots,k_m)$, as a sum of products of cluster
matrix-elements with up to $m-1$ partons (which together should
correctly approximate the full matrix element whenever there are
partons widely separated in rapidity), plus a remainder, ${\widetilde
  M}^2(k_1,\ldots,k_m)$. The first few steps of this iterative
definition give,
\begin{subequations}
\begin{align}
  {\widetilde M}^2(k_1) &\equiv M^2(k_1)\,,\\
  {\widetilde M}^2(k_1,k_2) &= M^2(k_1,k_2) - {\widetilde
    M}^2(k_1) {\widetilde M}^2(k_2)\,,\\
  {\widetilde M}^2(k_1,k_2,k_3) &= M^2(k_1,k_2,k_3) - {\widetilde
    M}^2(k_1) {\widetilde M}^2(k_2) {\widetilde M}^2(k_3) + 
  \nonumber\\
  & \qquad\qquad - ({\widetilde M}^2(k_1,k_2) {\widetilde M}(k_3)
  + 1\leftrightarrow3 + 2\leftrightarrow3)\,,\\
  {\widetilde M}^2(k_1,\ldots,k_m) &= \cdots \nonumber
\end{align}
\end{subequations}
The decomposition of the full $m$-parton matrix element into its
different clusterings is simply obtained by moving ${\widetilde
  M}^2(k_1,\ldots,k_m)$ to the right-hand-side and
${M}^2(k_1,\ldots,k_m)$ to the left-hand-side in the above equations.

So, for instance, if we have an event that consists of two widely
separated clusters, with respectively $m$ and $m'$ very collimated
particles, the matrix element is well approximated by the two-cluster
contribution, ${\widetilde M}^2(k_1,\ldots,k_m) {\widetilde
  M}^2(k_1,\ldots,k_{m'})$ (contributions with more than two clusters
are suppressed because they are unlikely to produce two very
collimated groups of particles). When those two clusters are instead
close in rapidity, this is no longer a good approximation to the full
matrix element; however there is additionally a single-cluster
term ${\widetilde M}^2(k_1,\ldots,k_{m+m'})$ and together with the
two-cluster term this reproduces the full matrix element.

This manner of viewing the clustering
  may perhaps be made clearer by the following explicit example, which
  refers to fig.~\ref{fig:lozengecluster},
  section~\ref{sec:towards-all-orders}. In
  the left-hand diagram of that figure, we can number
  the black emissions as 1 to 10, from left to right. The `natural'
  classification of the event as having 3 clusters corresponds to a
  contribution with matrix element ${\widetilde M}^2(k_1,k_2,k_3)
  {\widetilde M}^2(k_4,k_5,k_6) {\widetilde M}^2(k_7,\ldots,k_{10})$.
  However, part of the matrix element for the event comes from
  4-cluster contributions, such as ${\widetilde M}^2(k_1) {\widetilde
    M}^2(k_2,k_3) {\widetilde M}^2(k_4,k_5,k_6) {\widetilde
  M}^2(k_7,\ldots,k_{10})$, in which two clusters ($1$ and $2,3$)
  happen to be close in rapidity; similarly there are contributions to
  the matrix element that involve still larger numbers of clusters. 
  (There are no
  significant contributions with fewer than three clusters, such as
  ${\widetilde M}^2(k_1,\ldots,k_6) {\widetilde
    M}^2(k_7,\ldots,k_{10})$, because the cluster matrix elements are
  strongly suppressed when different emissions of a same cluster are
  widely separated in rapidity).  So there is no direct correspondence
  between the clusters being discussed here and those that would be
  defined, say, in a jet-algorithm --- instead any given event is
  amenable to a range of decompositions into clusters, and it is the
  sum of the partial matrix elements for all possible decompositions
  that yields the full matrix element for the event. Of course, an
  approximate correspondence is kept with the clusters of a jet
  algorithm, insofar as the largest contribution to the matrix element
  is likely to come from a cluster decomposition that resembles the
  jet clustering.

\subsection{Nature of higher order corrections from correlated emissions}
\label{sec:CorrelEmsn}

To help deal with the complexity of an all-order treatment including
multiple-emission correlations, it is useful to introduce a generating
functional $Z$ such that a state consisting of emissions
$k_1,\ldots,k_n$ is represented by $u(k_1)\ldots u(k_n)$. One then writes
the integrated distribution $\vProb(v)$ as
\begin{equation}
  \label{eq:vProb_from_Z}
  \vProb(v) = \sum_{n=0}^\infty \left.\prod_{i=1}^n\left(\int [dk_i]
      \frac{\delta}{\delta u(k_i)}\right) Z\right|_{u = 0}
  \Theta(v - V(\{\tilde p\},k_1,\ldots,k_n))\,.
\end{equation}
The all-order independent-emission contribution to the generating
functional is obtained by extracting the $\prod_{i=1}^n |M^2(k_i)|$
component of the $n$-gluon matrix element, $|M^2(k_1,\ldots,k_n)|$, and
is given by
\begin{equation}
  \label{eq:Zindep}
  Z^\mathrm{indep} = \sum_{n=0} \frac{1}{n!} \prod_{i=1}^n \int [dk_i]
  (u(k_i)-1) |M^2_{rc}(k_i)|\,.
\end{equation}
Next, for each $n$, one isolates all components of $|M^2(k_1,\ldots,k_n)|$
involving a single correlated gluon pair, which we label
$a,b$. Including also the component of the virtual corrections
corresponding to $|{\widetilde M}_{rc}^2(k_a,k_b)|$, and summing over
$n$, one obtains
\begin{equation}
  \label{eq:Z1correl}
  Z^\mathrm{1-correl} = \frac{Z^\mathrm{indep}}{2!} \int [dk_a][dk_b] 
  (u(k_a)u(k_b) - u(k_a+k_b)) |{\widetilde M}_{rc}^2(k_a,k_b)|\,,
\end{equation}
where $k_a+k_b$ is to be interpreted as a massless momentum with the
same transverse momentum and rapidity as the sum of $k_a$ and $k_b$
(\cnf section~\ref{sec:runn-coupl-effects}); the fact that
$M^2_{rc}(k_i)$ in eq.~(\ref{eq:Zindep}) includes running-coupling
effects, in the CMW scheme, ensures that virtual corrections to the
correlated emission can be accounted for simply by the $u(k_a+k_b)$
subtraction in eq.~(\ref{eq:Z1correl}). Note that we have introduced
running-coupling corrections also into the two-correlated gluon matrix
element, $|{\widetilde M}_{rc}^2(k_a,k_b)|$, though this is not
strictly necessary for the arguments that follow. Beyond
eqs.~(\ref{eq:Zindep}) and (\ref{eq:Z1correl}), there are additionally
all-order contributions with a single correlated triplet, two
correlated pairs, etc.

Reshuffling the virtual corrections in $Z^\mathrm{indep}$,
\begin{equation}
  \label{eq:Zindep-reshuffled}
  Z^\mathrm{indep} = e^{-R(v)} \sum_{n=0} \frac{1}{n!} \prod_{i=1}^n
  \int [dk_i] (u(k_i)- \Theta(v-V(\{\tilde p\},k_i))) |M^2_{rc}(k_i)|\,,
\end{equation}
one sees that the sum of the various contributions to $Z$ leads to
\begin{equation}
  \label{eq:vProbsumZ}
  \vProb(v)= e^{-R(v)} \left(\cF^\mathrm{indep} +
    \cF^\mathrm{1-correl} + \cdots\right)\,,
\end{equation}
where the independent-emission contribution to $\cF$ is (in a slightly
rewritten form compared to that given elsewhere in the article),
\begin{multline}
  \label{eq:cFindep}
  \cF^\mathrm{indep} = \exp\left(-\int_{\epsilon v}^v [dk]
    |M^2_{rc}(k)|\right) \sum_{n=0} \frac{1}{n!} \left( \prod_{i=1}^n
    \int_{\epsilon v} [dk_i]  |M^2_{rc}(k_i)|\right)\times\\
  \times \Theta(v - V(\{\tilde p\},k_1,\ldots,k_n)) \,.
\end{multline}
Here the limits on an integral over $[dk]$ are to be understood as
limits on $V(\{\tilde p\},k)$ and we have introduced a lower cutoff
$\epsilon v$ on (real and virtual) emissions, as is legitimate for
rIRC safe observables.  As discussed in the main text, the
single-logarithmic nature of $\cF$ follows from rIRC safety, since it
is the integral over a finite number of momenta near the
(single-logarithmic) boundary $V(\{\tilde p\},k)=v$ that contributes
to break the exact compensation between real and virtual corrections
in eq.~(\ref{eq:cFindep}) leading to $\cF \neq 1$.

The contribution to $\cF$ from the configurations with one correlated
pair can be written as 
\begin{multline}
  \label{eq:cF1correl}
  \cF^\mathrm{1-correl} = \exp\left(-\int_{\epsilon v}^v [dk]
    |M^2_{rc}(k)|\right) \times\\\times \sum_{n=0} \frac{1}{n!} \left(
    \prod_{i=1}^n
    \int_{\epsilon v} [dk_i]  |M^2_{rc}(k_i)|\right)
   \frac1{2!} \int [dk_a][dk_b] 
   |{\widetilde M}_{rc}^2(k_a,k_b)| \times\\
  \times \left[\Theta(v - V(\{\tilde p\},k_1,\ldots,k_n,k_a,k_b)) -
    \Theta(v - V(\{\tilde p\},k_1,\ldots,k_n,k_a+k_b)) \right] \,.
\end{multline}
It should be straightforward to see that for a continuously global
rIRC safe observable there can only be a contribution to
$\cF^\mathrm{1-correl}$ when $k_a$ and $k_b$ have similar rapidities,
 satisfy $V(\{\tilde p\},k_a)\sim V(\{\tilde p\},k_b) \sim v$,
and are not collinear to each other. As a result, the integral of the
correlated matrix element over the relevant phase space can at most
give a factor proportional to $\as^2 L$.  This will multiply a single
logarithmic function associated with the integral over the remaining
independent emissions.  Therefore the lowest-order correction to
$f(v)$ associated with non-independent emission is a factorised NNLL
contribution, as anticipated in section~\ref{sec:masterderiv}. The
corrections involving more than one correlated pair, or a correlated
triplet, etc., will also be factorised, and of still higher order.

We note that an equation of the form (\ref{eq:cF1correl}) would
serve as one of the building blocks were one to envisage an automated
NNLL resummation.

Finally we point out that for all the other classes of NNLL
corrections discussed only at fixed-order in the main text, it is
similarly possible to show that they too factorise at all orders.

\section{Incoming hard legs}
\label{sec:hardcollinear}

In this section we sketch a derivation of eq.~(\ref{eq:pdf-correct}),
which states that for processes with incoming legs, the resummation
entails a modification of factorisation scale of the parton density,
from $\mu_F$ to $\mu_F\,v^{1/(a+b_\ell)}$, because the limit on the
observable translates to a veto on emissions with $k_t \gtrsim
Q\,v^{1/(a+b_\ell)}$.

\subsection{Flavour non-singlet case}
For simplicity we consider a process with only one incoming hard leg,
say $\ell\!=\!1$ and initially we examine the flavour non-singlet
component\footnote{For an introduction to singlet and
non-singlet distributions, see chapter~4, section~3 of
\cite{ESWbook}.} of the cross section so as to have
only a single flavour channel. We have an incoming quark with momentum
${\tilde p}_1\!=\!X P$, whose parton density function (PDF) inside the
incoming hadron (of momentum $P$) is $q(X,\mu_0^2)$. We have
introduced a factorisation scale $\mu_0$, to be chosen smaller than any
transverse-momentum scale in the problem and will consider only
emissions (and virtual corrections) above that scale.

The integrated cross section for the observable to be smaller than $v$
is given by eq.~(\ref{eq:Sigmacut_resummed}). We write the Born
differential cross section directly in terms of the product of
the elementary hard cross section and the parton density,
\begin{equation}
  \label{eq:BornProduct}
  \frac{d\sigma_\subProc}{d\momConf} =
  \frac{d{\hat\sigma}_\subProc}{d\momConf} q_\subProc(x,Q^2)\,,
\end{equation}
where $x$ is determined by the Born configuration. We then rewrite
eq.~(\ref{eq:Sigmacut_resummed}) (for $n_i=1$) as
\begin{equation}
  \label{eq:Sigmacut_resummed_hat}
  \Sigma_\cH(v) = \sum_{\subProc} \int d\momConf\,
  \frac{d{\hat \sigma}_\subProc}{d\momConf} \,\vhProb_{\momConf,\subProc}(v)\,
  \cH(p_{2},\ldots,p_{n})\,,
\end{equation}
where, as compared to $\vProb_{\momConf,\subProc}(v)$,
$\vhProb_{\momConf,\subProc}(v)$ now includes the parton density as
follows,
\begin{multline}
  \label{eq:vhProb}
  \vhProb_{\momConf,\subProc}(v) = \int dX \> q_{\delta}(X,\mu_0^2) \>
  \sum_{n=0}^{\infty} \left( \prod_{i=1}^n
    \int [dk_i] \, |M^2_{rc}(k_i)| \Theta(k_{ti} - k_{t,i+1})
  \right)
  \times\\ \times
  \Delta(Q,k_1,\ldots,k_n,\mu_0)
  \Theta(v - V(\{\tilde p\},k_1, \ldots, k_n))
  \> \delta\Big(x - X \prod_i y_i \Big).
\end{multline}
This equation differs in a number of respects from that in the
outgoing case, eq.~(\ref{eq:MultipleIndepEmsn}). Firstly, the
combination of matrix
element and emission phase-space for an emission collinear to leg $1$
differs between the outgoing and incoming cases, and in the latter
it can be written
\begin{equation}
  \label{eq:M2incoming}
  [dk_i] M^2_{I,rc}(k_i) = \frac{\as \CF}{2\pi}\frac{d\phi_i}{2\pi}
  \frac{dk_{ti}^2}{k_{ti}^2} dy_i p_{qq}(y_i)\,,
\end{equation}
where $1-y_i$ is the fraction of the incoming momentum that is emitted
and $p_{qq}(y) = p_{gq}(1-y)$ and the suffix $I$ has been added just to
emphasise that the matrix element is for emission off an incoming
parton. To be able to relate the $y_i$ to the momentum fraction
$z^{(1)}_i$ defined in section~\ref{sec:single-emission}, it is
necessary to consider the emissions as radiated in some sequence. In
the flavour non-singlet case it is convenient to take the sequence as
being that of transverse momentum ordering, since this is standard when
discussing DGLAP evolution \cite{DGLAP} (we will, however, return to
the question
below).  This is the origin of the factors $\Theta(k_{ti} -
k_{t,i+1})$ and the disappearance of the $1/n!$ in
eq.~(\ref{eq:M2incoming}). Given the ordering, one can then write
\begin{equation}
  \label{eq:yi}
  z_i^{(1)} = \frac{1-y_i}{y_1 \dots y_i}\>.
\end{equation}
Additionally we define $k_{t,n+1}\equiv \mu_0$, so as to place an
explicit infrared cutoff on the transverse momentum of all emissions,
which serves also as the factorisation scale for the incoming parton
density $q_{\delta}(X,\mu_0^2)$. Finally, the virtual corrections
acquire a more complicated form than in
eq.~(\ref{eq:MultipleIndepEmsn}), because the available phase-space
changes after each emission. Writing just the part of the virtual
corrections associated with the incoming hemisphere, $2z^{(1)}_i x E_P >
k_{ti}$ (temporarily supposing, for example, that we are in the Breit
frame of DIS), we have
\begin{multline}
  \label{eq:DeltaIncoming}
  \ln \Delta(Q,k_1,\ldots,k_n,\mu_0) = \\
   = -\int [dk] |M^2_{I,rc}(k)|
   \Theta\Big((1-y) - \frac{k_{t}}{2xE_P}\prod_{i=1}^{m(k_t)}
   y_i\Big) \Theta(Q-k_t)\Theta(k_t-\mu_0)\,,\qquad\\
   m(k_t) \equiv \mbox{largest $m$ such that $k_{tm} > k_t$,}
\end{multline}
where $E_P$ is the energy of the incoming hadron.

To understand the operations that we now perform on
eq.~(\ref{eq:vhProb}), it is helpful to examine
fig.~\ref{fig:incomingLozenge}, an extension of
fig.~\ref{fig:lozengecluster} to the case with an incoming hadron.
Reading the figure from low to high transverse momenta (emissions
$n\ldots 1$) one sees that the maximal rapidity along the incoming leg
(thick yellow line on right-hand side) decreases at each hard
branching of the incoming parton, as the incoming parton's
longitudinal momentum is reduced from $XP$ at the $\mu_0$
factorisation scale, to $xP$ at the hard scale $Q$. This variation
with $k_t$ of the available rapidity is the origin of the variable
limit on $1-y$ in eq.~(\ref{eq:DeltaIncoming}).

\FIGURE{
  \includegraphics[width=0.75\textwidth]{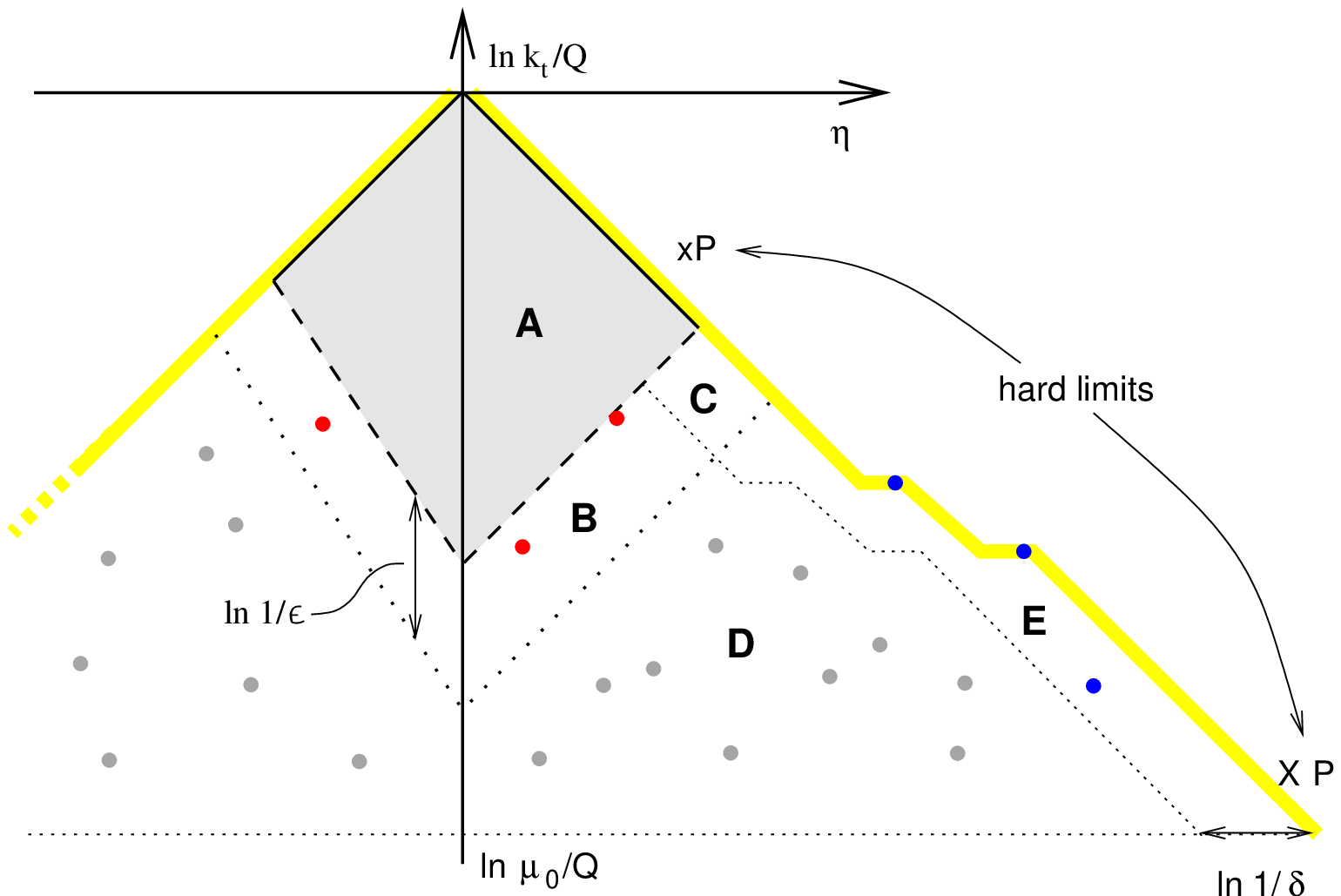}
  \caption{Analogue of fig.~\ref{fig:lozengecluster} for the case
    of an incoming leg (corresponding to positive rapidities). All
    emissions have already been clustered (as described in
    section~\ref{sec:towards-all-orders}). Emissions in region D
    contribute neither to the observable nor to the evolution of the
    PDF and can be neglected; emissions in E contribute to the PDF
    evolution but not to the observable; emissions in C contribute to
    both, but only at the NNLL level $\as^n L^{n-1} \ln \epsilon \ln
    \delta$; region B leads to the $\cF(R')$ function as in the case
    with only outgoing legs; and emissions are essentially forbidden
    in region A (except, potentially, close to the boundary with B),
    giving the usual double logarithmic $e^{-R(v)}$
    suppression.}
  \label{fig:incomingLozenge}
}

We subdivide the figure into a number of regions. As in
fig.~\ref{fig:lozengecluster}, we identify the region $V(k,\{\tilde
p\}) > v$, vetoed in the case of a single emission, now calling it
region A. 
We also define a second boundary $V(k,\{\tilde p\})=\epsilon v$, again
choosing $\epsilon \ll 1$, $\as \ln 1/\epsilon \ll 1$, below which it
is safe to neglect emissions when evaluating the $\Theta$-function for
the observable in eq.~(\ref{eq:vhProb}) (due to rIRC safety).

Additionally, we introduce a boundary parallel to the collinear limit,
$1-y = \delta$, and choose $\delta\ll1$, $\as \ln^2 1/\delta\ll 1$ and
also $\as \ln \epsilon \ln\delta \ll 1$. Emissions to the left of this
boundary will not significantly modify the argument of the
$\delta$-function on longitudinal momentum in eq.~(\ref{eq:vhProb}),
because they have $y_i$ sufficiently close to $1$ that it can be
ignored in the product of the $y_i$.

These two boundaries delineate a region D in
fig.~\ref{fig:incomingLozenge} in which emissions neither contribute
to the observable, nor  change the incoming parton momentum
fraction. Therefore one can sum over them to directly cancel the
corresponding part of the virtual corrections.

In the region $E$, the emissions do not contribute to the observable,
however they do contribute to the argument of the $\delta$-function in
eq.~(\ref{eq:vhProb}). So we make use of the fact that
\begin{multline}
  \label{eq:PDFevolve}
  q_\delta(x,\mu^2) = \int dX q_\delta(X,\mu_0^2) 
  \sum_{n=0}^{\infty} \left( \prod_{i=1}^n
    \int [dk_i] \, |M^2_{I,rc}(k_i)| \Theta(k_{ti} - k_{t,i+1})
    \Theta(1-y_i - \delta)
  \right) \times \\ \times\Theta(\mu - k_{t1})
  e^{-\int [dk] |M^2_{I,rc}(k)| \Theta(1-y -
    \delta) \Theta(\mu-k_t) \Theta(k_t - \mu_0)}
  \delta\Big(x - X \prod_i y_i \Big)\,,
\end{multline}
(as can be verified by noting that for $\mu=\mu_0$ one obtains
$q_\delta(x,\mu_0^2)$ and that taking a derivative with respect to
$\mu^2$, eq.~\eqref{eq:PDFevolve} yields the single-logarithmic
non-singlet DGLAP equation) and sum over all emissions in region E so
as to rewrite eq.~(\ref{eq:vhProb}) as
\begin{multline}
  \label{eq:vhProb_new_scale}
  \vhProb_{\momConf,\subProc}(v) = \int dX \> q_{\delta}(X,(\epsilon
  v)^{2/(a+b_1)}Q^2) \>
  \sum_{n=0}^{\infty} \left( \prod_{i=1}^n
    \int_{\epsilon v} [dk_i] \, |M^2_{I,rc}(k_i)| \Theta(k_{ti} - k_{t,i+1})
  \right)
  \times\\ \times
  \Delta_{\epsilon v}(Q,k_1,\ldots,k_n,\mu_0)
  \Theta(v - V(\{\tilde p\},k_1, \ldots, k_n))
  \> \delta\Big(x - X \prod_i y_i \Big),
\end{multline}
where the suffix $\epsilon v$, on both real and virtual corrections,
indicates, as elsewhere that we consider only contributions satisfying
$V(k,\{\tilde p\}) > \epsilon v$. Note that in arriving at
eq.~(\ref{eq:vhProb_new_scale}) we have additionally made use of the
fact that to within accuracy $\as^n L^{n-1}$ the slight mismatch
between region E and the region defined by $y > 1-\delta$, $(\epsilon
v)^{1/(a+b_1)}Q > k_t > \mu_0$ does not affect the single-logarithmic
reconstruction of
$q_{\delta}(X,(\epsilon v)^{2/(a+b_1)}Q^2)$ (it being a region of
phase space of order $\as \ln^2 \delta$). Note also that here our
integration variable $X$ is no longer the momentum fraction of the
parton extracted directly from the proton at scale $\mu_0^2$, but
rather the momentum fraction after it has radiated all the emissions
in E.

Next, we consider region C, where emissions can both modify the
incoming momentum fraction and modify the value of the observable, the
latter by an absolute amount of order $v$, or equivalently by a
relative factor of order $1$ (but no more, since the observable is
continuously global and rIRC safe). The effect of modifying the
momentum fraction essentially amounts to further DGLAP evolution up to
a scale of order $v^{1/(a+b_1)}Q$ (equivalent to a NNLL contribution
of order $\as \ln \epsilon (\as \ln v)^n$, which we include, and which
through eq.~(\ref{eq:PDFevolve}) also causes the longitudinal momentum
fraction for the PDF to become $x$), together
potentially with an effective $\order{\as^n \ln ^{n-1} v}$ coefficient
function, again NNLL (which we shall neglect).  Because of coherence,
the dynamics of multiple soft-collinear emission is independent of
both of these effects and so they will factorise from our result,
which to NLL accuracy therefore becomes,
\begin{multline}
  \label{eq:vhProb_nearly_there}
  \vhProb_{\momConf,\subProc}(v) = q_{\delta}(x,v^{2/(a+b_1)}Q^2) \>
  \sum_{n=0}^{\infty} \left( \prod_{i=1}^n
    \int_{\epsilon v} [dk_i] \, |M^2_{I,rc}(k_i)| 
  \right)\\
  \Delta_{\epsilon v}(Q,k_1,\ldots,k_n,\mu_0)
  \Theta(v - V(\{\tilde p\},k_1, \ldots, k_n))\,.
\end{multline}
In this formula, for the specific observable depicted in
fig.~\ref{fig:incomingLozenge}, one might worry that there are
emissions in region B, with transverse momenta of the same order of
magnitude as emissions in region E that have been removed. This could
conceivably cause problems because of the identification of
$z^{(1)}_i$ which depends on all emissions at higher scales. However,
the relevant part of B corresponds to soft emissions, and our
integration limits are directly on $z^{(1)}_i$, not on $y_i$ (since
they come from the boundary of the incoming hemisphere and from the
condition $V(k\{\tilde p\})=\epsilon v$). Using eq.~(\ref{eq:yi}), and
the fact that $1-y_i\ll 1$, we have $dy_i/(1-y_i) =
dz^{(1)}_i/z^{(1)}_i$ and so $[dk_i] \, |M^2_{I,rc}(k_i)| = [dk_i] \,
|M^2_{rc}(k_i)|$ (with $|M^2_{rc}(k_i)|$ as defined in
eq.~(\ref{eq:matrixelement})), we can integrate over these emissions
independently of the pattern of emissions in region E.

A similar argument can be applied to the part of region A that is
below transverse momentum scale $Q v^{\frac{1}{a+b_1}}$. However,
above that scale the virtual corrections need to be integrated up to
the hard collinear limit and some care is needed in evaluating
$\Delta$, eq.~(\ref{eq:DeltaIncoming}). Since we discuss transverse
momentum scales above $Q v^{\frac{1}{a+b_1}}$, we have that
$m(k_t)=0$.  Observing that
\begin{equation}
  \label{eq:y_zi_coincidence}
  \int_0^1 dy \, p_{qq}(y)\, \Theta\Big(1-y - \frac{k_t}{2xE_P}\Big)
  = \int^1_0 dz^{(1)}\, p_{gq}(z^{(1)}) \,\Theta\Big(z^{(1)} -
  \frac{k_t}{2xE_P}\Big)\,,
\end{equation}
one finds that the replacement $[dk_i] \, |M^2_{I,rc}(k_i)| \to [dk_i] \,
|M^2_{rc}(k_i)|$ can be used right up to the hard collinear limit.
Thus we may systematically rewrite eq.~(\ref{eq:vhProb_nearly_there})
in terms of $|M^2_{rc}(k)|$, removing also the ordering in $k_t$, to
obtain
\begin{multline}
  \label{eq:vhProb_there_nonsinglet}
  \vhProb_{\momConf,\subProc}(v) = q_{\delta}(x,v^{2/(a+b_1)}Q^2) \>
  \exp\left ( - \int_{\epsilon v} [dk] \,|M^2_{rc}(k)|\, \right)
  \times\\ \times
  \sum_{n=0}^{\infty} \frac{1}{n!} 
  \left( \prod_{i=1}^n
    \int_{\epsilon v} [dk_i] \, |M^2_{rc}(k_i)| 
  \right)
  \Theta(v - V(\{\tilde p\},k_1, \ldots, k_n))\,.
\end{multline}
While the precise sequence of our arguments relied on the fact that
$b_1\ge0$ (specifically, this led to region B and part of A being at
similar transverse momenta to region E),
eq.~(\ref{eq:vhProb_there_nonsinglet}) applies independently of the
value of $b_1$, as can be verified by repeating the analysis in the
case of $b_1 < 0$.

Apart from the overall PDF factor (and the removal of emissions below
$\epsilon v$), eq.~(\ref{eq:vhProb_there_nonsinglet}) is identical to
eq.~(\ref{eq:MultipleIndepEmsn}), and so all the manipulations of
section \ref{sec:mult-indep-emsn} may be repeated, to give
\begin{equation}
  \label{eq:vhProb_closed_nonsinglet}
  \vhProb_{\momConf,\subProc}(v) = q_{\delta}(x,v^{2/(a+b_1)}Q^2)\,
  e^{-R(v)} \, \cF\,,
\end{equation}
with $R(v)$ and $\cF$ as evaluated in that section. Therefore the
probability $\vProb(v)$ includes the correction factor quoted in
eq.~\eqref{eq:pdf-correct}.

\subsection{Flavour singlet case}
We finally discuss the extension of the above result to the flavour
singlet-case. We shall use the index $\delta$ to denote the incoming
parton flavour (as opposed to the hard process as a whole, as used
elsewhere in the paper). 

It would be tempting to simply extend eq.~(\ref{eq:vhProb}) so as to
have the appropriate flavour matrix structure.  However eq.~(\ref{eq:vhProb})
assumes an ordering in transverse momenta, while, as discussed in
section~\ref{sec:allorders} (see also chapter 5 of \cite{ESWbook}),
coherence actually implies that the colour factor for soft radiation
is determined by the combination of emissions at smaller
\emph{angles}, not smaller transverse momenta. 
It is therefore useful to rewrite eq.~\eqref{eq:vhProb} in terms of
angular ordered emissions as follows (restricting our attention just
to the incoming hemisphere),
\begin{multline}
  \label{eq:vhProb-NS-AO}
  \vhProb_{\momConf,\subProc}(v) = \int dX \> q_{\delta}(X,\mu_0^2) \>
  \sum_{n=0}^{\infty} \left( \prod_{i=1}^n
    \int [dk_i] \, |M^2_{rc}(k_i)| \Theta(\eta_{i+1} - \eta_{i})
   \overline \Delta(\eta_{i}, \eta_{i+1})
  \right)
  \times\\ \times
  \overline \Delta(0, \eta_1)
  \Theta(v - V(\{\tilde p\},k_1, \ldots, k_n))
  \> \delta\Big(x - X \prod_i y_i \Big),
\end{multline}
where in terms of angles virtual corrections become
\begin{equation}
  \label{eq:virtual-NS-AO}
  \ln {\overline \Delta}(\eta_i, \eta_{i+1}) = -\int
  [dk]
  M^2_{I, rc}(k)\Theta(\eta - \eta_i)\,\Theta(\eta_{i+1} - \eta)\,,
\end{equation}
where in the evaluation of ${\overline \Delta}(\eta_i, \eta_{i+1})$,
the rapidity of emission $k$ can be expressed in terms of $y$ as $\eta =
-\ln \frac{k_t}{2(1-y)xE_P} - \sum_{j=1}^i \ln y_j$.
We have now a relation between $\mu_0^2$ and the smallest angular
scale, $\eta_{n+1} = \ln 2xE_P/\mu_0 - \sum_{j=1}^n \ln y_j$ and we
continue to use eq.~(\ref{eq:yi}) to reconstruct the $z^{(1)}_i$ (and
so the $\eta_i$) from the $y_i$.

The extension of eq.~\eqref{eq:vhProb-NS-AO} to the flavour singlet
case requires that one replace the unit flavour matrix element in
eq.~\eqref{eq:vhProb-NS-AO} with the appropriate,
full flavour-singlet matrix structure
\begin{multline}
  \label{eq:vhProb_singlet_RIGHT}
  \vhProb_{\momConf,\subProc}(v) = \int dX \> 
  \sum_{n=0}^{\infty} \left( \prod_{i=1}^n
    \int [dk_i] \, |M^2_{rc,\delta_i\delta_{i+1}}(k_i)| \Theta(\eta_{i+1}
    - \eta_{i})  {\overline \Delta}(\eta_i,\eta_{i+1})
  \right)
  \times\\ \times
  \Theta(v - V(\{\tilde p\},k_1, \ldots, k_n))\, 
  {\overline \Delta}(0,\eta_1)\,
  \delta_{\delta\delta_1}
  \> q_{\delta_{n+1}}(X,\mu_0^2) \> \delta\Big(x - X \prod_i y_i \Big),
\end{multline}
where we now have an index $\delta_i$ to denote the flavour of the
exchanged parton at angular scales above that of emission $i$, and
$\delta_{n+1}$ is the flavour of the incoming parton, and we sum over
all flavours. 
The matrix element now has the structure
\begin{equation}
  \label{eq:matrix_element_singlet}
  [dk_i] \, |M^2_{rc}(k_i)|
   = \frac{\as(k_t^2)}{2\pi} \frac{d\phi_i}{2\pi}
   \frac{dk_{ti}^2}{k_{ti}^2}
   dy_i \left(
     \begin{array}{cc}
       \CF\, p_{qq}(y_i) & 2\nf T_R\, p_{qg}(y_i)\\
       \CF\, p_{gq}(y_i) & \CA [p_{gg}(y_i) + p_{gg}(1-y_i)]
     \end{array}
     \right),
\end{equation}
where the elementary splitting functions, $p_{\delta \delta'}$, are as
defined elsewhere (after eqs.~(\ref{eq:matrixelement_ell}) and
(\ref{eq:M2incoming}), and eqs.~(\ref{eq:gg_qg_mat_elements})) and the
left and right-hand columns act respectively on the quark singlet ($d
+ \bar d + u + \bar u + \ldots$) and gluon distributions. 

In term of angles, virtual corrections, which are diagonal in flavour
space, give
\begin{multline}
  \label{eq:virtual_array_eta}
  \ln {\overline \Delta}(\eta_i, \eta_{i+1}) = -\int
  \frac{dk_{t}^2}{k_{t}^2} \frac{\as(k_t^2)}{2\pi}  \,dy\,
  \Theta(\eta - \eta_i)\,\Theta(\eta_{i+1} - \eta)
   \times\\\times
  \left(
    \begin{array}{cc}
      \CF \,p_{qq}(y) & 0\\
      0  &  \CA\,p_{gg}(1-y) + \nf\TR\, p_{qg}(y)
    \end{array}\right),
\end{multline}
where, as above, $\eta = -\ln \frac{k_t}{2(1-y)xE_P} - \sum_{j=1}^i \ln
y_j$.

There are some subtleties to be aware of in the relation between
eq.~(\ref{eq:matrix_element_singlet}) and for example
(\ref{eq:gg_qg_mat_elements}) --- in the outgoing case of
eq.~(\ref{eq:gg_qg_mat_elements}) there is a symmetry between the two
partons coming out of a $g \to gg$ or $g \to q \bar q$ splitting (if
one neglects the difference between quark and antiquark, as is the
case for an event shape). In the incoming case, the symmetry is broken
by the fact that one of the descendants from the splitting will enter
the hard process while the other will go into the final state, and we
need to include separate contributions for the cases where it is the
first or the second of the splitting products that continues into the
hard process. It is for this reason that in the real matrix element
eq.~(\ref{eq:matrix_element_singlet}) (but not the virtual part) we
need to write $p_{gg}(y_i) + p_{gg}(1-y_i)$, and include a factor of
$2\nf$ in front of $p_{qg}$, rather than just $\nf$ (making use of the
explicit $y \leftrightarrow 1-y$ symmetry of $p_{qg}$), and similarly
that we need to include both the $p_{qq}$ \emph{and} the $p_{gq}$
splittings (whereas previously we had just one of them).

One can now repeat the flavour non-singlet analysis given above.
There are certain other small differences, for example in region E the
final DGLAP evolution that we obtain corresponds to angular-ordered
DGLAP evolution rather than transverse-momentum ordered (this is known
to give only subleading, $\as^n L^{n-1}$ differences, potentially
enhanced by $\ln^2 X/x$, such terms being in any case neglected in
eq.~(\ref{eq:matrix_element_singlet})). However the main thrust of the
analysis persists, one can still separate the different regions, and
one obtains a factorised $q_\delta(x,v^{2/(a+b_1)}Q^2)$ contribution
from E, $\cF$ from real (virtual) emissions in A and B (just B), $e^{-R(v)}$
from the integral over virtual terms in A, NNLL contributions from C
and terms suppressed by powers of $\epsilon$ or $\delta$ from D.
Furthermore the colour factor associated with $e^{-R(v)}$ and $\cF$ is
that of the parton flavour $\delta$ entering the hard process, since
all logarithmically enhanced flavour-changing branchings are in E
which is at smaller angles than regions A and B.

Our result for $\vhProb_{\momConf,\subProc}(v)$ is therefore given by
eq.~(\ref{eq:vhProb_closed_nonsinglet}) even in the singlet case. The
final step is to relate this to $\vProb_{\momConf,\subProc}(v)$
discussed elsewhere in the text. Comparing
eqs.~(\ref{eq:Sigmacut_resummed}), (\ref{eq:BornProduct}) and 
(\ref{eq:Sigmacut_resummed_hat}) one sees that
\begin{equation}
  \label{eq:vProf_with_PDFs}
  \vProb_{\momConf,\subProc}(v) = 
  \frac{\vhProb_{\momConf,\subProc}(v) }{q_\delta(x,Q^2)} 
  = \frac{q_\delta(x,v^{2/(a+b_1)}Q^2)}{q_\delta(x,Q^2)} e^{-\cR(v)}
  \cF\,,
\end{equation}
which is the result quoted in eq.~(\ref{eq:pdf-correct}).

\section{Further examples of rIRC unsafety}
\label{sec:further-rIRC-unsafe}

Here, with the help of some resolution thresholds in jet-clustering
algorithms, we illustrate three cases of IRC safe observables that are
rIRC unsafe. One example is devoted to each of the rIRC conditions of
section~\ref{sec:summary-master}.

\subsection[Jade jet algorithm: $E$-scheme]{Jade jet algorithm: $\boldsymbol E$-scheme}
\label{sec:JadeE}

Jet clustering algorithms are widely studied observables. They
typically involve a distance measure $y_{ij}$ between two
(pseudo)particles and a clustering sequence in which one searches for
the particle pair with the smallest $y_{ij}$, clusters it into a
single pseudoparticle and then repeats the clustering procedure until
all remaining pairs have $y_{ij} > y_{cut}$, where $y_{cut}$ is the
jet resolution parameter. From the point of view of this article, the
observable that is typically of interest is the distribution of the
value of $y_{cut}$ that demarcates the threshold between an $n$ and
an $n+1$-jet event.

Of particular interest is the family of JADE algorithms \cite{JADE}
because it represents the only example of an observable for which the
double logarithms have been found, analytically, not to exponentiate
\cite{JadeDL,CataniEtAlJets}. Taking the definition used in
\cite{JadeDL}, the distance measure is
\begin{equation}
  \label{eq:yijJADE}
  y_{ij} = \frac{(q_i + q_j)^2}{Q^2}\,,
\end{equation}
and the recombination scheme is the
$E$-scheme, $q_{ij} = q_{i} + q_{j}$. It is straightforward to show
that, in the two-jet limit, at the level of a single soft and
collinear emission, the $2$-to-$3$ jet threshold resolution, $y_{3}$
is identical to $\tau=1-T$,
\begin{equation}
  \label{eq:JADEy3coeffs}
  a = 1 =  b_\ell = d_\ell = g_\ell(\phi) = 1\,,\qquad\quad
  \ell = 1, 2\,.
\end{equation}
As above, we use ${\bar v}$ and $\xi_i$ to parametrise an emission
$\kappa_i({\bar v})$, giving
\begin{equation}
  \label{eq:ketadepy3Jade}
  \ln \frac{\kappa_{ti}({\bar v})}{Q} 
  = \left(1 - \frac{\xi_i}2\right) \ln {\bar v}\,, \qquad
  \eta_i({\bar v})  =  -\frac{\xi_i}2\ln {\bar v}\,.
\end{equation}
Let us now consider two emissions $\kappa_1({\bar v})$ and $\kappa_2({\bar v})$
collinear to the two different legs ($1$ and $2$ respectively). One
has (ignoring $y_{p_1p_2} \simeq 1$)
\begin{equation}
  \label{eq:y43JadeAllOpts}
  y_{\kappa_1 p_1} = y_{\kappa_2 p_2} = {\bar v}\,,\quad
  y_{\kappa_1 p_2} = {\bar v}^{1 - \xi_1}\,,\quad
  y_{\kappa_2 p_1} = {\bar v}^{1 - \xi_2}\,,\quad
  y_{\kappa_1 \kappa_2} =  {\bar v}^{2 - \xi_1 - \xi_2}\,,\quad
\end{equation}
and recombination will occur between $\kappa_1$ and $\kappa_2$ if
$\xi_1 + \xi_2 < 1$. If this is the case, then two recombinations are
now possible, with distance measures
\begin{equation}
  \label{eq:y32JadeAllOpts}
  y_{\kappa_{12} p_1} = y_{\kappa_1 p_1} + y_{\kappa_2 p_1} + y_{\kappa_1 \kappa_2}  \simeq
  y_{\kappa_2 p_1} \,,\quad
  y_{\kappa_{12} p_2} = y_{\kappa_1 p_2} + y_{\kappa_2 p_2} + y_{\kappa_1 \kappa_2}  \simeq
  y_{\kappa_1 p_2} \,,
\end{equation}
and as a result the three-jet resolution parameter will be
\begin{equation}
  \label{eq:y32JadeAllOpts2}
  y_3\left(\{\tilde p\},\kappa_1({\bar v}),\kappa_2({\bar v})\right) 
  \simeq {\bar v}^{1 - \min(\xi_1,
    \xi_2)}\,,\qquad\quad \xi_1 + \xi_2 < 1\,,\; \ell_1 \ne \ell_2\,.
\end{equation}
This does not scale as ${\bar v}$, so the limit
eq.~(\ref{eq:rIRClimit1}) is not finite, therefore the observable
fails the first rIRC test. 
One can verify that this breakdown of scaling as ${\bar v}$ occurs in
a double logarithmic region, in accord with the known result
\cite{JadeDL} that the E-scheme JADE jet-resolution distribution fails
to exponentiate at the double-logarithmic level. An interesting
demonstration of this point is in the evaluation of $\cF$, or more
specifically of its expansion in powers of $R'$, which starts at $R'^2$,
as discussed in appendix~\ref{sec:F2}. The $\bar v \to 0 $ limits of
$\cF$ and $\cF_2$ diverge. This divergence can be thought of as
somewhat analogous to the divergence of NLO corrections for an IRC
unsafe observable.

\FIGURE{
  \includegraphics[width=0.6\textwidth]{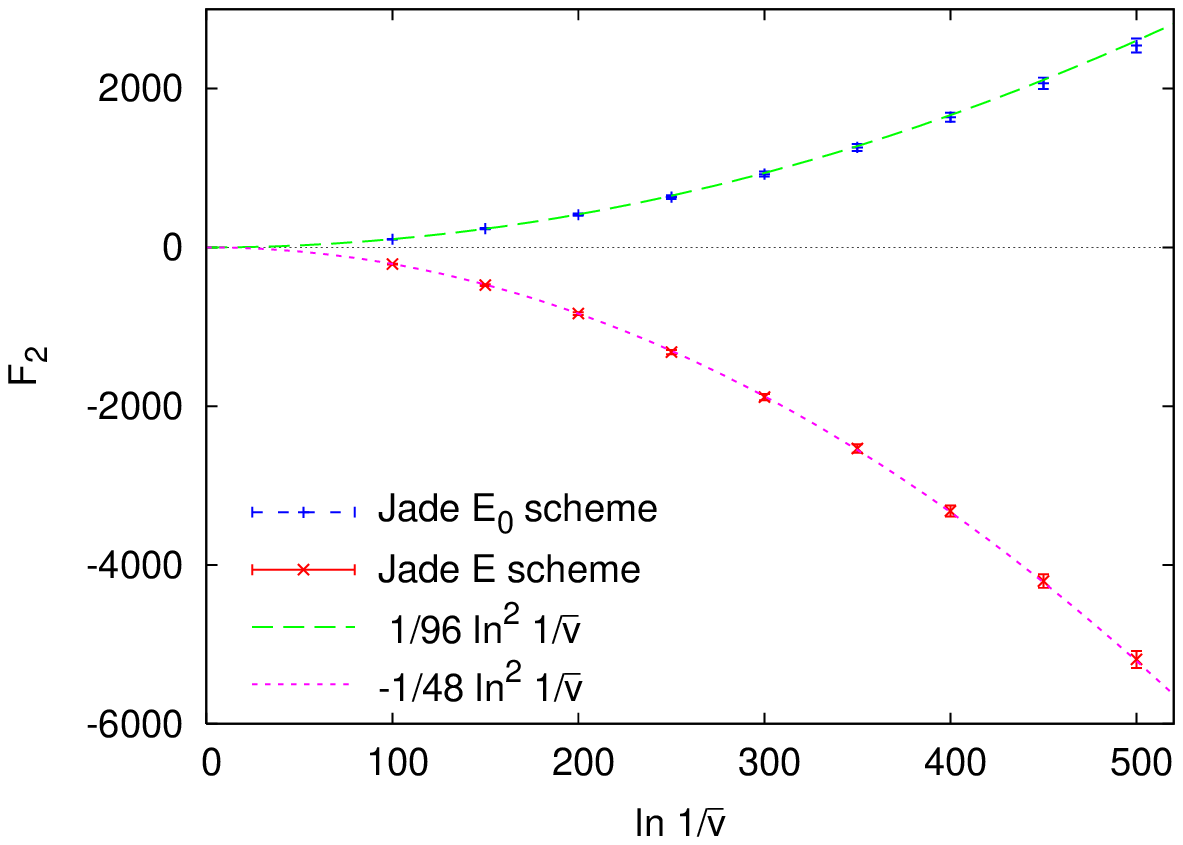}
  \caption{Calculation of $\cF_2$ for the Jade $y_3$ resolution
    parameter in the $E$ and $E_0$ recombination schemes. The points
    have been calculated by Monte Carlo evaluation of
    eq.~(\ref{eq:F2nonrIRC}).}
  \label{fig:JadeF2}
}

Additionally the nature of the divergence provides information about
the violation of exponentiation.  Figure~\ref{fig:JadeF2} shows
$\cF_2$ evaluated as a function of $\bar v$, using
eq.~(\ref{eq:F2nonrIRC}) (in whose derivation care has been taken to
retain all leading logarithmic dependence on $\bar v$). One sees that
$\cF_2$ diverges as $\ln^2 1/{\bar v}$, which is indicative of the
fact that the lack of rIRC safety is associated with
`multiple-emission' effects being relevant not at order $\as^2 L^2$,
but rather at order $\as^2 L^4$.

Using the following leading order, double logarithmic approximations
\begin{equation}
  \label{eq:JadeRpR}
  R(\bar v) = \frac{\as \CF}{\pi} \ln^2 \frac{1}{\bar v}\,,\qquad\quad
  R'(\bar v) = 2\frac{\as \CF}{\pi} \ln \frac{1}{\bar v}\,,
\end{equation}
and comparing to the double logarithmic result in \cite{JadeDL}, one
obtains that the deviation from exponentiation is expected to be of
the form
\begin{equation}
  \label{eq:JadeDLviolation}
  \vProb(v) = 1 - R(v) + \frac{5}{6} \frac{R^2(v)}{2} + \ldots\,.
\end{equation}
If one tries to account for this for by a function $\cF(R',v) = 1 +
\cF_2(v)R'(v)^2 + \ldots$ multiplying $e^{-R(v)}$, then $\cF_2(v) R'(v)^2$
should be equal to $-R(v)^2/12$, \ie
\begin{equation}
  \label{eq:JadeF2analytic}
  \cF_2(v) = -\frac{1}{48} \ln^2 \frac{1}{v}\,.
\end{equation}
This result is also plotted in fig.~\ref{fig:JadeF2} and coincides well
with the numerical evaluation based on eq.~(\ref{eq:F2nonrIRC}).

\subsection[Jade jet algorithms: $E_0$-scheme]{Jade jet algorithms: $\boldsymbol E_0$-scheme}
\label{sec:JadeE0}

As it happens, the $E$-scheme as defined above is rarely used
experimentally. More common is the $E_0$ scheme (see for example
\cite{DelphiOriented}) where particles are recombined according to
\begin{equation}
  \label{eq:E0recomb}
  E_{ij} = E_i + E_j\,,\qquad {\vec q}_{ij} = \frac{E_{ij}}{|\vec
    q_{i} + \vec q_{j} |} \left( \vec q_i + \vec q_j\right)\,.
\end{equation}
This has been studied analytically in \cite{CataniEtAlJets,Leder}.  One can
repeat the analysis of section~\ref{sec:JadeE}, and one finds that the
problem with the first rIRC condition disappears because
$\min_{\ell=1,2} y_{\kappa_{12} p_\ell} \simeq \max \left(y_{\kappa_1
    p_1}, y_{\kappa_2 p_2}\right)\simeq {\bar v}$ and so the limit
eq.~(\ref{eq:rIRClimit1}) is well-defined and finite. 

One still needs however to verify the other conditions,
eqs.~(\ref{eq:rIRClimit2}). A configuration that is of interest here
is that with two soft and collinear gluons, $\kappa_1({\bar v})$ and
$\kappa_2(\zeta_2 {\bar v})$ in the same hemisphere (say that containing leg
$1$). Note that we have reintroduced $\zeta_2$ ($\le 1$). Let us assume
$\kappa_2(\zeta_2 {\bar v})$ is much more collinear to the hard parton than
$\kappa_1({\bar v})$, $ (\zeta_2 {\bar v})^{\xi_2/2} \ll {\bar v}^{\xi_1/2}$. For the first
recombination, the various possible clusterings include,
\begin{equation}
  \label{eq:jadeE01stclust}
  y_{{\kappa_1} p_1} = {\bar v}\,,\quad y_{{\kappa_2} p_1} = \zeta_2
  {\bar v}\,,\quad
  y_{{\kappa_1} {\kappa_2}} = {\bar v} (\zeta_2 {\bar v})^{1 - \xi_2}\,,
\end{equation}
and when $\zeta_2 > {\bar v}^{\frac{1 - \xi_2}{\xi_2}}$, the first
recombination occurs between $\kappa_1$ and $\kappa_2$. Let us suppose
that this is the case. Then, as long as $\kappa_{t1} \ll \kappa_{t2}$,
\ie ${\bar v}^{1-\xi_1/2} \ll (\zeta_2 {\bar v})^{1-\xi_2/2}$, the energy and
transverse momentum of the $\kappa_{12}$ pseudo-particle will be
dominated by $\kappa_{2}$ and we will have
\begin{equation}
  \label{eq:jadeE02ndclust}
  y_{\kappa_{12}p_1} \simeq y_{\kappa_{2}p_1} = \zeta_2 {\bar v}\,.
\end{equation}
Let us now examine this result in the context of
eq.~(\ref{eq:rIRClimit2a}). To obtain the left-hand side we should
first take $ {\bar v}\to 0$. Our requirement on $\kappa_2(\zeta_2 {\bar v})$ being
more collinear than $\kappa_1$ simply implies $\xi_2 > \xi_1$,
as does the condition $\kappa_{t1} \ll \kappa_{t2}$; then the first
recombination is automatically $\kappa_1 \kappa_2$ and we obtain the
result that
\begin{equation}
  \label{eq:jadeE0LHScheck}
  \lim_{\zeta_2 \to 0 }\lim_{{\bar v}\to0} \frac{1}{{\bar v}} y_{3}(\{\tilde
  p\},\kappa_1({\bar v}),\kappa_2(\zeta_2 
  {\bar v})) = \Theta(\xi_1 - \xi_2) \,.
\end{equation}
In contrast the right-hand side of (\ref{eq:rIRClimit2a}) is
\begin{equation}
  \label{eq:jadeE0RHScheck}
  \lim_{{\bar v}\to0} \frac{1}{{\bar v}} y_{3}(\{\tilde p\},\kappa_1({\bar v})) = 1\,.
\end{equation}
Thus eq.~(\ref{eq:rIRClimit2a}) does not hold and the $E_0$ Jade
algorithm fails on the second of the recursive IRC safety conditions.
Despite the failure being on a different condition compared to the
observables discussed above, here too the nature of the violation is
such that exponentiation is broken at the level of $\as^2 L^4$
corrections.  As in the discussion for the $E$ scheme, one can evaluate
the second order contribution to $\cF$ as a function of $\ln \bar v$,
and compare it with the known analytical result of
\cite{CataniEtAlJets},
\begin{equation}
  \label{eq:JadeE0DLviolation}
  \vProb(v) = 1 - R(v) + \frac{13}{12} \frac{R^2(v)}{2} + \ldots\,,
\end{equation}
corresponding to $\cF_2(v) = \frac{1}{96} \ln^2
\frac{1}{v}$. The comparison is shown in figure~\ref{fig:JadeF2}
and, as for the $E$ scheme, one finds good agreement.

A final subtle, but non-trivial point to note here concerns the
requirement that $\kappa_{t2} \gg \kappa_{t1}$ for
eq.~(\ref{eq:jadeE02ndclust}) to hold --- though we have called our
condition recursive infrared collinear safety, in some cases the
limits that we take, notably here $\zeta_2 \to 0$, still leave the
`infrared and collinear' particle, $\kappa_2$, harder (larger
transverse momentum, larger energy, albeit smaller angle) than the
supposedly dominant contribution $\kappa_1$. This apparent
paradox is closely related to the fact that we use a single
quantity, the value of $V(\{\tilde p\},\kappa)$, to define the degree
to which an emission $\kappa$ is infrared and collinear. This controls
only some combination of the infrared and collinear limits, but not
the two independently (the remaining degree of freedom is set by
$\xi$). Thus two emissions which may be ordered according to one given
soft-collinear criterion, $V(\{\tilde p\},\kappa_2) < V(\{\tilde
p\},\kappa_1)$ are not necessarily ordered according to some other
criterion.

\subsection{Geneva jet algorithm}
\label{sec:Geneva}

Let us close this discussion of rIRC safety for jet algorithms by
examining the Geneva jet clustering algorithm \cite{Geneva}. It is
similar in spirit to the preceding algorithms, except that the
distance measure is given by
\begin{equation}
  \label{eq:GenevaDistance}
  y_{ij} = \frac{8}{9} \frac{E_iE_j(1-\cos \theta_{ij})}{(E_i + E_j)^2}\,,
\end{equation}
the essential change being the replacement of $Q^2$ in the denominator
with $(E_i+E_j)^2$. For events with two hard partons and one soft
collinear emission, this only changes the normalisation of the
$y_{ij}$ compared to the Jade family of algorithms, and one has
\begin{equation}
  \label{eq:GenevaCoeffs}
  a_\ell = b_\ell = g_\ell(\phi) = 1\,,\qquad d_\ell =
  \frac{16}{9}\,,\qquad \ell = 1, 2\,.
\end{equation}
The Geneva algorithm is interesting when a soft collinear
emission, $\kappa_1({\bar v})$, is split collinearly, $\kappa_1({\bar v}) \to
\{\kappa_{1_a},\kappa_{1_b}\}({\bar v}, \mu)$, with a small normalised pair
invariant mass, $\mu^2 = (\kappa_{1_a}+\kappa_{1_b})^2/\kappa_{t1}^2
\ll 1$, and fractions $z_{a}$ and $z_{b} = 1- z_{a}$ of the parent
momentum. The various possible recombinations include (assuming
$\kappa_1$ is collinear to leg $1$)
\begin{equation}
  \label{eq:Geneva1stClust}
  y_{\kappa_{1_a}p_1} = z_{a} y_{\kappa_{1}p_1} = z_{a} {\bar v}\,,\qquad
  y_{\kappa_{1_b}p_1} = z_{b} y_{\kappa_{1}p_1} = z_{b} {\bar v}\,,\qquad
  y_{\kappa_{1_a}\kappa_{1_b}} = \frac{16}{9}e^{-2\,\eta_1} \mu^2 \,.
\end{equation}
Whereas one would expect a `good' jet algorithm to first recombine
$\kappa_{1_a}$ and $\kappa_{1_b}$, what actually happens (as was first
observed in \cite{CataniEtAlJets}), for ${\bar v} \ll \mu^2 $, is that
for $z_a > z_b$ first $\kappa_{1_b}$ is recombined with $p_1$, and
then $\kappa_{1_a}$ is recombined with $p_1$ (inversely for $z_a <
z_b$), giving
\begin{equation}
  \label{eq:Genevay3res}
  V(\{\tilde p\},\{\kappa_{1_a},\kappa_{1_b}\}({{\bar v}}, \mu))  =
  \left\{
    \begin{array}{rc}
      {\bar v}\,,                 & \qquad \frac{16}{9}  e^{-2\,\eta_1} \mu^2 \lesssim \min(z_a, z_b) {\bar v}\,,\\
      \max(z_a, z_b) {\bar v}\,,  & \qquad \frac{16}{9} e^{-2\,\eta_1} \mu^2 \gtrsim \min(z_a, z_b) {\bar v}\,.
    \end{array}
  \right.
\end{equation}
As a result the two limits in eq.~(\ref{eq:rIRClimit2b}) differ, 
\begin{equation}
  \label{eq:GenevaBothLimits}
  \lim_{\mu\to0} \lim_{{\bar v}\to 0}\frac1{\bar v} V(\{\tilde
  p\},\{\kappa_{1_a},\kappa_{1_b}\}({{\bar v}}, \mu)) = \max(z_a,z_b)\,,\qquad
  \lim_{{\bar v}\to 0}\frac1{\bar v} V(\{\tilde
  p\},\kappa_{1}({\bar v}) ) = 1\,,
\end{equation}
and the observable fails on the second part of the second rIRC safety
criterion. The failure only occurs for hard collinear (non soft)
secondary splittings. Furthermore the two limits in
eq.~(\ref{eq:GenevaBothLimits}) differ by at most a factor of order $1$
(specifically by at most $1/2$). As a result it is possible to show
that the full resummed distribution for the Geneva $y_3$ resolution
parameter differs from the master formula (for which all elements are
well defined) by terms $\as^n L^n$. We note that in contrast to the
situation with the Jade algorithm, the rIRC unsafety of the Geneva
algorithm does not manifest itself through a divergent infrared
dependence in the integrals for $\cF$, because the integrations for
secondary collinear splitting have already been carried out
analytically, \emph{assuming} rIRC safety. Accordingly the $\cF$
function is well defined ($\cF(R')=1$).

\section{Infrared and collinear safety}
\label{sec:IRC}

In the automated approach discussed in this paper, we do not actually
explicitly test for the full infrared and collinear (IRC) safety of
the observable --- we rather assume that the user of the program is
able to correctly design and code IRC safe
observables.\footnote{Though many cases of IRC unsafety are actually
  caught out, for example by the rIRC safety tests or by requiring
  $a>0$ and $b_\ell >-a$.}

We believe though that it is instructive to discuss some aspects of
IRC safety, for two main reasons. Firstly, IRC safety turns out to be
somewhat more subtle than is usually reflected in `textbook'
discussions.  Secondly many of the issues that arise concerning IRC
safety are relevant also for rIRC safety, since the two conditions
have numerous similarities.

\subsection{Standard discussions of IRC safety}
\label{sec:standard-IRC-def}

The general definition of IRC safety is that it is the necessary and
sufficient condition that an observable has to satisfy in order for
its distribution to be calculable and finite, order-by-order within
perturbation theory.

For practical purposes however it is more convenient to attempt to
cast IRC safety in terms of certain properties of the observable's
functional dependence on the emission momenta, because IRC safety can
then be tested without explicitly calculating order by order
perturbative predictions for the observable.
An example of a definition (taken from p.~72 of \cite{ESWbook}) is
\begin{quote}
  For the [variable's distribution] to be calculable in perturbation
  theory, the variable should be infra-red safe, \ie insensitive to
  the emission of soft or collinear gluons.  In particular if ${\vec
    p}_i$ is any momentum occurring in its definition, it must be
  invariant under the branching
  \begin{equation}
    \label{eq:ESW3.39}
    {\vec p}_i \to {\vec p}_j + {\vec p}_k
  \end{equation}
  whenever ${\vec p}_j$ and  ${\vec p}_k$ are parallel or one of them
  is small.
\end{quote}
One notes that there are two parts to this definition, the first being
somewhat hand-waiving, the second appearing more precise.  In certain
other texts, only the second part is given, for example (from section
IV.A.2 of \cite{CTEQhandbook})
\begin{quote}
  [...] That is to say, the measurement should not distinguish between
  a final state in which two particles are collinear and the final
  state in which these two particles are replaced by one particle
  carrying the sum of the momenta of these collinear particles.
  Similarly, the measurement should not distinguish between a final
  state in which one particle has zero momentum and the final state in
  which this particle is omitted entirely.
  
  The argument that a cross section specified by functions $\cal S$
  with this property does not have infrared divergences may be
  understood as an extension of the KLN theorem [...]
\end{quote}

\subsection{Difficulties with standard definitions}
\label{sec:standard-IRC-difficult}

It is instructive to examine these (and other) definitions of IRC
safety for some `designer'\footnote{Just as designer clothes are those
  worn at fashion shows, but rarely in real life, designer observables
  are those discussed in theoretical articles, but rarely measured by
  real experimenters.} observables in $\ee$
processes. These will be constructed in terms of the $n$-jet threshold
resolution parameters, $y_n$, in the Durham jet algorithm
\cite{y3-kt_ee}.  Specifically, $y_n$ is the value of
$y_{\mathrm{cut}}$ below which one has an $n$ jet event, and above
which an $(n-1)$-jet event.  Individually, all the $y_n$ are IRC safe
observables.

Let us start by considering the following observable,
\begin{equation}
  \label{eq:y34combination}
  V = (1  + \Theta(y_{5} - y_{4}^2))\, y_{3}\,, \qquad (\mbox{if
  $y_4=0$: $V=y_3$})\,.
\end{equation}
It is non-zero starting from events with $3$ partons. If a fourth
parton is added then the observable is identical to $y_3$ and so
appears to be IRC safe. Adding a fifth parton and making it soft or
collinear to one of the other emissions, then $\Theta(y_{5} -
y_{4}^2)$ will be zero and again we will have the appearance of IRC
safety.

Now let us examine what happens if we integrate over the momenta of
partons $4$ and $5$, taking them to be ordered. Assuming that they are
emitted off different (hard) partons, we can approximate the phase
space for each of them as $dy_i /y_i \ln 1/y_i$. We also schematically
write the phase space and matrix element for the emission of the hard
gluon $y_3$ $dy_3 |M^2(y_3)|$. The mean value of $V$ then gets an
NNLO contribution which schematically has the form
\begin{subequations}
\begin{align}
  \label{eq:mean-y4y5}
  \langle V \rangle_\mathrm{NNLO} &\sim \as^3 \int dy_3 |M^2 (y_3)| \int^{y_3}
  \frac{dy_4}{y_4} \ln \frac{1}{y_4} \int^{y_4} \frac{dy_5}{y_5} \ln
  \frac{1}{y_5} \left[(1 + \Theta(y_{5} - y_{4}^2))y_3 - y_3\right]
  \\
  &\sim \as^3 \int dy_3 M^2 (y_3) y_3 \int^{y_3} \frac{dy_4}{y_4} \ln
  \frac{1}{y_4} \int^{y_4}_{y_4^2} \frac{dy_5}{y_5} \ln \frac{1}{y_5}\,,
\end{align}
\end{subequations}
where in the first line the rightmost term in square brackets accounts
for the combined one and two-loop virtual corrections. There is an infinite
region of phase
space for $y_4$ and $y_5$ where the real and virtual contributions do
not fully cancel. So even though the observable is insensitive to any
extra single arbitrarily soft or collinear emission (the condition
often used to characterise IRC safety, as in \cite{CTEQhandbook}), it
has a sensitivity to specific combinations of \emph{multiple} extra
arbitrarily infrared and collinear emissions, and this is sufficient
to make it IRC unsafe.

It would be interesting to find a definition that would correctly
identify eq.~(\ref{eq:y34combination}) as IRC unsafe, but that is more
precise than, say, the generic requirement of `insensitivity to the
emission of soft or collinear gluons' of \cite{ESWbook} and which
therefore can serve as a basis for automated testing of IRC safety. As
we shall see however, this is not a simple task.

\subsection{Search for a rigorous formulation of IRC safety}
\label{sec:nec-suff-IRC}

As a first attempt, let us consider the following definition, inspired
somewhat by the mathematical definition of a limit. First we introduce
some distance measure, which parametrises the degree of collinearity
of a pair of partons, or the softness of a parton (the distance
measure could be the relative $k_t$ of the pair, or their invariant
mass).
\newcommand{\VersionOne}{Version~1}%
\newcommand{\VersionOneText}{ Given almost any fixed set of partons (which we
  refer to as the `hard' partons) and any value $n$, then for any $x$,
  however small, there should exist an $\epsilon$ such that branching
  the partons so as create up to $n$ extra soft or collinear
  emissions, each emission being at a distance of no more than
  $\epsilon$ from the nearest `hard' parton, then the value of the
  observable does not change by more than $x$.}
\begin{quote}
  \underline{\VersionOne}
  
  \VersionOneText
\end{quote}
It is straightforward to see that the observable
eq.~(\ref{eq:y34combination}) violates this condition.

The issue of sensitivity to multiple soft or collinear emissions is
however not the only problem that arises when attempting to define a
general IRC safety condition. Also relevant for example is the
question of how quickly the effect of an emission disappears as it
is made soft or collinear.

If one defines
\begin{equation}
  \label{eq:V-slow-convergent}
  V = y_3\left(1 + \Theta(y_4) \ln^{-q}\frac{1}{y_4}\right)\,,\qquad q >0\,,
\end{equation}
then $V$ tends to $y_3$ in the limit $y_4\to0$ (as in
eq.~(\ref{eq:y34combination}), this and all subsequent observables are
defined to be $V=y_3$ if $y_4=0$.). Specifically, if we
take as our distance measure the squared relative transverse momentum
(normalised to the hard scale $Q$), then in our IRC definition given above,
however small an $x$ we choose, it suffices to take $\epsilon\equiv
y_4 < e^{-1/x^{1/q}}$ to ensure that any recombination will not change
$V$ by more than $xy_3$.

As we have already discussed, the phase space associated with a fourth parton,
expressed in terms of $y_4$ itself, goes roughly as $dy_4/y_4 \ln
1/y_4$ for each of the three harder partons to which parton 4 can be
collinear.  If one attempts to calculate the contribution to the
mean value from the integral over this phase space, including the
subtraction of the virtual terms, one finds an order $\as^2$
contribution of the form
\begin{equation}
  \label{eq:phasespace-bad-convergence}
  \langle V \rangle_\mathrm{NLO} \sim \as^2 \int dy_3 M^2 (y_3) \int^{y_3}
  \frac{dy_4}{y_4} \ln \frac{1}{y_4}
  \left[y_3\left(1+ \ln^{-q}\frac{1}{y_4}\right) - y_3\right]\,.
\end{equation}
This is divergent for $q<2$. At higher orders, since one effectively
includes extra logarithms in the numerator (for example from the
integrations that lead to the running of the coupling), one finds that
however large a value we take for $q$, there will be some fixed order
beyond which it is not possible to calculate the perturbative
corrections to the mean value of $V$.

This suggests therefore that any corrections to an observable from
extra emissions should vanish at least as fast as a \emph{power} of
the collinearity or softness of those emissions. This can be
incorporated into \VersionOne\ of our IRC definition, as follows:
\newcommand{\VersionTwo}{Version~2}
\begin{quote}
  \underline{\VersionTwo}

  \VersionOneText
  
  Furthermore there should exist a positive power $p$ such that for
  small $x$, $\epsilon$ can always be taken greater than $x^p$.
\end{quote}
One can straightforwardly verify that the observable of
eq.~(\ref{eq:V-slow-convergent}) is correctly classified as unsafe
with such a formulation of the IRC condition.

One of the patterns that the reader may see emerging from our
discussion so far is that for each definition or IRC safety, one is
able to design an observable that is incorrectly classified, requiring
that one further refine the definition. Unfortunately this is a major
difficulty, with even \VersionTwo\ of our  definitions suffering from
this problems. 

The difficulty can be illustrated with the following set of
observables,
\begin{align}
  \label{eq:theta24-not-enhanced}
  V &= y_3 ( 1 + \Theta(y_4 - |\cos \theta_{24}|))\,,\\
  \label{eq:theta23-not-enhanced}
  V &= y_3 ( 1 + \Theta(y_4 - |\cos \theta_{23}|))\,,\\
  \label{eq:theta23-enhanced}
  V &= y_3 \left( 1 + \frac{\Theta(y_4 - |\cos \theta_{23}|)}{y_4}\right)\,,
\end{align}
where $\theta_{ij}$ is the angle between jets $i$ and $j$ after
clustering to $\max\{i,j\}$ jets and with jets numbered such such that
$E_i>E_{i+1}$. 

The first observable is IRC safe, because the extra
$\Theta$-function term only contributes significantly in the
logarithmic integration over $y_{4}$ when $\cos \theta_{24}$ is close
to zero (a rare occurrence). It is however classified as IRC unsafe
according to \VersionTwo\ of our condition, because if one adds an
emission ($4$) such that it is exactly perpendicular to jet $2$ then
however soft it is, it changes the value of the observable by a factor
of $2$.

The second observable, eq.~(\ref{eq:theta23-not-enhanced}), is quite
similar, and in particular is also IRC safe. Unlike
eq.~(\ref{eq:theta24-not-enhanced}), it is correctly classified by
\VersionTwo\ of our IRC condition.  This is because $\epsilon$ is to
be found for a given fixed configuration of hard momenta (in
particular a given fixed value of $\theta_{23}$). It is not necessary
that the same $\epsilon$ be valid for all hard momenta.  Accordingly,
however close $\theta_{23}$ is to zero, one can always find an
appropriate value of $\epsilon$ for a given $x$. An exception occurs
for $\theta_{23}=0$, however this corresponds to a region of zero
measure in phase space, and is an allowed exception insofar as we
required that the condition be true for \emph{almost} any set of hard
partons.

The third observable, eq.~(\ref{eq:theta23-not-enhanced}), also passes
the test --- the presence of $y_4$ in the denominator does not change
one's ability to find a point at which the effect of the fourth
emission disappears. However it does change the integrability
properties, since the presence of the $1/y_4$ factor compensates the
reduced $\theta_{23}$ phase-space, leading to a divergent NLO
contribution,
\begin{subequations}
\begin{align}
  \label{eq:theta23-not-enhanced-integrated}
  \langle V \rangle_\mathrm{NLO} &\sim \as^2 \int \!\!dy_3 \,d\theta_{23} \,|M^2(y_3,
  \theta_{23})| \int^{y_3} \!\frac{dy_4}{y_4} \ln \frac{1}{y_4}
  \left[y_3 \left( 1 + \frac{\Theta(y_4 - |\cos \theta_{23}|)}{y_4}
    \right) \!-\! y_3\right]
  \\
  &\sim \as^2 \int dy_3 \,d\theta_{23} \,|M^2(y_3, \theta_{23})| \,\frac{y_3
     \ln (1/\cos \theta_{23})}{\cos \theta_{23}}\,.
\end{align}
\end{subequations}
For each of the mis-classifications identified above one could
envisage some work\-around that would solve the problem: for example
allowing a subset of soft and collinear emissions to violate the IRC
condition, as long as the subset's measure is sufficiently limited; or
requiring the observable's value to be bounded. 

But in the absence of a formal derivation of the resulting IRC condition,
a doubt will always persist as to its general validity. Such a formal
derivation might well be inspired by mathematical statements
concerning the properties required of multivariate functions in order
for them to be integrable. However that is beyond the scope of this
article.

Let us finally comment on \emph{recursive} IRC safety in the light of
this discussion. Eqs.~(\ref{eq:rIRClimit2}) for recursive IRC safety
are similar to formulations of normal IRC safety in terms of a single
emission that is made soft or collinear. The preamble to the
discussion of rIRC safety attempts instead to give a general
statement, somewhat analogous to that for IRC safety in
\cite{ESWbook}. The former is more understandable insofar as it
appears more precise. One should however be aware of its limitations.
For example in eq.~(\ref{eq:ignorelowvemissions}), we have explicitly
seen the need for the observable to be insensitive with respect to the
removal of multiple relatively much softer emissions.

\section{Divergences of $\cF$}
\label{sec:divergences-ff}

\subsection{General considerations}
\label{sec:gen-div-cF}

To obtain a more general understanding,  than was given in
section~\ref{sec:div}, of the contexts in which
divergences can appear, it is
useful to consider the case of the 
broadening with respect to the photon axis in the Breit-frame current
hemisphere of DIS, $B_{zE}$. From the analytical studies in
\cite{DSBroad}, one can write $B_{zE}$ in terms of the soft and
collinear emissions, $k_i$, as
\begin{equation}
  \label{eq:BzE-form}
  B_{zE} = \frac{1}{Q}\left(|{\vec k}_{t,\hc} + {\vec k}_{t,\hr}|
  + \sum_{i\in \hc} |{\vec k}_{t i}|\right)\,,\qquad
  {\vec k}_{t,\hr/\hc} = \sum_{i \in \hr/\hc} {\vec k}_{ti}\,,
\end{equation}
where $\hr$ and $\hc$ are the remnant and current hemispheres
(associated respectively with legs $1$ and $2$) and the notation
$\hr/\hc$ means either $\hr$ or $\hc$. For $B_{zE}$ to be small it is
necessary to suppress emissions on leg $2$ (since there are no
cancellations in $\sum_{i\in \hc}|{\vec k}_{t i}|$), while through a
cancellation in the 2-dimensional vector sum, emissions
on leg $1$ can contribute little overall to $B_{zE}$ even if
individually they have large transverse momenta. Accordingly, for
configurations in which the hardest emission is on leg $1$, there is a
small-$y$ contribution to $\cP(y)$ of the form
\begin{equation}
  \label{eq:cP-for-BzE}
  \cP(y) \sim y^2 \cdot y^{C_2 r'_2}\,,\qquad y \to 0\,,
\end{equation}
where the first factor is that associated with the cancellation in the
vector sum, while the second is associated with the Sudakov
suppression for emissions from leg $2$. 

More generally there are observables for which cancellations can occur
for emissions off a subset $s$ of the legs, while the complementary
subset of legs ($\bar s$) shows no cancellations. In such a case,
assuming in analogy with before that there is a power $p$ associated
with the structure of the cancellations on set $s$, then
\begin{equation}
  \label{eq:cP-for-s-sbar}
  \cP(y) \sim y^p \cdot y^{R'_{\bar s}}\,,\qquad y \to 0\,,\qquad
  R'_{s/\bar s} = \sum_{\ell \in {s/\bar s}} C_\ell r'_\ell\,.
\end{equation}
This gives a divergence at $R' = p + R'_{\bar s}$ or equivalently
$R'_{s} = p$. For the case of $B_{zE}$ this corresponds to $R'_s=2$ or
equivalently, $R' = 4$.

Such arguments can also be extended to cases where there are several
subsets of legs subject separately to cancellations, $s_1$, $s_2$,
\ldots, each being associated with a power $p_i$. In such a situation,
if the hardest emission is from a leg belonging to set $s_i$, then there is
a contribution to $\cP(y)$ for small $y$ of the form 
\begin{equation}
  \label{eq:cP-for-si-sbar}
  y^{p_i} \cdot y^{R'-R'_{s_i}}\,,
\end{equation}
in which a cancellation occurs on set $i$ and Sudakov suppression is
responsible for limiting the contributions on all other legs. This
would lead to a divergence in $\cF$ when $R'_{s_i} = p_i$.  There are
also situations in which a cancellation occurs additionally on a
second set, $s_j$, giving a contribution to $\cP(y)$ that goes as
\begin{equation}
  \label{eq:cP-for-sij-sbar}
  y^{p_i + p_{j}} \cdot y^{R'-R'_{s_i} - R'_{s_j}}\,.
\end{equation}
This leads to a divergence in $\cF$ when $R'_{s_i} + R'_{s_j} = p_i +
p_j$. The argument can be extended to situations in which cancellations
occur on any number of subsets of legs with cancellations.

The divergence that limits the calculation of $\cF$ is that which occurs at
smallest value of the overall $R'$. One can show that it is determined
by contributions of the form eq.~(\ref{eq:cP-for-si-sbar}) in which
the cancellations occur within a single set. Thus the position of the
divergence of $\cF$ is given by the solution of $R'_{s_i} = p_{i}$
that corresponds to the smallest $R'$ (we recall that for a fixed
colour configuration, the $\{R'_{s_i},R'_{\bar s}\}$ are not
independent quantities, but rather all depend on $\lambda =\beta_0 \as L$).

\subsection{Speed of Monte Carlo convergence}
\label{sec:analysis-divergent-cF}

As was mentioned in section~\ref{sec:div},
in many cases the divergence occurs at a value of the overall $R'$
that is sufficiently large that one can ignore it for phenomenological
purposes. However it turns out that problems arise in the Monte Carlo
determination of $\cF$ at smaller, relevant, values of $R'$. To better
appreciate the issue we 
consider the \emph{variance} for the calculation of $\cF$,
\begin{equation}
  \label{eq:cF-variance}
  \sigma_\cF = \int_0^\infty dy \,\frac{d\cP(y)}{dy}\, e^{-2R' \ln y}
  - \cF^2\,,
\end{equation}
The statistical error on the Monte Carlo integration with $N$ events
is given by $\sqrt{\sigma_\cF/N}$. Considering the general case,
introduced in section~\ref{sec:gen-div-cF}, of an observable with subsets
$s_i$ of legs each having a zero associated with a `power behaviour'
$p_i$, one can show that $\sigma_\cF$ diverges for the smallest value
of $R'$ for which there is a solution to any of the equations $p_i =
R'_{s_i} + R'$. For an observable where all legs are simultaneously
involved in the cancellation ($R' \equiv R'_s$) this just corresponds
to $R' = p/2 \equiv R'_c/2$. In contrast when there are different
subsets of legs with and without cancellations the variance usually
diverges earlier, at $R'<R'_c/2$ --- for example for $B_{zE}$,
eq.~(\ref{eq:BzE-form}), one can show that it diverges for $R' = 4/3$
(whereas $R'_c = 4$).

The divergence of the variance is a standard characteristic of Monte
Carlo integration when dealing with integrands with singularities of
the form $1/\sqrt{y}$ and stronger. It does not imply that Monte Carlo
methods cannot be used --- the result of the integration still
converges, but since $\sigma_\cF$ grows with $N$, the error on $\cF$,
$\sqrt{\sigma_\cF/N}$, converges more slowly than $1/\sqrt{N}$.
Specifically the error on an integral of the form $\int_0^1dy/y^a$
converges as $N^{a-1}$ for $a>1/2$, when $y$ is generated uniformly
between $0$ and $1$.

For values of $R'$ close to the point where the variance diverges,
this is not too serious a problem, however if one wishes to
investigate the structure of $\cF$ closer to the divergence of $\cF$
itself then the slow Monte Carlo convergence becomes a significant
issue. A standard solution is to perform a Jacobian transformation on
the integration so as to increase the number of points in the vicinity
of the divergence.  Because of the complexity of the probability
distribution eq.~(\ref{eq:ProbRatio}), it is highly non-trivial to do
this for an arbitrary observable.

However for many observables of practical relevance, the cancellations
that are observed tend to fall into a limited number of classes (such
as the 2-dimensional vector sum discussed in section~\ref{sec:div}).
Given knowledge of which sets of legs have cancellations, as well as
the class of cancellation, improvements can be obtained.\footnote{The
  information could also be used to provide analytical improvements
  beyond the point of the divergence, as in \cite{DSBroad}. One should
  however also be aware of the complication of dynamically
  discontinuous globalness \cite{DiscontGlobal} which arises in many
  such cases.}

To analyse possible cancellations the program proceeds through various
steps. It first  considers configurations with
two emissions, off legs $\ell_1$ and $\ell_2$ respectively. For each
combination of $\ell_1$ and $\ell_2$ (which can be equal) it
establishes whether there can be cancellations that lead to a zero of
the observable --- we call this a `common zero' of legs $\ell_1$ and
$\ell_2$. The legs are then classified into subsets of legs such that,
the legs from two different subsets never have a `common zero', and such that
if a subset contains more than one leg, then each leg in that subset
has a common zero with at least one of the other legs in the
subset. In this manner one determines the subsets $s_{(i)}$ and $\bar s$
of section~\ref{sec:gen-div-cF}.

For each subset ($s$) the program then examines various hypotheses
concerning the origin of the zeroes. The hypotheses can be formulated
as the requirement that the value of the observable be unchanged under
the replacement of all emissions $k_i \in s$ by a suitably chosen
single emission $K$ (while emissions $k_i \notin s$ are not modified).
An example replacement is that corresponding to the two-dimensional
vector sum,
\begin{equation}
  \label{eq:replaceveckt}
  {\vec K}_t = \sum_{i\in s} {\vec k}_{ti}\,,
\end{equation}
(where the choice of $\eta_K$ is free). This tends to be relevant only
for observables where for $\ell \in s$, $b_\ell=0$ and $g_\ell(\phi)=1$.

Certain observables (for example the transverse momentum of a
Drell-Yan pair) are fully described by this condition. But the
resulting divergence (at $R'_c=2$) significantly limits the region of
validity of the calculation and one is better off using the transform
methods of \cite{Bonciani:2003nt} for performing the resummation. In
many other cases (such as $B_{zE}$) the vectorial cancellation
applies only to a subset of legs. The resulting $R'_c$ is therefore
larger, and there is a significant region in which the observable's
distribution is formally well-predicted, but the Monte Carlo
calculation is poorly convergent.

Having established that some simple form for the replacement is valid,
such as eq.~(\ref{eq:replaceveckt}), one can then obtain significant
improvements in the convergence of the Monte Carlo calculation of
$\cF$, essentially by generating not the $k_i\in s$, but rather
directly the replacement emission $K$, with the appropriate
analytically calculated distribution. As discussed in detail
in~\ref{app:MCDetails}, this gives one the freedom to introduce a
Jacobian in the calculation of $\cF$ (which otherwise is quite
difficult to do), which vastly improves the Monte Carlo convergence.

Such methods hold not only for the cancellation
eq.~(\ref{eq:replaceveckt}), but also for other classes of zeroes, for
example in observables that are sensitive to cancellations in a single
component of the transverse momentum and to cancellations from legs
that individually are additive but combine together with different
signs. Full details are given in appendix~\ref{app:MCDetails}.

It is to be kept mind that there exists a small
number of observables with multiple-emission zeroes for which the
detailed analytical origin of the zeroes has not been understood (for
example the $\ee$ oblateness) or does not fall into any of the above
classes.  For such observables a Jacobian improvement is not
available, so that while $\cF$ remains calculable there is a
region of $R'$ in which the Monte Carlo convergence is rather poor.

\subsection{Details of MC analysis}
\label{app:MCDetails}

The details of the Jacobian-improvement method are as follows. With
the naive Monte Carlo approach to 
calculating $\cF$, the number of events in a given interval $\delta y$
of $y$ (in the notation of section~\ref{sec:div}) is $\delta y d\cP/dy$,
the corresponding weight being $y^{-R'}$. Let us first consider how to
improve the convergence when all legs have a simple common zero ($R'_s
= R'$, $R'_{\bar s} = 0$) and for which we can calculate $d\cP/dy$
analytically. We are then free to generate $y$ with some alternative
distribution $d{\tilde \cP}/dy$ and then for each event include an extra
weight $w(y)$ such that $w d{\tilde \cP}/dy = d\cP/dy$. The results for
$\cF$ and its variance are then
\begin{equation}
  \label{eq:cF-variance-with-weight}
  \cF = \int_0^\infty dy \frac{d{\tilde \cP(y)}}{dy}\, w(y) y^{-R'}\,,\qquad
  \sigma_\cF = \int_0^\infty dy \frac{d{\tilde \cP(y)}}{dy} \, w^2(y)
  y^{-2R'} - \cF^2\,.
\end{equation}
Since one can use any form for $d\tilde \cP/dy$, by making it
sufficiently peaked at small $y$ one can ensure that the variance
converges for $R' < p$. Specifically, for a distribution $\cP(y) \sim
y^p$ at small $y$, one can take $\tilde \cP(y) = y^{p-R'}$, implying a
weight function $w(y) \sim y^{R'}$. One immediately sees that the
combination $w(y) y^{-R'}$ is independent of $y$ for small $y$,
ensuring that the variance remains under control even for values of
$R'$ approaching $p$.

Of course if we are able to calculate $d\cP/dy$ analytically, then we
are probably also in a position to obtain $\cF$ with only marginally
more work and there is no need for any Monte Carlo integration!
However there are observables for which we are able to calculate to
$d\cP/dy$ analytically for emissions off only the subset $s$ of legs
with a common zero, but not necessarily for the full situation
including emissions off the remaining legs (subset $\bar s$). 

To explain the situation that then arises, let us denote by $y_s$
($y_{\bar s}$) the value of the observable with just the emissions off 
the subset $s$ ($\bar s$), with integrated probability distributions
$\cP_s(y_s)$ and $\cP_{\bar s}(y_{\bar s})$ separately for $y_s$ and
$y_\sbar$. The function $\cF$ is then given by,
\begin{equation}
  \label{eq:cF-with-yssbar}
  \cF = \int dy_s dy_\sbar \frac{d\cP_s(y_s)}{dy_s}
  \frac{d\cP_\sbar(y_\sbar)}{dy_\sbar}
  [y(y_s,y_\sbar,\ldots)]^{-R'}\,,
\end{equation}
where the rescaled value $y$ of the observable has been written as a
function $y(y_s,y_\sbar,\!\ldots)$ of $y_s$, $y_\sbar$ and other
(unspecified) degrees of freedom such as correlations between
emissions in $s$ and $\sbar$ over which we integrate implicitly. The
function $y(y_s,y_\sbar,\ldots)$ typically has the property that for
$y_s\ll y_\sbar$, $y\simeq y_\sbar$, while for $y_s\gg y_\sbar$,
$y\simeq y_s$, so we can understand the behaviour of $\cF$ by modelling
it with $y = \max\{y_s,y_\sbar\}$.

The configurations that are responsible for the divergence in $\cF$
are those whose hardest emission is in $s$, since if the
hardest emission is in $\sbar$, then $y_\sbar$ is bound to be of order
$1$ (as is $y$). Accordingly, for small $y_s$ and $y_\sbar$,
\begin{equation}
  \label{eq:dPs-dPsbar}
  \cP_s(y_s) \sim y_s^p\,,\qquad\quad \cP_\sbar(y_\sbar) \sim
  y_\sbar^{R'_\sbar}. 
\end{equation}
This, together with the model $y = \max\{y_s,y_\sbar\}$ is the origin
of eq.~(\ref{eq:cP-for-s-sbar}).

To improve the Monte-Carlo convergence of the integral, we can as
before generate $y_s$ with a modified probability distribution, as
above, ${\tilde \cP}_s(y_s) = y_s^{p - R'_s}$ and a weight function $w_s(y_s)
\sim y_s^{R_s'}$. This improves the Monte Carlo convergence a little,
the variance diverging for $R' > p$ rather than $R'+R'_s>p$, but the
situation is still problematic.

To further solve the problem one should also modify the generation of
$y_\sbar$. We have no analytical information about the form of
$\cP_\sbar(y_\sbar)$ other than that in eq.~(\ref{eq:dPs-dPsbar}) ---
however the origin of the behaviour in eq.~(\ref{eq:dPs-dPsbar}) is
simply Sudakov suppression, essentially associated with the
probability distribution of the generation of the hardest emission in
$\sbar$. This generation of the hardest emission is actually
straightforward to modify, so that, given a value for $y_s$ we can
generate $y_\sbar$ with an integrated distribution ${\tilde
  \cP}_\sbar(y_\sbar) \sim y_\sbar^{-1}$ down to $y_\sbar \sim y_s$ and
a corresponding weight function $w_\sbar(y_\sbar) \sim
y_\sbar^{R'_\sbar + 1}$. It is straightforward to show, within our
model for the observable, that the variance then remains integrable for all
values of $R'$ up to the divergence of $\cF$ itself.

In situations in which no method is available for modifying the
generated distribution of $y_s$, we can actually still modify the
distribution of $y_\sbar$, and we find that the variance remains
integrable up to $R'_s = (p+1)/2$.  This does
not in general take us all the way to the position of the divergence
of $\cF$ but still represents a significant improvement. 

We close this section by providing the exact forms of $\cP_s(y_s)$ for
certain common classes of cancellations. These are used in the
program.

\subsubsection{Two-dimensional vector sum}
\label{sec:2-dimensional-vector}

Let us first consider an observable having a set $s$ of legs with a
common zero, such that the replacement emission is one whose
transverse momentum is the vector sum of all the actual emissions in
$s$, \cnf eq.~(\ref{eq:replaceveckt}). For simplicity we will assume
that all the legs in the set have $b_\ell=0$,
$g_\ell(\phi) = 1$, and a common value of $d_\ell$, however the result
can be applied more generally.

If we use $k_{t0}$ to denote the largest of the transverse momenta of
the actual emissions (if emission $1$ is in $s$ then $k_{t0} = ({\bar
  v} /d_\ell)^{1/a}Q$), for a given value of $R'_s = \sum_{\ell \in s}
C_\ell r'_\ell$, the distribution of the value of the replacement
transverse momentum $K_t$ will be
\begin{subequations}
\begin{align}
  \label{eq:replacement_Kt_vecsum}
  \frac{d\cP(K_t)}{dK_t} &= 
  K_t \int \frac{d^2 \vec b \, d\phi_{k_{t0}}}{4\pi^2}
  e^{i \vec b. {\vec K}_t} e^{-i \vec b. {\vec k}_{t0}}
  \sum_{m=0}^{\infty} \frac{(R'_s)^{m}}{m!} \prod_{i=1}^{m} 
  \int \frac{d^2k_{ti}}{2\pi k_{ti}^2}
  \left(e^{-i \vec b .{\vec k}_{ti}} - 1 \right)
  \\
  &=
  K_t \int_0^\infty bdb \,J_0(b K_t)\,
  J_0(b k_{t0})\, \exp\left(R'_s \int_0^{k_{t0}} \frac{dk_t}{k_t}(J_0(b
    k_t) - 1)\right),
\end{align}
\end{subequations}
as can be derived using standard analytical resummation techniques,
such as those of section~\ref{sec:thrust}.

\subsubsection{One-dimensional signed sum}
\label{sec:1-dimensional-sum}

Various observables involve direct differences between the effect of
emissions in different regions. For example the absolute difference
between the two squared jet masses $\rho_D = |\rho_1 - \rho_2|$ in
$\ee$, where we recall that the squared jet masses have the property
that individually they are additive (like the thrust). Let us extend
this temporarily to be any \emph{signed} difference, $V_D = V_1 -
V_2$, where the $V_\ell$ are additive observables separately for the two
legs.  Therefore we can write
\begin{equation}
  \label{eq:VD-def}
  V_D(\{\tilde p\},
  \kappa_1(\zeta_1 \bar v), \ldots,\kappa_m(\zeta_m \bar
  v)) = \bar v \left(
    \sum_{\forall\, i,\, \ell_i = 1} \zeta_i - \sum_{\forall\, i,\,
      \ell_i = 2} \zeta_i 
  \right).
\end{equation}
Given $R_\ell' = C_\ell r'_\ell$ and allowing for the possibility that
$R'_1 \ne R'_2$ (of relevance to subsequent applications) we can write
the following expression for the distribution of $y = \lim_{\bar
  v\to0} V_D/{\bar v}$,
\begin{equation}
  \label{eq:VD-prob-def}
  \frac{d\cP(y)}{dy}
  = 
  \int \frac{d\nu}{2\pi i} e^{\nu y} \left[\frac{R'_1 e^{-\nu} +
    R'_2 \,e^{\nu}}{R'_1 + R'_2}\right]
  \prod_{\ell=1}^2
  \left(\sum_{m=0}^\infty \frac{(R'_\ell)^m}{m!} \prod_{i=1}^m \int_0^1
    \frac{dy'}{y'} \left(e^{(-1)^\ell\nu y'}-1\right) \right)\,,
\end{equation}
where the factor in square brackets accounts for the fact that there
is an emission with $\zeta=1$ on either leg $1$ or leg $2$; the factor
$(-1)^\ell$ is simply a compact notation for the fact that the
contributions from leg $1$ ($2$) enter with a positive (negative) sign
in eq.~(\ref{eq:VD-def}). This gives
\begin{equation}
  \label{eq:VD-prob-res}
  \frac{d\cP(y)}{dy}
  = 
  \int \frac{d\nu}{2\pi i} e^{\nu y} \frac{R'_1 e^{-\nu} +
    R'_2 \,e^{\nu}}{R'_1 + R'_2}
  e^{-R'_1 E(\nu) - R'_2 E(-\nu)}\,,\qquad
  E(z) = \int_0^z \frac{dt}{t}(1 - e^{-t})\,.
\end{equation}
The probability distribution for the absolute value of $y$ is then
simply
\begin{equation}
  \label{eq:VD-absdiff}
  \frac{d\cP(|y|)}{d|y|} =  
  \int \frac{d\nu}{2\pi i} e^{\nu y} \left( \frac{R'_1 e^{-\nu} +
    R'_2 \,e^{\nu}}{R'_1 + R'_2}
  e^{-R'_1 E(\nu) - R'_2 E(-\nu)} +  \frac{R'_1 e^{\nu} +
    R'_2 \,e^{-\nu}}{R'_1 + R'_2}
  e^{-R'_1 E(-\nu) - R'_2 E(\nu)}\right)\,.
\end{equation}
Quantities that are amenable to this kind of analysis can arise not
only from differences between contributions from two different legs,
but also within a single leg $\ell$, for example in sums of a single
component of transverse momentum. 
In hadronic-dijet production this occurs for instance for a thrust
minor distribution based on particles only in restricted phase-space
region $\cR$
\begin{equation}
\label{eq:tmin-ind}     
T_{m} \equiv \frac{\sum_{i \in \cR} |q_{xi}|}
{Q_{\perp,\cR}}\,, \qquad \qquad Q_{\perp,\cR}= \sum_{i\in \cR} q_{\perp i}\,, 
\end{equation}
where the $x$ direction is defined as that perpendicular to the beam
and to the global transverse thrust axis, which together define the
event plane and $q_{\perp i}$ denote momenta transverse to the beam
direction. 
In such cases $R'_1$ and $R'_2$ are each replaced by $R'_\ell/2$.

\section{Specific $\boldsymbol{e^+e^-}$ observables}
\label{sec:ee-observables}

In this section we present some theoretically and phenomenologically
interesting issues which arise from the study of (new) observables in
the simple environment of $\ee$ collisions.
\subsection{BKS observables: limiting cases}
\label{sec:bks-observables}

Following \cite{BKS03} we consider the three-jet observables\footnote{In
  the original definition $x$ is named $a$, this would however  
cause confusion with our coefficient $a$ parameterising the dependence
on the transverse momentum.}
\begin{equation} \label{eq:BKS.def}
\tau_x \equiv 
\frac{\sum_i E_i |\sin \theta_i|^x \left( 1-|\cos\theta_i|\right)^{1-x}}
{\sum_i |\vec q_i| }\>,
\end{equation}
where the $\theta_i$ are the angles with respect to the thrust axis. The
adjustable parameter $x$ allows one to control the importance of the
soft-large angle and hard collinear region. For the observable to be
IRC safe $x$ should be in the range $-\infty <x < 2$, 
the value of $x=0$ giving the thrust, while $\tau_1$ corresponds to the
total broadening~\cite{BroadDef} (to within a factor of two).

It is well known that the perturbative resummation of the thrust and
broadening distributions are different in that in the thrust case hard
parton recoil can be neglected at NLL (giving an additive observable,
$\cF = e^{-\gae R'}/\Gamma(1+R')$), while this is not the case for the
broadening distribution (which has a more complicated form for $\cF$).
As was shown in \cite{BKS03,BS03}, the additivity property actually
holds for all values of $x<1$. Since we know that a transition occurs
at $x=1$ it is interesting to examine what happens beyond that point,
for $1<x<2$, especially since this region of $x$ was not studied in
\cite{BKS03,BS03}.  We therefore show here two observables,
$\tau_{1/2}$ and $\tau_{3/2}$ (though we have also studied other
values of $x$).

We establish numerically that both observables satisfy all
applicability conditions of Sec.~\ref{sec:summary-master}. The
properties with respect to a single emission are parametrised by the
coefficients in Table~\ref{tab:BKSsingle-emission}.
\TABLE{
\hspace{+1.2cm}$\tau_{1/2}$\hspace{6.8cm}$\tau_{3/2}$\\
\vspace{.3cm}
\begin{tabular}{| c | c | c | c | c |}
 \hline
 leg $\ell$ & $a_{\ell}$ & $b_{\ell}$ & $g_{\ell}(\phi)$ & $d_{\ell}$
\\
 \hline
 \hline
1 &  1.000 & $  0.500$ & $1.000$ & 1.000  \\
 \hline
2 &  1.000 & $  0.500$ & $1.000$ & 1.000  \\
 \hline
 \end{tabular}
\hspace{0.5cm}  
 \begin{tabular}{| c | c | c | c | c | c |}
 \hline
 leg $\ell$ & $a_{\ell}$ & $b_{\ell}$ & $g_{\ell}(\phi)$ & $d_{\ell}$
\\
 \hline
 \hline
1 &  0.500 & $  0.000$ & $1.000$ & 1.000  \\
 \hline
2 &  0.500 & $  0.000$ & $1.000$ & 1.000  \\
 \hline
 \end{tabular}
 \caption{Leg parametrisation coefficients for $\tau_{1/2}$ (left) 
   and $\tau_{3/2}$ (right).}
 \label{tab:BKSsingle-emission} 
}
  
For $\tau_{1/2}$ the results are consistent with $a=1$, $b=1-x$, as
derived in \cite{BKS03,BS03} for $x\le1$. The multi-emission
properties of $\tau_{1/2}$ are also found to be consistent with
additivity, again as expected, a consequence of the fact that the sum
in the numerator of eq.~(\ref{eq:BKS.def}) is dominated by the soft
and collinear emissions rather than by the recoiling hard partons.

Instead for $\tau_{3/2}$ we see that the analytical dependence of $a$
and $b$ on $x$ must change. Examining a range of values of $x$ reveals
that for $1<x<2$ one has $a=2-x$, $b=0$. The multi-emission structure
is also interesting in that the program reveals that for each leg
$\ell$ separately, the observable remains unchanged under the
replacement of all emissions with a single emission having transverse
momentum $\sum_{i\in\ell} \vec k_{ti}$. Both of these features are a
consequence of the fact that with $x>1$, it is the recoiling hard
partons that dominate the sum in the numerator of
eq.~(\ref{eq:BKS.def}).

In \cite{BS03} it was argued that the class of observables in
eq.~\eqref{eq:BKS.def} is particularly interesting from a
non-perturbative point of view. Non-perturbative corrections to these
observables were shown (under certain assumptions) to follow a scaling
rule which allows one to relate $1/Q$ power-suppressed
non-perturbative corrections of an observable with a given value of
$x$, to one with a different value of $x$, for instance the thrust,
whose non-perturbative corrections have been extensively studied. What
is remarkable about this scaling is that it holds for all moments of
the shape function, not just for the first moment as is usually the
case \cite{DokWebDists} when relating perturbative corrections of
different event shapes.

This scaling rule breaks down for $x\ge1$, the $x=1$ (broadening) case
being known to have a more complicated power correction structure
\cite{BroadNP-rev}. Actually, even for $x$ approaching $1$ from below
it is likely to be difficult to test the scaling rule in detail: the
first moment of the power correction scales as $1/(1-x)$ \cite{BS03},
but the broadening is known to have a first moment enhanced by
$1/\sqrt{\as}$. This suggests that the scaling must actually start to
break down for $1-x \sim \sqrt{\as}$.

Furthermore, perturbatively, at NLL there is a discontinuous change in
the structure of $\cF$ when going from $x<1$ to $x=1$. Given that
abrupt transitions at one order are usually associated with divergent
corrections at higher orders, for $x\to1$ we expect the NNLL terms to
be enhanced by factors related to $\ln(1-x)$, meaning that predictions
at any fixed resummed order may be unreliable for $x$ close to $1$.

This has prompted us to search for a class of observables having
identical perturbative and non-perturbative properties to the BKS
class for $x<1$, but with a smoother transition through $x=1$.

\subsection{Fractional moments of energy-correlations}
\label{sec:FEC}

Given the above arguments, and inspired by \cite{FoxWolfram}, we
modify the definition of the observables in eq.~\eqref{eq:BKS.def} to
be
\begin{equation} 
\label{eq:FCx.def} 
FC_{x} \equiv \sum_{i \ne j} \frac{E_i E_j |\sin\theta_{ij}|^{x}
(1-|\cos\theta_{ij}|)^{1-x}}{(\sum_i E_i)^2} 
\Theta\left[({\vec q}_i\cdot \vec n_T) ({\vec q}_j \cdot \vec n_T)\right]\>,  
\end{equation}
where the sum runs over all particles in the event, $\theta_{ij}$
denotes the angle between particle $i$ and $j$ and $\vec n_T$ is the
thrust axis.

As for the BKS class, these observables are IRC-safe for all values of
$x< 2$, and they vanish in the two-jet limit. The $\Theta$-function in
the definition
serves to eliminate recoil corrections that would otherwise have
entered in the term of the sum that involves both hard partons. Its
particular argument is designed so as to ensure the observable is
non-zero for all large-angle 3-jet configurations.

For $x <1$ one can verify that both the NLL and non-perturbative
properties are identical to those of the BKS class. It is therefore
most interesting to show results of numerical studies at the BKS
transition point, $x=1$ and beyond, for $x>1$.
We consider here then as examples $FC_{1}$ and $FC_{3/2}$.

The numerical analysis of these observables allows us (as usual!) to
establish immediately that they satisfy all applicability conditions
needed to achieve NLL accuracy in the resummation. 
The dependence on a single emission is associated with the
coefficients in Table~\ref{tab:FCsingle-emission}.
\TABLE{
\hspace{+1.2cm}$FC_{1}$\hspace{6.8cm}$FC_{3/2}$\\
\vspace{.3cm}
\begin{tabular}{| c | c | c | c | c |}
 \hline
 leg $\ell$ & $a_{\ell}$ & $b_{\ell}$ & $g_{\ell}(\phi)$ & $d_{\ell}$
\\
 \hline
 \hline
1 &  1.000 & $  0.000$ & $1.000$ &    1.000  \\
 \hline
2 &  1.000 & $  0.000$ & $1.000$ &    1.000  \\
 \hline
 \end{tabular}
\hspace{0.5cm}  
 \begin{tabular}{| c | c | c | c | c | c |}
 \hline
 leg $\ell$ & $a_{\ell}$ & $b_{\ell}$ & $g_{\ell}(\phi)$ & $d_{\ell}$
\\
 \hline
 \hline
1 &  1.000 & $  -0.500$ & $1.000$ &    1.000  \\
 \hline
2 &  1.000 & $  -0.500$ & $1.000$ &    1.000  \\
 \hline
\end{tabular}
\caption{Leg parametrisation coefficients for $FC_{1}$ (left) 
  and $FC_{3/2}$ (right).}
\label{tab:FCsingle-emission}
}
Of particular interest here, is that $FC_{3/2}$ constitutes an example
of an observable whose $b_\ell$ coefficients are negative. This means
that, for fixed transverse momenta, collinear emissions are more
important than large angle ones. We are not aware of any other
observables (other than trivial modifications of
eq.~(\ref{eq:FCx.def})) that have this property. It turns out that for
all $x<2$, $a=1$ and $b=1-x$, \ie there is a continuous transition
through $x=1$.

The other interesting property of these observables is that they are
all additive, independently of the value of $x$. This suggests that
the perturbative prediction will remain well-behaved across the whole
range of $x$, allowing one in particular to examine the region around
$x=1$.

The additivity also has interesting consequences for the
non-perturbative properties of these observables.
Specifically for all event shapes for which leading $1/Q$ power
corrections have been computed, it turns out that non-perturbative
corrections can be parametrised in terms of one single parameter,
which in the dispersive approach~\cite{DMW} can be expressed in terms
of the average value of the coupling constant below an infrared
matching scale $\mu_I$
\begin{equation}
\label{eq:alpha0}
\alpha_0 = \frac{1}{\mu_I}\int_0^{\mu_I} dk\, \alpha_s(k)\>. 
\end{equation}
(After merging perturbative and non-perturbative results, the answer
does of course not depend on the value of $\mu_I$.)
Testing the universality pattern of non-perturbative emissions reduces
then to verifying that $\alpha_0$ extracted from fits to distributions of
different observables has the same value.

As with the BKS observables, for $x<1$, the coefficient of the power
correction will go as $1/(1-x)$. However because of their additive
nature, it is to be expected that $FC_x$ observables will maintain
this behaviour up to a somewhat larger value of $x$, possibly giving a
$\ln Q /Q$ rather than a $1/Q$ corrections in the limit $x=1$.
Following the arguments of \cite{Milan} (originally applied to the
jet-broadening, for which a more sophisticated analysis subsequently
turned out to be necessary) this would suggest that non-perturbative
corrections to $FC_{1}$ will depend both on $\alpha_0$ and on a higher
moment of the coupling
\begin{equation}
  \label{eq:alpha0p}
  \alpha'_0 = \frac{1}{\mu_I}\int_0^{\mu_I} dk\,
  \alpha_s(k)\ln\frac{k}{\mu_I}\>.
\end{equation}
The observables with $x>1$ would also be interesting to study from the
non-perturbative point of view: $b=1-x<0$ implies that the
non-perturbative correction will come dominantly from the collinear
region (as opposed to the large-angle region, as is usually the case
for event shapes), potentially involving a fractional moment of the
coupling. This is a region which has not so far received much
attention in analytical studies of non-perturbative effects in
final-state observables and deserves to be further investigated.



\begin{thebibliography}{99}

\bibitem{CSS}
J.~C.~Collins, D.~E.~Soper and G.~Sterman,
\npb{250}{1985}{199}.

\bibitem{CTTW}
S.~Catani, L.~Trentadue, G.~Turnock and B.~R.~Webber,
\npb{407}{1993}{3}.

\bibitem{Bonciani:2003nt}
R.~Bonciani, S.~Catani, M.~L.~Mangano and P.~Nason,
\plb{575}{2003}{268}
[hep-ph/0307035].

\bibitem{Herwig}
G.~Abbiendi, I.~G.~Knowles, G.~Marchesini, M.~H.~Seymour, L.~Stanco
and  B.~R.~Webber,    
\cpc{67}{1992}{465};
G.~Corcella {\it et al.},
\jhep{01}{2001}{010}
[hep-ph/0011363].

\bibitem{Pythia}
T.~Sj\"ostrand,
\cpc{82}{1994}{74};
T.~Sj\"ostrand, P.~Eden, C.~Friberg, L.~L\"onnblad, G.~Miu, S.~Mrenna
and E.~Norrbin,
\cpc{135}{2001}{238}
[hep-ph/0010017].

\bibitem{FrixWeb}
S.~Frixione and B.~R.~Webber,
\jhep{06}{2002}{029}
[hep-ph/0204244].

\bibitem{CataniSeymour}
S.~Catani and M.~H.~Seymour,
\npb{485}{1997}{291}
[Erratum \ibid{B 510}{1997}{503}]
[hep-ph/9605323];
\plb{378}{1996}{287}
[hep-ph/9602277].


\bibitem{Disaster}
D.~Graudenz,
hep-ph/9710244.

\bibitem{NLOJET}
Z.~Nagy,
\prl{88}{2002}{122003}
[hep-ph/0110315].

\bibitem{MCFM}
J.~Campbell and R.~K.~Ellis,
\prd{65}{2002}{113007}
[hep-ph/0202176].

\bibitem{CMW} 
S.~Catani, B.~R.~Webber and G.~Marchesini,
\npb{349}{1991}{635};
Yu.~L.~Dokshitzer, V.~A.~Khoze and S.~I.~Troyan,
\prd{53}{1996}{89} 
[hep-ph/9506425].



\bibitem{NG1} 
M.~Dasgupta and G.~P.~Salam,
\plb{512}{2001}{323} 
[hep-ph/0104277];
\jhep{03}{2002}{017}
[hep-ph/0203009].


\bibitem{thr_res}
S.~Catani, G.~Turnock, B.~R.~Webber and L.~Trentadue,
\plb{263}{1991}{491}.

\bibitem{mh_ee}
S.~Catani, G.~Turnock and B.~R.~Webber,
\plb{272}{1991}{368}.


\bibitem{cpar_res}
S.~Catani and B.~R.~Webber,
\plb{427}{1998}{377}
[hep-ph/9801350].


\bibitem{CTWbroad}
S.~Catani, G.~Turnock and B.~R.~Webber,
\plb{295}{1992}{269}.

\bibitem{DLMSBroadPT}
Y.~L.~Dokshitzer, A.~Lucenti, G.~Marchesini and G.~P.~Salam,
\jhep{01}{1998}{011}
[hep-ph/9801324].

\bibitem{y3-kt_ee}
S.~Catani, Yu.~L.~Dokshitzer, M.~Olsson, G.~Turnock and B.~R.~Webber,
\plb{269}{1991}{432}.

\bibitem{CatDokWeb}
S.~Catani, Y.~L.~Dokshitzer and B.~R.~Webber,
\plb{322}{1994}{263}.

\bibitem{JetsDisSchmell}
G.~Dissertori and M.~Schmelling,
\plb{361}{1995}{167}.

\bibitem{BSZ} 
A.~Banfi, G.~P.~Salam and G.~Zanderighi,
\jhep{01}{2002}{018} 
[hep-ph/0112156].

\bibitem{BKS03}
C.~F.~Berger, T.~Kucs and G.~Sterman,
\ijmpa{18}{2003}{4159} 
[hep-ph/0212343];
\prd{68}{2003}{014012}
[hep-ph/0303051].

\bibitem{ADS}
V.~Antonelli, M.~Dasgupta and G.~P.~Salam,
\jhep{02}{2000}{001}
[hep-ph/9912488].

\bibitem{DSBroad}
M.~Dasgupta and G.~P.~Salam,
\epjc{24}{2002}{213}
[hep-ph/0110213].

\bibitem{eeKout} 
A.~Banfi, G.~Marchesini, Yu.~L.~Dokshitzer and G.~Zanderighi,
\jhep{07}{2000}{002} 
[hep-ph/0004027]; 
\jhep{05}{2001}{040} 
[hep-ph/0104162].

\bibitem{KoutZ0}
A.~Banfi, G.~Marchesini, G.~Smye and G.~Zanderighi,
\jhep{08}{2001}{047}
[hep-ph/0106278].

\bibitem{KoutDIS}
A.~Banfi, G.~Marchesini, G.~Smye and G.~Zanderighi,
\jhep{11}{2001}{066}
[hep-ph/0111157].

\bibitem{AzimDIS}
A.~Banfi, G.~Marchesini and G.~Smye,
\jhep{04}{2002}{024}
[hep-ph/0203150].

\bibitem{GRthrust}
E.~Gardi and J.~Rathsman,
\npb{609}{2001}{123}
[hep-ph/0103217].

\bibitem{GRmass}
E.~Gardi and J.~Rathsman,
\npb{638}{2002}{243}
[hep-ph/0201019].

\bibitem{GardMan}
E.~Gardi and L.~Magnea,
\jhep{08}{2003}{030}
[hep-ph/0306094].

\bibitem{BergerMagnea}
C.~F.~Berger and L.~Magnea,
hep-ph/0407024.


\bibitem{JADE}
W.~Bartel {\it et al.}  [JADE Collaboration],
\zpc{33}{1986}{23}.


\bibitem{JadeDL}
N.~Brown and W.~J.~Stirling,
\plb{252}{1990}{657}. 

\bibitem{CataniEtAlJets}
S.~Catani, B.~R.~Webber, Y.~L.~Dokshitzer and F.~Fiorani,
\npb{383}{1992}{419};
S.~Catani,
CERN-TH-6281-91,
in Erice 1991, Proceedings, QCD at 200-TeV, p.~21.

\bibitem{Leder}
G.~Leder,
\npb{497}{1997}{334}
[hep-ph/9610552].

\bibitem{DiscontGlobal}
M.~Dasgupta and G.~P.~Salam,
\jhep{08}{2002}{032}
[hep-ph/0208073].

\bibitem{MP} 
D.~H.~Bailey, ``A Portable High Performance
Multiprecision Package'', NASA Ames RNR Technical Report RNR-90-022; 
``A Fortran-90 Based Multiprecision System'', RNR Technical Report
RNR-94-013. 

\bibitem{BottsSterman}
J.~Botts and G.~Sterman,
\npb{325}{1989}{62};


\bibitem{KS}
N.~Kidonakis and G.~Sterman,
\plb{387}{1996}{867}
\npb{505}{1997}{321}
[hep-ph/9705234].

\bibitem{KOS}
N.~Kidonakis, G.~Oderda and G.~Sterman,
\npb{531}{1998}{365} 
[hep-ph/9803241]. 

\bibitem{Oderda}
G.~Oderda,
\prd{61}{2000}{014004}
[hep-ph/9903240].

\bibitem{KidonakisOwens}
N.~Kidonakis and J.~F.~Owens,
\prd{63}{2001}{054019}
[hep-ph/0007268].

\bibitem{BSZhh} A.~Banfi, G.~P.~Salam and G.~Zanderighi, 
hep-ph/0407287. 

\bibitem{BS03}
C.~F.~Berger and G.~Sterman,
\jhep{09}{2003}{058}
[hep-ph/0307394].

\bibitem{qcd-caesar.org} Analyses and resummed results for a large
  number of observables are available from
  \texttt{http://qcd-caesar.org/}~.

\bibitem{BSZ03}
A.~Banfi, G.~P.~Salam and G.~Zanderighi,
\plb{584}{2004}{298} 
[hep-ph/0304148].


\bibitem{ThrustDef} 
E.~Farhi,
\prl{39}{1977}{1587}.

\bibitem{ml_ee}
S.~J.~Burby and E.~W.~Glover,
\jhep{04}{2001}{029} 
[hep-ph/0101226].


\bibitem{FoxWolfram}
A.~V.~Manohar and M.~B.~Wise,
\plb{344}{1995}{407} 
[hep-ph/9406392].


\bibitem{JetratesHH}
S.~Catani, Y.~L.~Dokshitzer, M.~H.~Seymour and B.~R.~Webber,
\npb{406}{1993}{187}; 
S.~D.~Ellis and D.~E.~Soper,
\prd{48}{1993}{3160} 
[hep-ph/9305266]. 

\bibitem{ESWbook}
R.~K.~Ellis, W.~J.~Stirling and B.~R.~Webber,
``QCD And Collider Physics,''
Cambridge University Press, 1996.

\bibitem{Coherence}
V.~S.~Fadin,
\yf{37}{1983}{408} [\sjnp{37}{1983}{245}];\\
%
B.~I.~Ermolaev and V.~S.~Fadin,
\jetpl{33}{1981}{269}
[Pisma Zh.\ Eksp.\ Teor.\ Fiz.\  {\bf 33} (1981) 285];\\
%
A.~H.~Mueller,
\plb{104}{1981}{161};\\
%
Y.~L.~Dokshitzer, V.~S.~Fadin and V.~A.~Khoze,
\zpc{15}{1982}{325};\\
A.~Bassetto, M.~Ciafaloni and G.~Marchesini,
\prep{100}{1983}{201}.

\bibitem{KodairaTrentadue}
J.~Kodaira and L.~Trentadue,
\plb{112}{1982}{66};
J.~Kodaira and L.~Trentadue,
\plb{123}{1983}{335}.

\bibitem{CataniTrentadue}
S.~Catani and L.~Trentadue,
\npb{327}{1989}{323}.

\bibitem{Milan1}
Y.~L.~Dokshitzer, A.~Lucenti, G.~Marchesini and G.~P.~Salam,
\npb{511}{1998}{396}
[Erratum \ibid{B 593}{2001}{729}]
[hep-ph/9707532].

\bibitem{Moch:2002sn}
S.~Moch, J.~A.~M.~Vermaseren and A.~Vogt,
\npb{646}{2002}{181}
[hep-ph/0209100];
\npb{688}{2004}{101}
[hep-ph/0403192];
hep-ph/0404111.

\bibitem{Berger:2002sv}
C.~F.~Berger,
\prd{66}{2002}{116002}
[hep-ph/0209107].

\bibitem{CollinsSoper}
J.~C.~Collins and D.~E.~Soper,
\npb{193}{1981}{381}
[Erratum \ibid{B 213}{1983}{545}];
J.~C.~Collins and D.~E.~Soper,
\npb{197}{1982}{446}.


\bibitem{DGLAP}
V.N.~Gribov and L.N.~Lipatov, 
\sjnp{15}{1972}{438};
G.~Altarelli and G.~Parisi, 
\npb{126}{1977}{298};
Yu.L.~Dokshitzer, 
\jetp{46}{1977}{641}.

\bibitem{RakowWebber}
P.~E.~L.~Rakow and B.~R.~Webber,
\npb{187}{1981}{254}.


\bibitem{ParPet}
G.~Parisi and R.~Petronzio,
\npb{154}{1979}{427}.

\bibitem{ExpMath} See for example J.~Borwein and D.~Bailey,
\emph{Mathematics by Experiment: Plausible Reasoning in the 21st
  Century}, A.~K.~ Peters, Natick, Massachussetts, 2003. Also:
   \texttt{http://www.expmath.info}~.

\bibitem{ExpertSyst}
P.~Jackson. ``Introduction to Expert Systems'', 
Harlow, England, Addison Wesley Longman, 1999. 

\bibitem{ERT}
R.~K.~Ellis, D.~A.~Ross and A.~E.~Terrano,
\npb{178}{1981}{421}.


\bibitem{Nagy03}
Z.~Nagy,
\prd{68}{2003}{094002}
[hep-ph/0307268].

\bibitem{TriRad}
W.~B.~Kilgore and W.~T.~Giele,
{\it High energy physics}, Osaka 2000, Vol. 1, 502
[hep-ph/0009193];
\prd{55}{1997}{7183}
[hep-ph/9610433].

\bibitem{D0Thrust}
I.~A.~Bertram  [D0 Collaboration],
\appol{B 33}{2002}{3141}.

\bibitem{BSZPrep} A.~Banfi, G.~P.~Salam and G.~Zanderighi, work in progress.

\bibitem{BMS}
A.~Banfi, G.~Marchesini and G.~Smye,
\jhep{08}{2002}{006}
[hep-ph/0206076].

\bibitem{ApplebySeymour}
R.~B.~Appleby and M.~H.~Seymour,
\jhep{12}{2002}{063}
[hep-ph/0211426];
\jhep{09}{2003}{056}
[hep-ph/0308086].

\bibitem{BanfiDasgupta}
A.~Banfi and M.~Dasgupta,
\jhep{01}{2004}{027}
[hep-ph/0312108].

\bibitem{WeigertNG}
H.~Weigert,
\npb{685}{2004}{321}
[hep-ph/0312050].

\bibitem{KraussRodrigo}
F.~Krauss and G.~Rodrigo,
\plb{576}{2003}{135}
[hep-ph/0303038].

\bibitem{DelphiOriented}
P.~Abreu {\it et al.}  [DELPHI Collaboration],
\epjc{14}{2000}{557}
[hep-ex/0002026].

\bibitem{Geneva}
S.~Bethke, Z.~Kunszt, D.~E.~Soper and W.~J.~Stirling,
\npb{370}{1992}{310}
[Erratum \ibid{B 523}{1998}{681} [hep-ph/9803267]].

\bibitem{CTEQhandbook}
R.~Brock {\it et al.}  [CTEQ Collaboration],
``Handbook of perturbative QCD: Version 1.0,''
\rmp{67}{1995}{157}.

\bibitem{BroadDef}
P.~E.~L.~Rakow and B.~R.~Webber,
\npb{191}{1981}{63}.

\bibitem{DokWebDists}
Y.~L.~Dokshitzer and B.~R.~Webber,
\plb{404}{1997}{321}
[hep-ph/9704298].

\bibitem{BroadNP-rev}
Y.~L.~Dokshitzer, G.~Marchesini and G.~P.~Salam,
\epjd{1}{1999}{3} 
[hep-ph/9812487].


\bibitem{DMW} Yu.~L.~Dokshitzer, G.~Marchesini and B.~R.~Webber,
\npb{469}{1996}{93}  [hep-ph/9512336]; 
see also M.~Beneke,
\prep{317}{1999}{1} 
[hep-ph/9807443].

\bibitem{Milan}
Y.~L.~Dokshitzer, A.~Lucenti, G.~Marchesini and G.~P.~Salam,
\jhep{05}{1998}{003} 
[hep-ph/9802381].


\end{thebibliography}
\end{document}